\RequirePackage{fix-cm} 
\documentclass[oneside,phd]{snuphyathesis} 

\usepackage{amsmath,amssymb,amsfonts}
\usepackage{hyperref}
\usepackage{cite}
\allowdisplaybreaks 

\usepackage[titles]{tocloft} 
\makeatletter 
\if@snu@ko
	\renewcommand\cftchappresnum{제~}
	\renewcommand\cftchapaftersnum{~장}
	\renewcommand\cftfigpresnum{그림~}
	\renewcommand\cfttabpresnum{표~}
\else
	\renewcommand\cftchappresnum{Chapter~}
	\renewcommand\cftfigpresnum{Figure~}
	\renewcommand\cfttabpresnum{Table~}
\fi
\makeatother 
\newlength{\mytmplen}
\ifpdf
	\input glyphtounicode\pdfgentounicode=1 
	\usepackage[pdftex]{graphicx}
\else
	\usepackage[dvipdfmx]{graphicx}
\fi

\usepackage{array}
\usepackage{multirow}
\usepackage[table]{xcolor}
\usepackage{ctable}
\usepackage{booktabs}		

\newcommand{\eq}{\begin{equation}}
\newcommand{\eqe}{\end{equation}}
\newcommand{\eqa}{\begin{eqnarray}}
\newcommand{\eqae}{\end{eqnarray}}
\newcommand{\nn}{\nonumber}
\newcommand{\bn}{\begin{enumerate}}
\newcommand{\en}{\end{enumerate}}

\newcommand{\eqc}[1]{(\ref{#1})}

\def\identity{{\rlap{1} \hskip 1.6pt \hbox{1}}}
\def\iden{\identity}

\def\IC{\mathbb{C}}

\def\IP{\mathbb{P}}
\def\IR{\mathbb{R}}

\def\CA{{\cal A}}

\def\CD{{\cal D}}

\def\CF{{\cal F}}

\def\CM{{\cal M}}

\def\CO{{\cal O}}

\def\CR{{\cal R}}

\def\a{\alpha}
\def\b{\beta}
\def\g{\gamma}
\def\e{\epsilon}
\def\ve{\varepsilon}
\def\z{\zeta}
\def\th{\theta}

\def\k{\kappa}
\def\l{\lambda}
\def\m{\mu}
\def\n{\nu}

\def\r{\rho}

\def\s{\sigma}

\def\t{\tau}

\def\w{\omega}
\def\G{\Gamma}
\def\D{\Delta}

\def\L{\Lambda}
\def\S{\Sigma}

\def\O{\Omega}

\def\half{\frac{1}{2}}

\def\p{\partial}

\def\identity{{\rlap{1} \hskip 1.6pt \hbox{1}}}

\def\tr{{\rm tr}}

\newcommand{\bfig}{\begin{figure}}
\newcommand{\efig}{\end{figure}}

\def\la{{\langle}}
\def\ra{{\rangle}}
\def\abs#1{{\left| #1 \right|}}

\def\bl#1\el{\begin{align} #1 \end{align}}
\def\bg#1\eg{\begin{gather} #1 \end{gather}}
\newcommand{\fig}[1]{figure \ref{#1}}
\def\bld#1\eld{\begin{aligned} #1 \end{aligned}}
\def\bgd#1\egd{\begin{gathered} #1 \end{gathered}}

\newcommand{\bra}[1]{\langle{#1}|}
\newcommand{\ket}[1]{|{#1}\rangle}

\newcommand{\fbra}[1]{ ({#1} |}
\newcommand{\fket}[1]{ | {#1} )}

\newcommand{\sbra}[1]{ [{#1} |}
\newcommand{\sket}[1]{ | {#1} ]}

\newcommand{\fsl}[1]{{\ooalign{\(#1\)\cr\hidewidth\(/\)\hidewidth\cr}}}

\renewcommand{\tt}{\texttt}
\renewcommand{\bf}{\textbf}

\newcommand{\AB}[1]{\langle #1 \rangle}
\newcommand{\SB}[1]{[ #1 ]}
\newcommand{\MixLeft}[3]{\langle #1 | #2 | #3 ]}
\newcommand{\BS}[1]{\boldsymbol{#1}}
\newcommand{\RAB}[1]{| #1 \rangle}
\newcommand{\LAB}[1]{\langle #1 |}

\newcommand{\LSB}[1]{[ #1 |}

\newcommand{\RN}[1]{%
  \textup{\uppercase\expandafter{\romannumeral#1}}%
}

\newcommand*{\mathcolor}{}
\def\mathcolor#1#{\mathcoloraux{#1}}
\newcommand*{\mathcoloraux}[3]{%
  \protect\leavevmode
  \begingroup
    \color#1{#2}#3%
  \endgroup
}

\usepackage{lipsum} 
\usepackage[toc]{appendix}

\title{Quantum point particle approximation of \\ spinning black holes and compact stars}
\title*{회전하는 블랙홀과 밀집성의 양자 점입자 근사}

\author{김정욱}
\author*{김~정~욱} 

\studentnumber{2017-36622}

\advisor{이상민}
\advisor*{이~상~민}

\graddate{JULY 2020}

\submissiondate{2020~년~7~월}

\approvaldate{2020~년~7~월}

\committeemembers%
{이원종}%
{이상민}%
{강궁원}%
{김~~~석}%
{김형도}

\begin{document}
\pagenumbering{Roman}
\makefrontcover
\makeapproval

\cleardoublepage
\pagenumbering{roman}

\begin{abstract}
\begin{flushright}
\vskip -16pt
Jung-Wook Kim\\
Department of Physics and Astronomy\\
The Graduate School\\
Seoul National University
\end{flushright}
\noindent
Gravitational wave observatories targeted for compact binary coalescence, such as LIGO and VIRGO, require various theoretical inputs for their efficient detection. One of such inputs are analytical description of binary dynamics at sufficiently separated orbital scales, commonly known as post-Newtonian dynamics. One approach for determining such two-body effective Hamiltonians is to use quantum scattering amplitudes.

This dissertation aims at an improved understanding of classical physics of spinning bodies in quantum scattering amplitudes, for application to the problem of effective two-body Hamiltonians. The main focus will be on spin-induced higher-order multipole moments
. In this dissertation results for the first post-Minkowskian order (linear in Newton's constant $G$ and to all orders in relative momentum $p^2$) Hamiltonian that is valid for arbitrary compact spinning bodies to all orders in spin is presented. Next, obstruction and prospects for the formulation's extension to second post-Minkowskian order is discussed, based on an equivalent loop order quantum field theory computations.

This dissertation is based on the works~\cite{Chung:2018kqs,Chung:2019duq,Chung:2019yfs,Chung:2020rrz}.
\end{abstract}

\tableofcontents
\listoffigures
\listoftables

\cleardoublepage
\pagenumbering{arabic}

\chapter{Introduction}
On 11th February 2016, LIGO and Virgo collaborations announced the first direct observation of gravitational waves~\cite{Abbott:2016blz}. Together with electromagnetic waves and neutrinos, three quarters of the fundamental interactions ever known to humanity has now become the eyeglasses through which we observe the skies.

The signal observed by the observatories has been named as the event GW150914, which has been identified as coming from the coalescence of two black holes. The signal increases in amplitude and frequency in about 8 cycles from 35 to 150 Hz, with its duration about 0.2 seconds. Initial estimate for the total mass of the binary was $M = m_1 + m_2 \gtrsim 70$ solar masses ($M_{\odot}$) in the detector frame, bounding the sum of Schwarzschild radii as $2GM/c^2 \gtrsim 210$ km. For comparison, equal Newtonian point masses orbiting at a frequency of 75 Hz (half the frequency of gravitational wave) would be separated $\simeq 350$ km apart. The only known objects compact enough to reach such an orbital frequency without contact are black holes~\cite{Abbott:2016blz}, and their masses were estimated to be $35.8^{+5.3}_{-3.9} \, M_{\odot}$ and $29.1^{+3.8}_{-4.3} \, M_{\odot}$ in the source frame at the time of detection~\cite{TheLIGOScientific:2016wfe}.

The collaborations utilised two search methods to detect gravitational waves. The first search method, generic transient search, makes minimal assumptions on gravitational waves~\cite{TheLIGOScientific:2016uux}. Importantly, the method does not make any assumptions on the waveform of the gravitational waves. The search begins by removing all possible noise events, known as glitches, and identifies the remaining signal that cannot be attributed to background noise as the signal of gravitational waves. The approach can be succinctly summarised through the words of Sherlock Holmes; ``when you have eliminated the impossible, whatever remains, however improbable, must be the truth''\footnote{Sir Arthur Conan Doyle, \textit{The Sign of the Four}.}.

The second search method, binary coalescence search, assumes that the gravitational waves are generated by binary systems evolving according to Einstein's theory of gravitation~\cite{TheLIGOScientific:2016qqj}. The search begins by generating a table of expectations for gravitational waves called templates, and calculates how likely a given template correctly describes the signal that has been detected. The templates are generated using waveform models, which combines results from post-Newtonian approach with black hole perturbation theory and numerical relativity into effective-one-body formalism. Also, the method provides estimates for the properties of the merger event, which can be used to study various astrophysical problems such as tidal responses of neutron stars~\cite{Flanagan:2007ix,TheLIGOScientific:2017qsa}. Because this search method relies on how accurate our expectation is for the gravitational wave signals, it is important to have an accurate input for the waveform models.

The gravitational waves that will be detected by the observatories can be divided into two parts; the \emph{coalescence phase} corresponds to the last few cycles before merger of the two bodies, while the \emph{inspiral phase} corresponds to two compact bodies slowly approaching each other as they spiral around one another. It is expected that several thousand cycles of the inspiral phase could be detected by future interferometors, allowing precision measurements for the wave's phase $\Phi = 2 \pi \int f dt$~\cite{Cutler:1992tc}. Therefore precise measurement for the wave's phase during the inspiral phase qualifies as a precision test of general relativity.

The inspiral phase of binary coalescence can be described by post-Newtonian (PN) dynamics, which allows semi-analytical treatment of the binary's motion. PN dynamics aim at a systematic approximation of general relativity where deviations from Newtonian gravity are given as perturbative series of relativistic corrections. While 1.5 PN dynamics are considered sufficient for the waveform templates when \emph{searching} for the signals, more accurate waveforms will be needed when \emph{extracting properties} of the event~\cite{Cutler:1992tc}; for example, while the \emph{chirp mass} $\CM \equiv (m_a m_b)^{3/5} (m_a + m_b)^{-1/5}$ is expected to be determined to $0.1\% - 1\%$ accuracy when effects up to 1.5PN are considered, the uncertainty for the \emph{reduced mass} $\m \equiv m_a m_b/(m_a + m_b)$ is expected to be as large as $50\%$ at the same PN order~\cite{Cutler:1994ys}. The large discrepancy between estimated accuracy of both variables is due to the spin of binary constituents, and assuming smallness of individual spins ($\lesssim 0.01 \, m_i^2$ in geometrised units, respectively) allows $\m$ to be determined to $1\%$ accuracy~\cite{Cutler:1994ys}.

Therefore it is desirable to have a better understanding of spin effects in PN dynamics. In fact, initial estimates for the properties of the event GW150914~\cite{TheLIGOScientific:2016wfe} have been refined in later investigations~\cite{Abbott:2016izl} by including precession effects from spins of the binary constituents in waveform generators, improving consistency of estimates from different waveform models used to infer the properties of the merger.

This dissertation makes contact with one of the inputs of precision waveform models for gravitational wave detection; effects of spin on PN 
dynamics of gravitating two-body systems. The main focus of the dissertation will be understanding how scattering amplitudes of quantum field theory contains classical physics of spinning bodies, especially on the effects of spin-induced higher-order multipoles. Improved understanding of the problem can then be applied to the problem of PN dynamics of spinning compact binaries.

While various approaches have been employed to obtain PN dynamics\footnote{The interested reader may consult the reviews such as~\cite{Blanchet:2013haa,Schafer:2018kuf,Barack:2018yvs}.}, this dissertation will review and refine the techniques inspired from ``quantum gravity''; effective field theory (EFT) approach and quantum scattering amplitudes approach, mostly weighted towards the latter. In the EFT approach, the effective action is evaluated by integrating out gravitons exchanged between classical point particle sources~\cite{Goldberger:2004jt}. In the quantum scattering amplitudes approach, the effective interaction Hamiltonian is evaluated as an inverse problem of Born approximation for quantum scattering amplitudes~\cite{Iwasaki:1971vb}. Modern quantum scattering amplitudes approach offers an advantage over more traditional approaches in that only \emph{physical degrees of freedom} are considered in the computations.

A novel feature will be treatment of generic spin multipoles in quantum scattering amplitudes approach. A spinning star will be deformed from a spherical shape and develop multipole moments, and its spin-induced $2^n$-pole moment will be proportional to symmetric and traceless product of $n$ spin vectors $(S^\mu)^n$. The effects of such moments can be studied by considering spinning particles and identifying spin multipole matrix elements as classical spin-induced moments~\cite{Holstein:2008sx,Vaidya:2014kza}. The analyses were limited to hexadecapole ($S^4$) order, however, as inefficient Lagrangian description for massive higher spin fields~\cite{Singh:1974qz,Singh:1974rc} stymied extensions beyond this order: A spin-$s$ particle can only possess up to spin $2^{2s}$-pole moment, so describing beyond hexadecapole moment ($S^{>4}$) requires particles with spins $s>2$. The obstruction can be circumvented by working directly with on-shell states of massive higher spin particles~\cite{Chung:2018kqs}.

The rest of the dissertation is organised as follows. Chapter~\ref{chap:prelim} provides the background materials for this dissertation. Chapter~\ref{chap:tree} focuses on tree-level amplitudes and classical physics of black holes and compact stars that can be obtained from tree amplitudes, and presents the first post-Minkowskian order effective Hamiltonian \eqc{V-1PM-master} which includes spin effects to all orders. Chapter~\ref{chap:loop} extends the discussion to one-loop amplitudes for black holes, divided into three topics; the conceptual difference between massive higher-spin and lower-spin particles; eligibility of running massive higher-spin particles inside loops for classical physics; and the contributions to the one-loop amplitude relevant for the second post-Minkoskian order effective Hamiltonian.

Natural units $\hbar = c = 1$ are adopted throughout this dissertation, but these constants will appear explicitly in some discussions involving non-relativistic and/or classical limits. Materials of this dissertation are based on the works~\cite{Chung:2018kqs,Chung:2019duq,Chung:2019yfs,Chung:2020rrz}.

\chapter{Preliminaries} \label{chap:prelim}

\section{Effective field theory for binary dynamics}
\subsection{Overview of effective field theory for gravitational radiation from stellar binaries}
\bfig
\centering
\includegraphics[width=\linewidth,trim={0 9.9cm 4.45cm 0},clip]{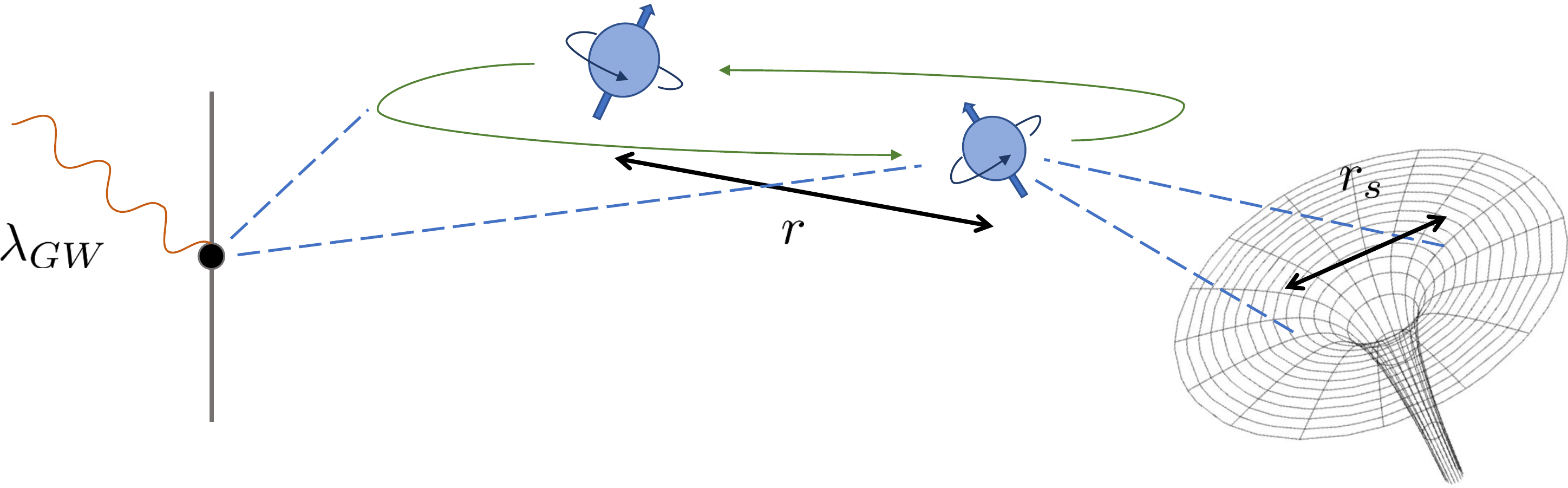}
\caption{Scale separation $\l_{GW} \gg r \gg r_s$ of a compact binary system emitting gravitational waves. $\l_{GW}$ is the wavelength of radiated graviational waves, $r$ is the separation between the gravitating bodies, and $r_s$ is the size of stellar objects. Image reproduced from ref.\cite{Porto:2016zng}.} \label{fig:scale}
\efig
Hierarchy of scales exists in a binary system of compact stars as illustrated in \fig{fig:scale}; typical wavelength of radiated gravitational waves $\l_{GW}$ is much larger than the typical separation between the binary constituents $r$, which is in turn much larger than the size of the stellar object $r_s$. This serves as the foundation of effective field theory (EFT) approach to post-Newtonian (PN) expansion of general relativity (GR), introduced by Goldberger and Rothstein~\cite{Goldberger:2004jt}. Spin and multipole moments in this context were first considered by Porto~\cite{Porto:2005ac}, and the formulation introduced by Levi and Steinhoff~\cite{Levi:2015msa} will be the starting point for the construction in this dissertation. Application of EFT to effective spinning two-body Hamiltonian can be found in refs.\cite{Porto:2005ac,Porto:2006bt,Porto:2008tb,Porto:2008jj,Levi:2008nh,Perrodin:2010dy, Porto:2010tr,Levi:2010zu,Levi:2014gsa,Levi:2015msa,Levi:2017kzq,Levi:2019kgk}; consult the reviews~\cite{Porto:2016pyg,Levi:2018nxp} for a more complete list of references.

This section aims to give only a conceptual overview on EFT approach, and detailed description is beyond the scope of this dissertation; the reader is referred to reviews and lecture notes on the subject~\cite{Goldberger:2007hy,Foffa:2013qca,Porto:2016pyg,Levi:2018nxp} for a more complete overview.

The conceptual starting point is the partition function of gravitational path integral, which can be schematically written as
\bl
Z &= \int [\CD h] [\CD \Phi] e^{ i \left( S_{\text{mat}}[\Phi, h] + S_{\text{grav}} [h] \right) } \,.
\el
The field $\Phi$ denotes matter fields and $h$ denotes graviton fields, defined by $g_{\m\n} = \eta_{\m\n} + \k h_{\m\n}$ with $\k = M_{\text{Pl}}^{-1} = \sqrt{32\pi G}$. For full rigour ghost field and gauge-fixing action needs to be included, but they are irrelevant for understanding the conceptual framework.

First we choose to disregard the physics on scales shorter than the scale of stellar objects $\le r_s$. All modes with wavelengths $\l \le r_s$ are ``integrated out'', leaving behind point particles moving along classical worldlines $x_i$ and worldline operators having the interpretation of multipole moments and/or internal excitations. This process can be schematically written as follows.
\bl
\int [\CD h_{\le r_s}] [\CD \Phi] e^{ i \left( S_{\text{mat}}[\Phi, h] + S_{\text{grav}} [h] \right) } &\simeq e^{i \left( \sum_i S_{\text{pp}}[x_i, h_{>r_s}] + S_{\text{grav}} [h_{>r_s}] \right)} \,.
\el
$S_{\text{pp}}[x_i, h_{>r_s}]$ is the effective action of the point-particle on the worldline $x_i$ over the graviton field background $h_{>r_s}$. In practice, RHS of the above equation is ususally the starting point for EFT approach where all worldline operators consistent with constraints of the problem are written down, ordered by relevance to the problem.

We can now initiate our study on the problem of compact binary dynamics, which is the main problem of interest in this dissertation. For this purpose the remaining modes of the graviton field $h_{>r_s}$ are decomposed into two groups; radiation modes $h_{\text{rad}}$ that propagate out to infinity and potential modes $h_{\text{pot}}$ that bind the stellar objects together. ``Integrating out'' potential modes results in the effective action $\G_{\text{eff}}[x_i, h_{\text{rad}}]$ of the radiating binary system.
\bl
\int [\CD h_{\text{pot}}] \left( e^{i \sum_i S_{\text{pp}}[x_i, h_{>r_s}] } \right) e^{i S_{\text{grav}} [h_{>r_s}] } &= e^{i \G_{\text{eff}}[x_i, h_{\text{rad}}]} 
\,.
\el
Diagrammatic tools of quantum field theory (QFT) can be utilised when effecting the ``integration over potential modes'' $[\CD h_{\text{pot}}]$, resulting in Feynman graphs and Feynman rules for non-relativistic general relativity (NRGR). The hard UV cut-off $r_s$ naturally regulates the UV divergences encountered in na\"ive computation of Feynman graphs at this stage, in contrast to other approaches where gravitational sources are literally understood as point sources.

The conservative part of the effective two-body dynamics, which will be the focus of this dissertation, is encoded in the effective action $\G_{\text{eff}}$ with radiation modes $h_{\text{rad}}$ set to zero. When computing the Feynman graphs for the effective action $\G_{\text{eff}}$, a parametrisation for the graviton field specialised for computing PN dynamics~\cite{Kol:2007bc} is usually employed, which generically leads to higher time derivatives at high perturbation orders. Such higher time derivative terms are eliminated by iterating lower perturbation order equations of motion, which is equivalent to redefinition of the coordinates\footnote{Iterating equations of motion is equivalent to adding multiples of Euler-Lagrange equations into the action. On the other hand, Euler-Lagrange equation is the coefficient of linear order coordinate variations of the action. Therefore the former can be traded for the latter and be interpreted as redefinition of the coordinates~\cite{Schafer:1984mr}.}. Finally, the effective Hamiltonian for the binary system is obtained by performing a Legendre transform on the effective action $\G_{\text{eff}}$ with higher time derivatives removed.

At the last stage of EFT approach, the multiple worldlines of gravitating bodies are matched onto multipoles of a single worldline describing the whole radiating system. The multipole moments of the whole radiating system is computed by solving for the motions of the constituents of the binary and adding up stress tensor contributions from the worldlines of point particles and potential graviton modes in the form of Landau-Lifshitz pseudo-tensor which accounts for nonlinearities of gravity. The worldline multipole moment operators of the radiating system is then matched onto the multipole moments of the computed total stress tensor. The so-called \emph{tail effects}, the effects caused by deformed geometry from flat space due to gravitating sources, can be incorporated in this picture leading to \emph{radiative multipole moments} or \emph{renormalised moments} which depends logarithmically on the (IR regulating) scale $\mu$. The scale dependence comes from the fact that gravitation is a long-range force, and renormalisation group methods can be applied for their computations.

\subsection{The point particle effective action} \label{sec:ppEFT}
The building blocks for the amplitudes will be constructed by matching onto effective action for point particles, which in turn will be used for computing PN dynamics. The following effective action for a relativistic spinning particle has been used by Porto~\cite{Porto:2005ac} for EFT computations.
\bl
S&=\int d\sigma \;\left\{-m\sqrt{u^2}-\frac{1}{2}S_{\mu\nu}\Omega^{\mu\nu} + \CO_{MM} \right\} \,. \label{eq:EFTmin}
\el
Here $u^\mu := \frac{dx^\mu}{d\sigma}$, $S_{\m\n} := J_{\m\n} - (x_\m p_\n - p_\m x_\n)$ is the rank-2 spin tensor, and $\O^{\m\n} := e^\m_A \frac{D e^{A\n}}{D\s}$ is the angular velocity. $e^{\m}_A (\s)$ is the tetrad attached to the worldline of the particle which parametrises the orientation of the body. The first two terms of the action are universal and referred to as \emph{minimal coupling} in the literature, but to avoid confusion they will be referred to as \emph{minimal terms} in this dissertation. Since the spin variable is defined as the total angular momentum $J_{\m\n}$ minus the orbital angular momentum $L_{\m\n} = x_\m p_\n - p_\m x_\n$, the variable is accompanied by gauge redundancies that corresponds to the freedom of chosing the centre of the body ($x^\m \to \tilde{x}^\m = x^\m + \delta x^\m$)~\cite{Levi:2015msa,Steinhoff:2015ksa}. The gauge-fixing conditions are known as \emph{spin supplementary conditions} (SSC), one of the popular choices being the covariant condition\footnote{This condition is also known as Tulczyjew condition~\cite{tulczyjew1959motion} in the literature.} $p^\mu S_{\mu\nu}=0$.

As mentioned previously, the point particle effective action contains worldline operators $\CO_{MM}$ that parametrise structures swept under the rug while ``integrating out'' short distance modes. We will limit our interest to spin-induced multipole moments $L_{SI}$, which parametrise the deformations from centrifugal effects of the body's spin\footnote{These are the multipole moments of a star in isolation which has settled down to an axisymmetric equilibrium state. Other examples of $\CO_{MM}$ include; multipole moments due to tidal deformations and dissipative effects~\cite{Goldberger:2005cd,Porto:2007qi,Chakrabarti:2013lua,Chakrabarti:2013xza,Steinhoff:2016rfi}.}. This dissertation's construction is based on the following parametrisation introduced by Levi and Steinhoff~\cite{Levi:2015msa}.
\bl
\bld
L_{SI} &= \sum_{n=1}^{\infty} \frac{(-1)^n}{(2n)!} \frac{C_{\text{ES}^{2n}} }{m^{2n-1}} D_{\m_{2n}} \cdots D_{\m_{3}} \frac{E_{\m_{1} \m_{2}} }{\sqrt{u^2}} S^{\m_{1}} S^{\m_{2}} \cdots S^{\m_{2n-1}} S^{\m_{2n}}
\\ &\phantom{=} + \sum_{n=1}^{\infty} \frac{(-1)^n}{(2n+1)!} \frac{C_{\text{BS}^{2n+1}} }{m^{2n}} D_{\m_{2n+1}} \cdots D_{\m_{3}} \frac{B_{\m_{1} \m_{2}} }{\sqrt{u^2}} S^{\m_{1}} S^{\m_{2}} \cdots S^{\m_{2n}} S^{\m_{2n+1}}\,.
\eld \label{eq:EFTSI}
\el
Here, $S^\m := - \frac{1}{2 \sqrt{p^2}} \e^{\m\n\l\s} p_\n S_{\l\s}$ is the spin vector which is naturally identified with the Pauli-Lubanski pseudo-vector, $C_{\text{ES}^{2n}}$ and $C_{\text{BS}^{2n+1}}$ (collectively $C_{\text{S}^n}$) are \emph{Wilson coefficients} normalised to unity for Kerr black holes (BH), $E$ and $B$ are the electric and magnetic components of the Weyl tensor\footnote{The vacuum Einstein equation reduces to $R_{\m\n} = 0$, therefore the Riemann tensor is equal to the Weyl tensor $R_{\m\n\l\s} = C_{\m\n\l\s}$ on this background.} defined as
\bl
\bld
E_{\m\n} &:= R_{\m\a\n\b} u^{\a} u^{\b}
\\ B_{\m\n} &:= \half \e_{\a\b\g\m} R^{\a\b}_{~~\delta \n} u^{\g} u^{\delta}\,,
\eld
\el
and the covariant derivatives act on the Riemann tensors. Because the electric and magnetic components of the Weyl tensor are traceless, the trace part of the spin products in \eqc{eq:EFTSI} is immaterial and the spin products can be considered as spin-induced multipoles. The above parametrisation for the multipole moments have the advantage that they are unaffected by gauge redundancies of spin variables.

\section{Analyticity properties of the S-matrix} \label{sec:anal_S-mat}
The probability amplitude for a scattering process $\a \to \b$ is defined as the overlap between the asymptotic in-state $\ket{\a}$ and the asymptotic out-state $\ket{\b}$. The \emph{S-matrix} $S$ is defined as the matrix whose elements in free particle basis correspond to the corresponding probability amplitudes.
\bl
\left\langle \b \middle| \a \right\rangle_{\hskip-37pt out \hskip24pt in} = \phantom{{}_{ffe}} \left\langle \b \middle| S \middle| \a \right\rangle_{\hskip-52pt free \hskip33pt free}
\el
The S-matrix can be evaluated by the following \emph{Dyson series},
\bl
S &= \iden + \sum_{n=1}^{\infty} \frac{(-i)^n}{n!} \int_{-\infty}^{+\infty} \prod_{i=1}^n dt_i T \left\{ H_{in} (t_1) \cdots H_{in} (t_n) \right\} \nn
\\ &= T \left\{ e^{-i \int_{-\infty}^{+\infty} H_{in}(t) dt} \right\} \label{eq:SDyson}
\\ H_{in}(t) &:= e^{i H_0 t} H_{in} e^{-i H_0 t} \,,
\el
where $T$ is the time-ordering operator and $H_{in}$ is the interaction Hamiltonian in the interaction picture. The free Hamiltonian in the interaction picture is denoted as $H_0$.

It is customary to define the \emph{T-matrix} as $S = \iden + i T$ and call delta-stripped elements $M$ of the T-matrix as \emph{scattering amplitudes}, or simply \emph{amplitudes}.
\bl
\left\langle \b \middle| S \middle| \a \right\rangle = i \left\langle \b \middle| T \middle| \a \right\rangle = i (2\pi)^4 {\delta}^{(4)} (P_\b - P_\a) M_{\a \to \b} \,. \label{eq:SADef}
\el
The subscript `free' has been suppressed in the above equation. It is assumed that $\a \neq \b$ and $P_{\a(\b)}$ denotes the four-momentum of the state $\ket{\a}$($\ket{\b}$), respectively. For abbreviation, we also introduce the notation $\tilde{\delta} = 2\pi \delta$.

Another common practice for studying analyticity properties of the S-matrix is to invoke \emph{crossing symmetry} and consider all particles as incoming(outgoing). Conversion from incoming to outgoing particle is done by flipping the sign of four-momentum $p^\m \to - p^\m$ and helicity $h \to - h$, and vice versa.

\subsection{Simple poles and one-particle states}
\bfig
\centering
\includegraphics[width=0.8\linewidth]{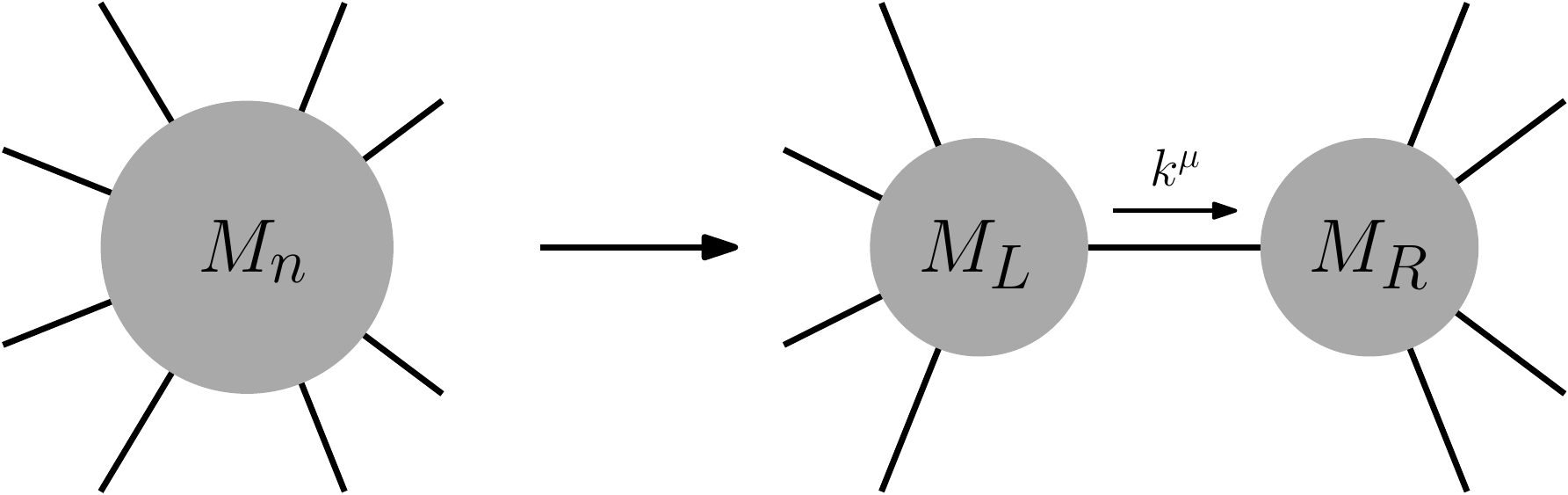}
\caption{Graphical representation of factorisation for a general on-shell scattering amplitude $M_n$. The internal line corresponds to an on-shell particle with momentum $k^\m$.} \label{fig:factorisation}
\efig
\emph{Polology} is a statement about pole structures of S-matrix elements. It states that when an intermediate one-particle state go on-shell with momentum $k^\m$, the amplitude possesses a pole at the on-shell condition $k^2 - m^2 = 0$ and the residue is given as a product of two subamplitudes $M_L$ and $M_R$.
\bl
M_n \to \frac{M_L \times M_R}{- k^2 + m^2 - i0^+} + \cdots \,. \label{eq:factDef}
\el
This property commonly called \emph{factorisation} can be graphically represented as in \fig{fig:factorisation}. It is important to note that the one-particle state that becomes on-shell on this pole does \emph{not} need to be an elementary particle, e.g. pions. The property can be understood as the result of projecting over intermediate one-particle state as in \fig{fig:polpf}.
\bfig
\centering
\includegraphics[width=0.8\linewidth]{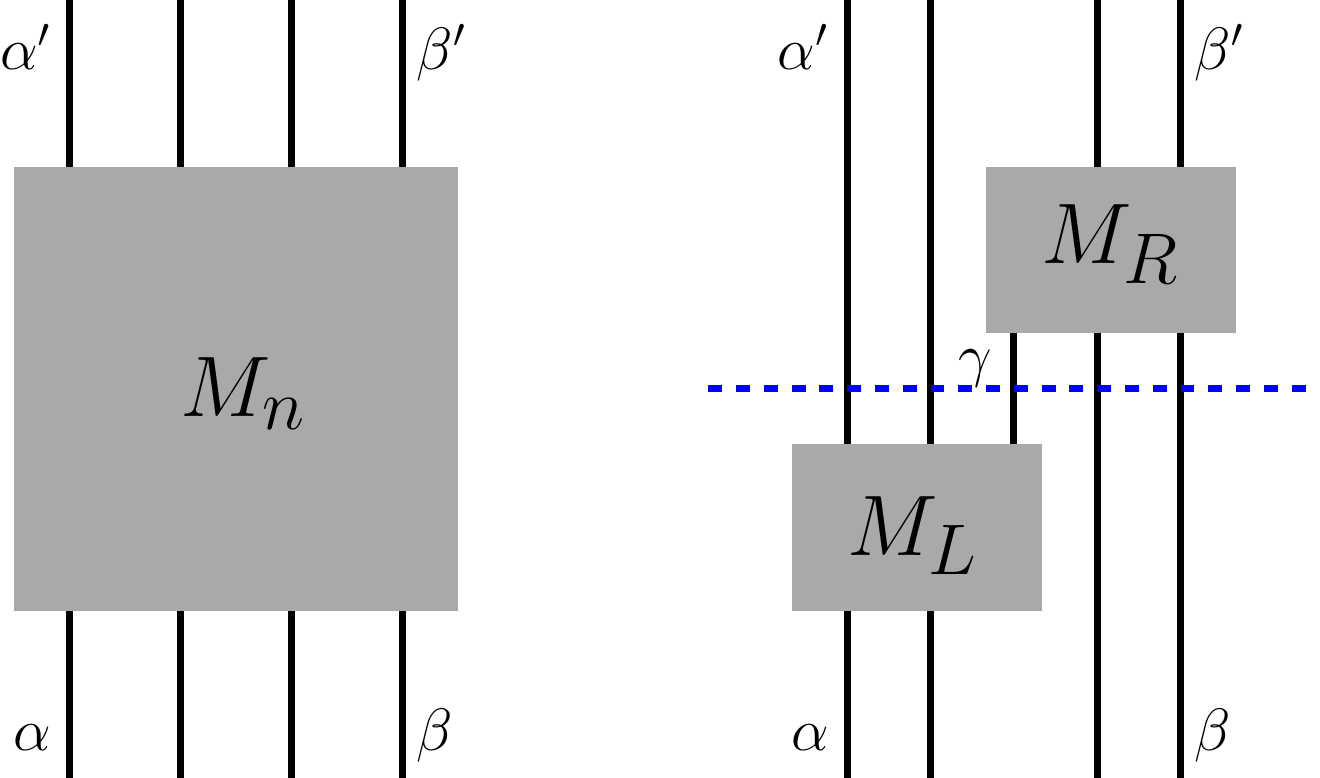}
\caption{The factorised amplitude $M_L \times M_R$ on the right side is obtained from the full amplitude $M_n$ by projecting over the state $\ket{\a'\b\g}$, inserted on the dashed line $t=0$. Time flows from bottom to top.} \label{fig:polpf}
\efig

Its proof can be sketched as follows. As depicted in \fig{fig:polpf}, consider the amplitude for the process $\a\b \to \a'\b'$. Define the ``half S-matrices'' $S_+$ and $S_-$ by the modified Dyson series \eqc{eq:SDyson}
\bg
S_+ = T \left\{ e^{-i \int_{0}^{+\infty} H_{in}(t) dt } \right\} \,,\, S_- = T \left\{ e^{-i \int_{-\infty}^{0} H_{in}(t) dt } \right\} \,.
\eg
Decompose the S matrix as $S = S_+ S_-$ and insert the projection operator onto the state $\a'\b\g$, where $d\text{LIPS}$ is an abbreviation for \emph{Lorentz Invariant Phase Space}. Ellipsis denotes terms irrelevant for our purpose.
\bl
\bld
\bra{\a'\b'} S \ket{\a\b} &= \bra{\a'\b'} S_+ S_- \ket{\a\b}
\\ &= \int d\text{LIPS}_{\a'\b\g} \bra{\a'\b'} S_+ \ket{\a'\b\g} \bra{\a'\b\g} S_- \ket{\a\b} + \cdots \,. \label{eq:fact1}
\eld
\el
Insert the definition \eqc{eq:SADef} into the above formula, with a minor assumption that $S_{\pm}$ are essentially equivalent to $S$ for scattering amplitudes.
\bl
\bld
\bra{\a'\b'} S \ket{\a\b} &= i \tilde{\delta}^{(4)} (P_{\a'\b'} - P_{\a\b}) M_{\a\b \to \a'\b'}
\\ \int d\text{LIPS}_{\a'} \bra{\a'\b'} S_+ \ket{\a'\b\g} &= i \tilde{\delta}^{(4)} (P_{\b'} - P_{\b} - k) M_{\b\g \to \b'}
\\ \int d\text{LIPS}_{\b} \bra{\a'\b\g} S_- \ket{\a\b} &= i \tilde{\delta}^{(4)} (P_{\a'} + k - P_{\a}) M_{\a \to \a'\g}
\eld \label{eq:fact2}
\el
The four-momentum of one-particle state $\ket{\g}$ has been denoted as $k^\m$. Following \fig{fig:polpf} we denote $M_{\a\b \to \a'\b'} = M_n$, $M_{\b\g \to \b'} = M_L$, and $M_{\a \to \a'\g} = M_R$. The LHS of \eqc{eq:fact1} follows from the definition.
\bl
\bra{\a'\b'} S \ket{\a\b} = i \tilde{\delta}^{(4)} (P_{\a'\b'} - P_{\a\b}) M_n \,. \label{eq:fact3}
\el
Performing the integral over $d\text{LIPS}_\g$, the RHS of \eqc{eq:fact1} becomes
\bl
\bgd
i M_L \times i M_R \int \frac{d^4 k}{(2\pi)^4} \tilde{\delta}( k^2 - m^2) \tilde{\delta}^{(4)} (P_{\b'} - P_{\b} - k) \tilde{\delta}^{(4)} (P_{\a'} + k - P_{\a})
\\ = i \tilde{\delta}^{(4)} (P_{\a'\b'} - P_{\a\b}) M_L \times M_R \times \left[ 2 \pi i \delta( k^2 - m^2) \right]
\egd \label{eq:fact4}
\el
Since amplitudes are analytic functions, delta needs to be substituted by an equivalent analytic function. The following relation does the job.
\bl
2 \pi i \delta(x) = \frac{\mp 1}{x \pm i0^+}
\el
Choosing the upper sign\footnote{This sign choice is consistent with time ordering.} and combining \eqc{eq:fact1}, \eqc{eq:fact2}, \eqc{eq:fact3}, and \eqc{eq:fact4} yields \eqc{eq:factDef}. For a more detailed derivation, consult ref.\cite{Weinberg:1995mt} where a proof for time-ordered correlation function in momentum space is given.

\subsection{The optical theorem and Cutkosky rules}
In a unitary theory, the S-matrix satisfies the following unitarity condition.
\bl
S^\dagger S = \iden \,.
\el
Decomposing the S-matrix as $S = \iden + i T$, we find the following equation for the T-matrix $T$.
\bl
i (T - T^\dagger) = - T^\dagger T \,.
\el
This equation is the analogue of the optical theorem in quantum mechanics, where imaginary part of the forward scattering amplitude is related to the total cross-section. In theories with perturbative parameters, the equation relates the imaginary part of the T-matrix $i (T - T^\dagger)$ at a given perturbation order to ``forward scattering'' at lower perturbation orders $-T^\dagger T$. For example, assume $g$ as the perturbation parameter and expand the T-matrix as a power series in $g$; $T = \sum_{n \ge 1} g^n T_n$. Taking a matrix element of the above equation gives the following equation, where $\sum_c$ is a sum over on-shell intermediate states $\ket{c}$.
\bl
i \left( \bra{b} T_n \ket{a} - \bra{b} T_n^\dagger \ket{a} \right) = - \sum_{m \le n} \sum_c \bra{b} T_{n-m}^\dagger \ket{c} \bra{c} T_m \ket{a} \,. \label{eq:cut}
\el
An amplitude possess an imaginary part when it develops a branch cut, and the discontinuity across the brach cut is given as the imaginary part of the amplitude. Therefore, the above equation can be understood as relating discontinuities of an amplitude to lower perturbation order amplitudes. Note that the intermediate state $\ket{c}$ is not required to consist of elementary particles, similar to the case of polology.

The equation \eqc{eq:cut} can be computed diagrammatically as in \fig{fig:cutrule} using the cutting rules introduced by Cutkosky~\cite{Cutkosky:1960sp}. 
\bfig
\centering
\includegraphics[width=0.8\linewidth]{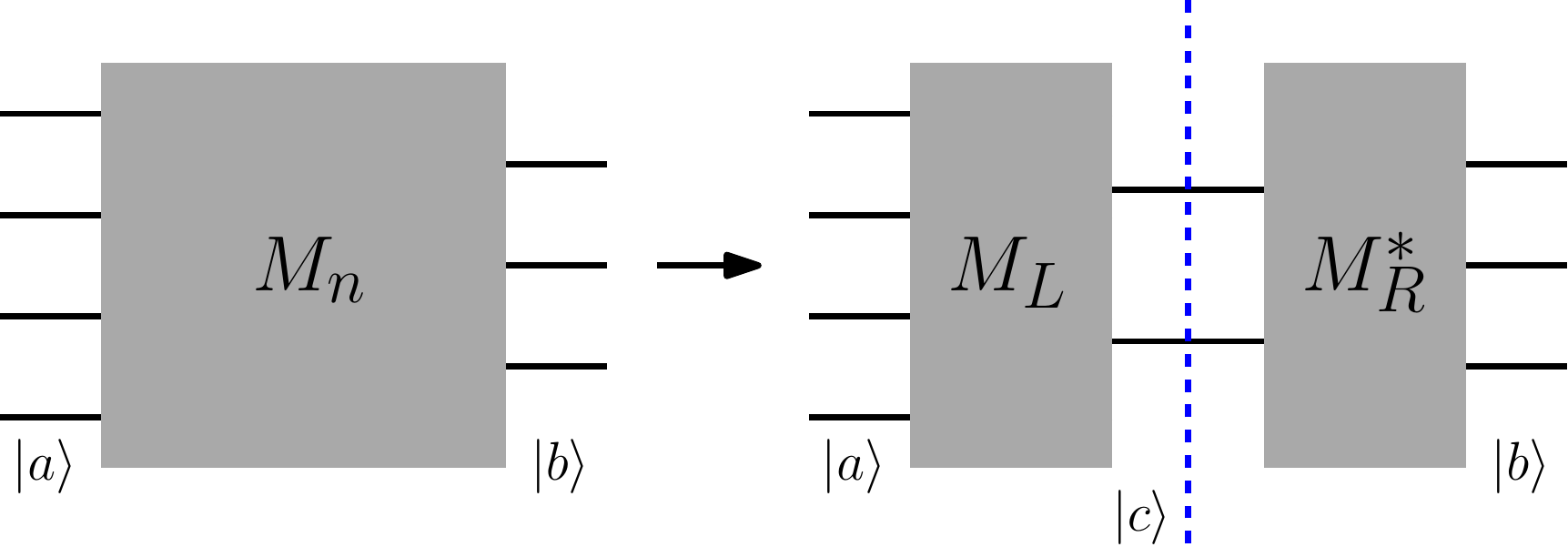}
\caption{Diagrammatic representation of the RHS of \eqc{eq:cut}. The subamplitudes are given as $M_L = \bra{c} T \ket{a}$ and $M_R^\ast = \bra{b} T^\dagger \ket{c} = \bra{c} T \ket{b}^\ast$, where momentum conservation conditions were dropped for brevity. Time flows from left to right.} \label{fig:cutrule}
\efig
The dashed line represents the \emph{cut}, which divides the Feynman graph $M_n$ into two parts; $M_L$ and $M_R^\ast$. The subdiagram denoted as $M_L$ is evaluated using usual Feynman rules, while the subdiagram denoted as $M_R^\ast$ is evaluated using complex-conjugated Feynman rules. The propagators that has been ``cut'', or the propagators that intersects the dashed line are substituted to on-shell conditions.
\bl
\frac{-i}{-k^2 + m^2 - i0^+} \to 2 \pi {\delta} (k^2 - m^2) \th (k^0) \label{eq:cutrule}
\el
In essence, the cutting rules relate the discontinuity of the amplitude to internal propagators going on-shell.

The optical theorem allows construction of the imaginary part of the T-matrix from lower perterbation order T-matrix elements. Therefore if it is possible to construct the real part of the T-matrix from the imaginary part, then it would be possible to iterate the process to any perturbation order. This is the basis of the so-called \emph{S-matrix programme} vigorously pursued in early 1960's. The programme came to a dead end due to insufficient generality of \emph{dispersion relations}, which was the main tool for reconstructing the real parts from the imaginary parts~\cite{Veltman:1994wz}. 

\subsection{Generalised unitarity and scalar integral coefficients} \label{sec:GenUnit}
A generic one-loop amplitude in dimensional regularisation $D = 4 - 2\e$ can be represented as the schematic form
\bl
M_n^{(1)} &= \int \frac{d^D l}{(2 \pi)^D} \frac{N(l ; p_i ; \ve_i)}{\prod_j D_j} \,,
\el
where $D_j = l_j^2 - m_j^2 + i0^+$ is the inverse Feynman propagator with $l_j^\m = l^\m + q_j^\m$ some linear shift of the loop momentum $l^\m$. The numerator $N(l ; p_i ; \ve_i)$ is a polynomial function of loop momentum $l^\m$ and external kinematic data; momenta $p_i^\m$ and polarisations $\ve_i$. A simple algebraic manipulation rewrites loop momentum dependence in the numerator as linear combinations of inverse propagators;
\bl
\bld
2 (q \cdot l) &= (l + q)^2 - l^2 - q^2
\\ &= [(l + q)^2 - m_1^2] - [l^2 - m_0^2] + m_1^2 - m_0^2 - q^2 \,.
\eld
\el
This is the basis of \emph{Passarino-Veltman reduction}~\cite{Passarino:1978jh} which rewrites a generic one-loop amplitude as a sum over \emph{scalar integrals} $I_{N}$. In general, any one-loop amplitude in dimensional regularisation can be written as a linear combination of scalar integrals and a remnant $\CR$ from regularisation procedure called \emph{rational terms}.
\bl
M_n^{(1)} &= c_{4;j} I_{4;j} + c_{3;j} I_{3;j} + c_{2;j} I_{2;j} + c_{1;j} I_{1;j} + \CR + \CO (D-4) \label{eq:SIexp}
\el
The coefficients $c_{i;j}$ and the rational term $\CR$ are rational functions of external kinematic data. The scalar integrals\textemdash traditionally referred to as tadpoles($I_1$), bubbles($I_2$), triangles($I_3$), and boxes($I_4$)\textemdash are defined as
\bl
I_N (p_i^2; s_{ij} ; m_i^2) &= \frac{\m^{4-D}}{i \pi^{\frac{D}{2}}} \frac{\G(1-2\e)}{\G^2(1-\e) \G(1+\e)} \int d^D l \prod_{i=1}^{N} \frac{1}{(l+q_{i-1})^2 - m_i^2 +i0^+} \,,
\el
where $q_n := \sum_{i=1}^n p_i$, $q_0 = 0$, and $s_{ij} = (p_i + p_j)^2$. Tables for scalar integrals can be found in refs.\cite{tHooft:1978jhc,Denner:1991qq,Ellis:2007qk}, therefore the problem of computing one-loop amplitudes reduces to finding the right coefficients for scalar integrals and computing the rational terms.

\emph{Generalised unitarity} aims to construct the integrands for loop integrals based on non-analytic structures and collinear/soft divergences of the amplitude implied by the optical theorem; if two expressions for the integrand results in same non-analytic structures and divergences, the loop amplitude computed from them must be the same~\cite{Bern:1994zx,Bern:1994cg}. For this purpose we can consider a generalisation of the cutting rules for the optical theorem.

The optical theorem for one-loop amplitude relates the discontinuity of the amplitude to the bisecting cut which divides the full amplitude into two tree amplitudes. Instead, consider cutting only a single propagator using the rule \eqc{eq:cutrule}; this is called the \emph{generalised cut}. Successive appication of generalised cuts can reveal more information of discontinuities than the original optical theorem, which can be enough to compute the one-loop amplitude without doing loop integrals by fixing the coefficients of the scalar integral expansion \eqc{eq:SIexp}. This is a well-studied procedure in the literature~\cite{Britto:2004nc,Britto:2005ha,Forde:2007mi,Britto:2007tt,Kilgore:2007qr,Mastrolia:2009dr}.

The general idea is based on a parametrisation for the one-loop integrand of the schematic form~\cite{Ossola:2006us,Ellis:2007br,Giele:2008ve}
\bl
\bld
M_n^{(1)} &= \int \frac{d^D l}{(2 \pi)^D} \left[ \frac{\D_5 (l)}{D_1 D_2 D_3 D_4 D_5} + \frac{\D_4 (l)}{D_1 D_2 D_3 D_4} + 
\cdots + \frac{\D_1 (l)}{D_1} \right] \,,
\eld
\el
where parametrisations for the numerators $\D_i (l)$ have been chosen so that only the constant part $\D_i(0)$ contributes to the loop integral.
\bl
\int \frac{d^D l}{(2 \pi)^D} \frac{\D_i (0) + l_\m \D_{i,1}^\m + l_\m l_\n \D_{i,2}^{\m\n} + \cdots}{D_1 \cdots D_i} = \int \frac{d^D l}{(2 \pi)^D} \frac{\D_i (0)}{D_1 \cdots D_i} + \CR \,.
\el
Here $\CR$ refers to the rational terms in \eqc{eq:SIexp}, which arises from loop momentum integrations along the $2\e$-dimension directions in dimensional regularisation $D = 4 - 2\e$. The above parametrisation implies that \emph{the scalar integral coefficients are determined by the residue of the propagator poles}, and the residues in turn can be determined by applying generalised cuts\footnote{The positive energy condition $\th(k^0)$ appearing in RHS of \eqc{eq:cutrule} is dropped when evaluating generalised cuts.}~\cite{Bern:2002zk}. Since the pentagon integral $\int \frac{\D_5}{D_1 \cdots D_5}$ reduces to other scalar integrals in the limit $D \to 4$, a sketch of the procedure will be presented for the box integral and the triangle integral based on the presentation in ref.~\cite{Ellis:2011cr}.

\bfig
\centering
\includegraphics[width=0.8\linewidth]{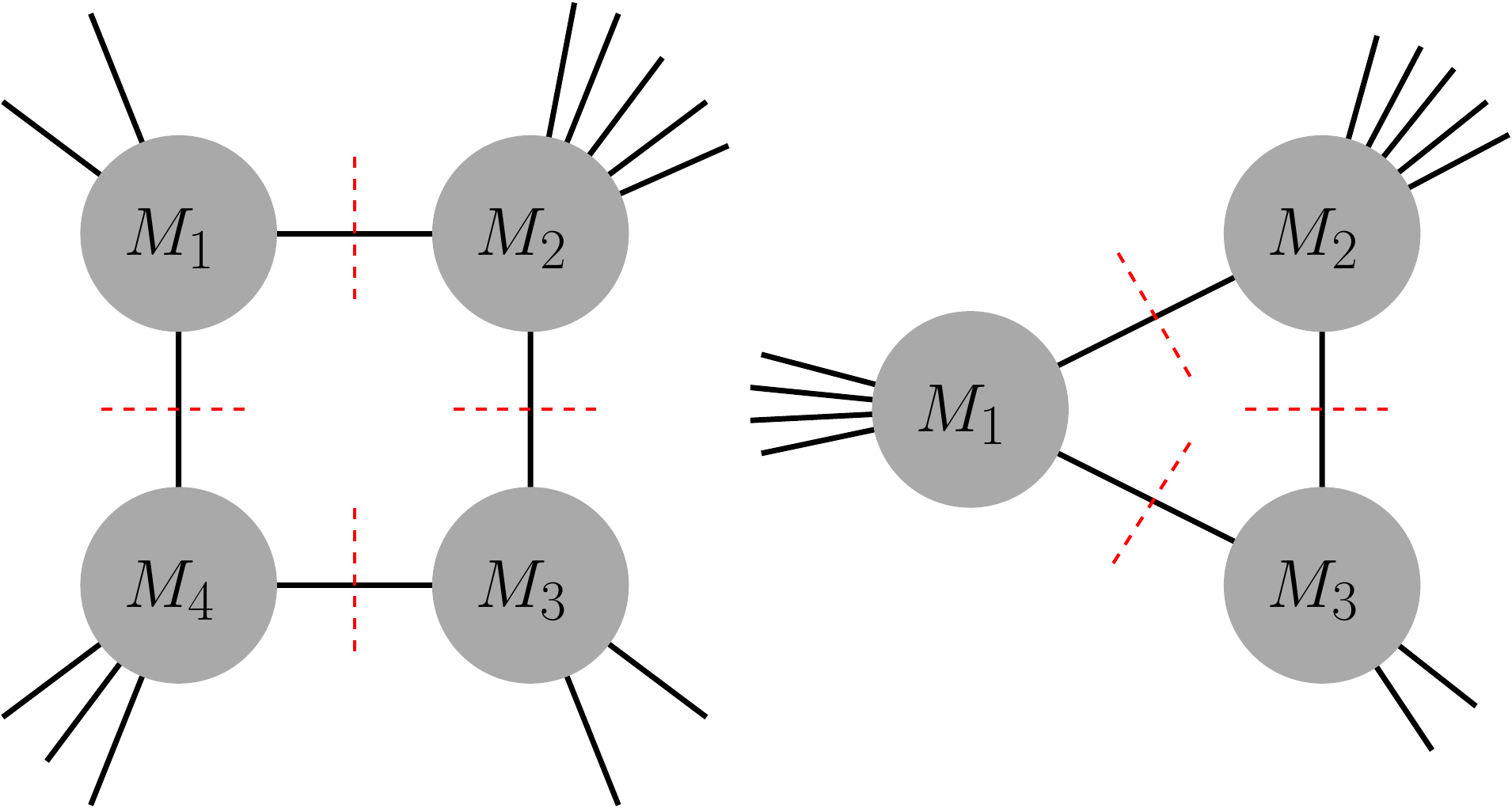}
\caption{LHS: The quadrupole cut where four internal propagators are substituted by corresponding four on-shell conditions. RHS: The triangle cut obtained by substituting three internal propagators to corresponding on-shell conditions.} \label{fig:fourcut}
\efig
Consider a four-particle cut where four internal propagators are substituted by four on-shell conditions as on the LHS of \fig{fig:fourcut}.
\bl
D_1 D_2 D_3 D_4 \to \prod_i^4 2 \pi i \delta( D_i ) \,.
\el
Generically there are two (complex) solutions $l_\pm^\m$ for the loop momentum $l^\m$ that satisfy the above on-shell condition. The residue $\D_4 (0)$ is determined as the average of the products of resulting tree amplitudes $M_1 M_2 M_3 M_4$, summed over on-shell states on the cut~\cite{Britto:2004nc}.
\bl
\D_4 (0) &= \half \sum_{l = l_\pm} \sum_{\text{int. states}} M_1 (l) M_2 (l) M_3 (l) M_4 (l) \,. \label{eq:quadcut_coeff}
\el
This result is justified because adding up all Feynman graphs that share the common cut propagators will result in a product of sums over Feynman subgraphs in each grey blob of \fig{fig:fourcut}. The sum over Feynman subgraphs inside each blob is simply the subamplitude, since each blob only contains on-shell external legs\footnote{When massive particles are running inside the loop complications arise due to self-energy and wavefunction renormalisation contributions. This problem has been addressed in refs.\cite{Ellis:2008ir,Britto:2011cr,Badger:2017gta}.}.

For the triangle cut as on the RHS of \fig{fig:fourcut}, the situation is more involved. The loop momentum $l^\m$ is first regarded as a point in $\IC \IP^4$, the manifold $\IC^4$ including the point at infinity. Imposing the triple-cut condition,
\bl
D_1 D_2 D_3 \to \prod_i^3 2 \pi i \delta( D_i ) \,,
\el
the loop momentum $l^\m$ originally in $\IC\IP^4$ is now restricted on a complex curve having the topology\footnote{Sometimes the solution to triple-cut condition branches into two sets joining at a point, having the topolgy of $\IC\IP^1 \oplus \IC\IP^1$. In this case each $\IC\IP^1$ branch is examined independently and the average over the two branches is taken.} $\IC\IP^1$. Parametrising the curve by the complex variable $y$, the residue of the loop integrand on this curve is schematically
\bl
\sum_{\text{int. states}} M_1 M_2 M_3 &= \frac{\D_4}{D_4} + \D_3 \,, \label{eq:triangleres}
\el
where $y$ dependence has been suppressed and $\frac{\D_4}{D_4}$ is the box integral contribution. For a good parametrisation $l(y) = a_0 y + a_1 y^{-1} + a_2$, this contribution will decompose into two simple poles at finite $y=y_\pm$.
\bl
\frac{\D_4 (y)}{D_4 (y)} = \frac{R_+}{y-y_+} + \frac{R_-}{y-y_-} \,. \label{eq:boxontriangle}
\el
The triangle coefficient $\D_3(0)$ is the ``constant part'' of the residue \eqc{eq:triangleres} on the curve described by $y$. Since the box contribution reduces to sum of simple pole contributions located at finite $y_\pm$, this constant piece can be evaluated using a residue integral\footnote{The decomposition \eqc{eq:boxontriangle} may contain a constant contribution if $\D_4 (l)$ contains linear terms. This contribution can be removed by considering a complex conjugate parametrisation $\bar{y}$ and taking an average over residues of $y=\infty$ and $\bar{y}=\infty$.} at $y=\infty$~\cite{Forde:2007mi}.
\bl
\D_3(0) = \frac{1}{2 \pi i} \oint_{y=\infty} \frac{dy}{y} \left\{ \frac{\D_4}{D_4} + \D_3 \right\} \,. \label{eq:tricut_coeff}
\el

Determination of bubble coefficients follows in a similar vein, the only difference being the topology of the solution for on-shell conditions; the bubble coefficient $\D_2(0)$ is now a constant piece of the function defined on a complex surface $\simeq \IC\IP^2$ rather than a curve. This subject is beyond the scope of this dissertation, and the reader is referred to the references~\cite{Forde:2007mi,Britto:2007tt,Kilgore:2007qr,Mastrolia:2009dr}.

Various reviews on generalised unitarity and techniques for computing one-loop amplitudes can be found in the literature, such as refs.\cite{Ellis:2011cr,Bern:1996je,Britto:2010xq,Bern:2011qt,Carrasco:2011hw}.

\section{The on-shell formalism for the S-matrix}
Discussions in section~\ref{sec:anal_S-mat} have persistently alluded to the conclusion that properties of the S-matrix can be determined from considering only the \emph{physical} asympototic states. Therefore, if physical asymptotic states can be described without redundancies, difficulties in computation of S-matrix elements can be greatly reduced. This is accomplished by \emph{on-shell variables} such as spinor-helicity variables.

This section reviews spinor-helicity formalism and its generalisation to massive and arbitrary spin case developed by Arkani-Hamed, Huang, and Huang~\cite{Arkani-Hamed:2017jhn}. A popular approach to spinor-helicity formalism is to introduce them as ``square root'' of null momenta as in ref.\cite{Henn:2014yza}, but this section will motivate it from representation theory perspective as in ref.\cite{elvang2015scattering}. Reader's familiarity with Wigner's little group classification scheme~\cite{Wigner:1939cj,Bargmann:1948ck} is assumed, which is well-reviewed by Weinberg in ref.\cite{Weinberg:1995mt}.

\subsection{The spinor-helicity formalism}
Weyl spinors are the smallest nontrivial unitary representations of the 4d Lorentz group $SO(3,1) \simeq SL(2,\IC)$ corresponding to spin-$\half$ representations. Given a null momentum $p^\mu$, denote the on-shell chiral/left-handed spinor\footnote{The spinors considered in this section are $c$-number valued, just like mode functions of spinor fields; mode operators carry fermion statistics in the mode expansion of spinor fields.} having helicity $-\half$ as $\l_\a = \ket{p}_\a$ and on-shell anti-chiral/right-handed spinor having helicity $+\half$ as $\bar{\l}^{\dot\a} = \sket{p}^{\dot\a}$. The outer product relation of Dirac spinors is inherited to Weyl spinors as the relation
\bl
\sum_{s} u(p;s) \bar{u}(p;s) = \fsl{p} + m && \Rightarrow &&
\left\{ 
\begin{aligned}
\ket{p}_\a \sbra{p}_{\dot{\a}} &= p_\m \s^\m_{\a\dot{\a}}
\\ \sket{p}^{\dot{\a}} \bra{p}^{\a} &= p_\m \bar{\s}^{\m \dot{\a} \a}
\end{aligned}
\right. \,,
\el
where $\fsl{p} = p_\m \g^\m$ is the Feynman slash notation, $\sbra{p}_{\dot\a} = \e_{\dot\a \dot\b} \sket{p}^{\dot\b}$, $\bra{p}^\a = \e^{\a\b} \ket{p}_\b$, $\s^\m = (\iden, \vec{\s})$, and $\bar{\s}^\m = (\iden, -\vec{\s})$. The above relation implies that Weyl spinors $\ket{p}$ and $\sket{p}$ are ``square root'' of the momentum $p^\m$, and spinor brackets such as $\la kq \ra = \e_{\a\b} \bra{k}^\a \bra{q}^\b$ and $[kq] = \e^{\dot\a \dot\b} \sbra{k}_{\dot\a} \sbra{q}_{\dot\b}$ are ``square root'' of Mandelstam invariants. For example,
\bl
(k + q)^2 = 2 k \cdot q = \la kq \ra [qk] \,.
\el

On the other hand, the representation for a massless one-particle state with definite helicity and any spin\footnote{Continuous spin representation will not be considered in this dissertation.} can be constructed as a symmetric product representation of Weyl spinors $\ket{p}$ or $\sket{p}$. Spinor-helicity is a formalism for writing Weyl spinors that maximises the strength of these two properties by assigning double roles to the spinors;
\bn
\item Wavefunctions for massless particles of general spins with definite helicity.
\item Spinor brackets as ``square root'' of Mandelstam invariants.
\en
In spinor-helicity formalism, the spinors of the particle with momentum $p_i^\m$ are simply written as $\ket{i}$ and $\sket{i}$. An elegant example that demonstrates the power of spinor-helicity formalism is the colour-ordered tree-level MHV amplitude of Yang-Mills theory, valid to all multiplicities $n$;
\bl
A_n [1^+,2^+, \cdots, i^-, \cdots, j^-, \cdots, n^+] &= \frac{\la ij \ra^4}{\la 12 \ra \la 23 \ra \cdots \la n-1 , n \ra \la n 1\ra} \,. \label{eq:MHV}
\el
The details and conventions used in this dissertation are summarised in the appendix.

\subsection{Spinor-helicity for massive particles}
Weyl spinors are usually reserved for describing massless particles, but it is not forbidden to use them for massive particles. The massive spinor-helicity formalism of ref.\cite{Arkani-Hamed:2017jhn} can be understood as simply ``uplifting'' the $SU(2)$ non-relativistic spinors to $SL(2,\IC)$ spinors. For massive spin-$\half$ particles the Weyl spinors $\ket{p^I}$ and $\sket{p^I}$ carry an extra $SU(2)$ index $I$ commonly called the \emph{little group index}, which denotes the $SU(2)$ spinor it was continued from.

This ``uplifting'' is not unique as there are two choices for continuation; chiral or anti-chiral. However, this ambiguity is immaterial for amplitudes; the amplitude depends on particle's state, and the particle's state is completely specified at the particle's rest frame by its little group representation~\cite{Wigner:1939cj,Bargmann:1948ck}. Thus the real input for amplitudes of massive particles is the non-relativistic $SU(2)$ index rather than the full-relativistic $SL(2,\IC)$ index, which translates to the language of mathematics as ``an amplitude is a tensor in little group space''. Therefore, Weyl spinors can be used for describing massive particles as long as none of their $SL(2,\IC)$ indices remain as free indices in the full expression of the amplitude~\cite{Arkani-Hamed:2017jhn}.

Furthermore, the chiral and anti-chiral Weyl spinors are not independent variables. The Dirac equation relates chiral spinors to anti-chiral spinors and vice versa,
\bl
(\fsl{p} - m) u(p;I) = 0 && \Rightarrow && 
\left\{
\begin{aligned}
p_\m \bar{\s}^{\m \dot{\a} \a} \ket{p^I}_\a &= m \sket{p^I}^{\dot\a}
\\ p_\m \s^\m_{\a \dot{\a}} \sket{p^I}^{\dot{\a}} &= m \ket{p^I}_\a
\end{aligned}
\right. \,,
\el
implying all $SU(2)$ indices can be exclusively attached to chiral (or anti-chiral) spinors without loss of generality. This property can be used to constrain an amplitude from purely kinematical considerations, similar in spirit to the bootstrap programme for CFTs.

In non-relativistic theories a spin-$s$ representation is constructed as the symmetric product representation of $2s$ spin-$\half$ representations, having $2s$ symmtrised $SU(2)$ indices $I_1 \cdots I_{2s} =: \{ I_{2s} \}$. Therefore an amplitude involving a massive spin-$s$ particle will contain a symmetrised set of $SU(2)$ indices. A convention that hides explicit symmetrisation procedure can yield compact expressions, which is the motivation for introducing the \textbf{bold} notation~\cite{Arkani-Hamed:2017jhn}. Continuing with the convention for writing spinors of the particle with momentum $p_i^\m$ by the index $i$, the spinors are written as $\ket{\mathbf{i}} = \ket{i^I}$ and $\sket{\mathbf{i}} = \sket{i^I}$, and little group indices over same-indexed spinors are implicitly symmetrised. The details and conventions used in this dissertation are summarised in the appendix.

\subsection{Little group constraints of amplitudes}
Focusing on a single external leg described by one-particle state $\ket{p;\{I\}}$ and suppressing information of other particles, the amplitude can be abstractly written as the overlap between $\ket{p;\{I\}}$ and some other abstract vector $\ket{\Psi}$ independent of $\ket{p;\{I\}}$;
\bl
M_n &= \left\langle {\Psi} \middle| {p;\{I\}} \right\rangle \,.
\el
Therefore the amplitude must have the same little group transformation properties as the one-particle state. This property serves as a kinematic constraint on possible amplitudes, often called \emph{little group constraints}. For massive particles the constraint is rather trivial; the amplitude with $i$-th external leg having spin $s_i$ must be a homogeneous polynomial of $2s_i$ bolded spinors $\ket{\bf{i}}$ and $\sket{\bf{i}}$. For massless particles the constraint is more involved as Mandelstam invariants can be decomposed into spinor brackets, and inverse powers of spinors can appear in the amplitude.

The little group of massles particles is the complexified $U(1)$ group\footnote{The full little group of massless particles with real momenta is the isometry group of the Euclidean plane $ISO(2)$, but only its subgroup $SO(2) = U(1)$ is realised for unitary representation. Momenta of particles will be often complexified for studying analytic structures, and in such cases the group also becomes complexified.}, acting on the spinors as
\bg
\ket{p} \to e^{- i \th} \ket{p} \,,\, \sket{p} \to e^{+ i \th} \sket{p} \,.
\eg
This implies an amplitude $M_n$ with $i$-th leg having helicity $h_i$ obeys the following scaling relations, because this scaling is realised by little group transformations on the $i$-th particle's wavefunction.
\bl
\left. \begin{aligned}
\ket{i} &\to t^{-1} \ket{i}
\\ \sket{i} &\to t^{+1} \sket{i}
\end{aligned} \right\} && \Rightarrow && M_n \left(\cdots,i^{h_i},\cdots \right) \to t^{2 h_i} M_n \left(\cdots,i^{h_i},\cdots \right)
\el
Note that this constraint is nontrivially satisfied by the MHV amplitude \eqc{eq:MHV}; except for $i$-th and $j$-th legs, other legs satisfy the above constraint by inverse powers of the spinors.

\subsection{Kinematical constraints for three-point amplitudes} \label{sec:3pt}
In modern understanding of amplitudes the three-point amplitude can be used as the building block from which one can bootstrap oneself the S-matrix, using relations of amplitudes between different multiplicities known as \emph{recursion relations}~\cite{Benincasa:2007xk}. One of the most well-known recursion relations are \emph{BCFW recursion relations}~\cite{Britto:2004ap,Britto:2005fq} which constructs higher multiplicity tree amplitudes from lower multiplicity tree amplitudes using complex analysis. The recursion relations are known to uniquely determine the Yang-Mills theory as the solution to recursively constructable gluon amplitudes satisfying certain criteria~\cite{Rodina:2016mbk}.

The inputs for bootstrapping the S-matrix\textemdash the three-point amplitudes\textemdash are severely constrained by kinematics. The momentum conservation condition\footnote{All particles are defined as incoming in this section.} $p_1 + p_2 + p_3 = 0$ and the on-shell conditions $p_i^2 = m_i^2$ imply that all possible Mandelstam invariants are constant, meaning that the amplitude only depends on the wavefunction factors such as spinors. Therefore the little group constraints described in the previous section is the only nontrivial kinematic costraint that determines the amplitude.

For massless scattering the constraints are so powerful that they are essentially determined from kinematical considerations alone~\cite{Benincasa:2007xk}; denoting the helicities of each particle as $h_1$, $h_2$, and $h_3$, kinematics determine the three-point amplitude up to a coupling constant as
\bl
M_3(1^{h_1},2^{h_2},3^{h_3}) &= \left\{
\begin{aligned}
\la 12 \ra^{+d_3} \la 23 \ra^{+d_1} \la 31 \ra^{+d_2} && d_0 \le 0
\\ [ 12 ]^{-d_3} [ 23 ]^{-d_1} [ 31 ]^{-d_2} && d_0 \ge 0
\end{aligned} \right.
\el
where $d_i = 2 h_i - d_0$ and $d_0 = h_1 + h_2 + h_3$. These power dependencies can be determined from little group scaling discussed in the previous section.

For massive particles the kinematics is not as strongly constraining as in the massless case. The amplitude that will be most relevant to this dissertation is the two massive-one massless amplitude where massive particles have equal mass $m$ and spin $s$. In this case, kinematic considerations reduce the possible parameters of the three-point amplitude to $2s+1$ variables~\cite{Arkani-Hamed:2017jhn}. When the massless particle has positive helicity the natural basis is
\bl
M_3^{h,s,s} &= (mx)^h \left[ g_0 \frac{\la \bold{2} \bold{1} \ra^{2s}}{m^{2s}} + g_1 x \frac{\la \bold{2} \bold{1} \ra^{2s-1} \la \bold{2} 3 \ra \la 3 \bold{1} \ra}{m^{2s+1}} + \cdots + g_{2s} x^{2s} \frac{\la \bold{2} 3 \ra^{2s} \la 3 \bold{1} \ra^{2s}}{m^{4s}} \right]\,,\label{eq:poshel3pt}
\el
while the natural basis for negative helicity is its ``complex conjugate''
\bl
M_3^{-\abs{h},s,s} &= \frac{m^{\abs{h}}}{x^{\abs{h}}} \left[ \bar{g}_0 \frac{[ \bold{2} \bold{1} ]^{2s}}{m^{2s}} + \frac{\bar{g}_1}{x} \frac{[ \bold{2} \bold{1} ]^{2s-1} [ \bold{2} 3 ] [ 3 \bold{1} ]}{m^{2s+1}} + \cdots + \frac{\bar{g}_{2s}}{x^{2s}} \frac{[ \bold{2} 3 ]^{2s} [ 3 \bold{1} ]^{2s}}{m^{4s}} \right] \,.\label{eq:neghel3pt}
\el
The $x$-factor introduced in ref.\cite{Arkani-Hamed:2017jhn} is the proportionality factor of the massless spinor that carries little group weights of the massless particle
\eq
x\lambda_3^\alpha = \tilde{\lambda}_{3\dot{\alpha}} \frac{p_{1}^{\dot{\alpha}\alpha}}{m}\,, \label{eq:x_def}
\eqe
and it can also be described using traditional polarisation vectors~\cite{Chung:2018kqs}
\eq
mx=\frac{1}{\sqrt{2}}\varepsilon^{(+)}\cdot(p_1-p_2) \,.
\eqe
The \emph{minimal coupling} is defined by setting all parameters $g_i = 0$ ($\bar{g}_i = 0$) except for $g_0$($\bar{g}_0$), which gives the amplitude the best high-energy behaviour~\cite{Arkani-Hamed:2017jhn}.
\bg
M_{3,min}^{+h,s,s} = g_0 x^h \frac{\la \bold{2} \bold{1} \ra^{2s}}{m^{2s-h}} \,,\quad M_{3,min}^{-\abs{h},s,s} = \bar{g}_0 x^{-\abs{h}} \frac{[ \bold{2} \bold{1} ]^{2s}}{m^{2s-\abs{h}}} \,. \label{eq:mincoup_def}
\eg

\section{Two-body effective Hamiltonians from quantum scattering amplitudes}
\subsection{Classification of perturbative general relativity}
The governing equations of GR are nonlinear partial differential equations, and solving the equations exactly to understand a physical process of interest is often an unnecessarily laborious work, if possible at all. A more practical approach is to start from a sufficiently valid description that is much easier to solve than the full Einstein's equations, and add effects from GR as perturbative corrections. Such approaches can be classified into two categories, which is summarised in the table \ref{table:PNPM}.

\begin{table}[]
\centering
\begin{tabular}{c|cc}
classification & \phantom{a}post-Newtonian (PN)\phantom{a} & \phantom{a}post-Minkowskian (PM) \\ [0.5ex] \hline 
\begin{tabular}{@{}c@{}} unperturbed\\ theory \end{tabular} & \phantom{\Big|}Newtonian gravity\phantom{\Big|} & special relativity \\ [3ex] 
\begin{tabular}{@{}c@{}} unperturbed\\ Hamiltonian \\ (0 PN/PM) \end{tabular} & $\displaystyle \sum_{i=1}^2 \frac{p_i^2}{2m_i} - \frac{G m_1 m_2}{r}$ & $\displaystyle \sum_{i=1}^2 \sqrt{p_i^2 + m_i^2}$ \\ [4ex] 
\begin{tabular}{@{}c@{}} expansion\\ parameter \end{tabular} & $\displaystyle \frac{1}{c^2}$ & $G$ \\ [3ex] 
small numbers & $\displaystyle \frac{G \m}{rc^2}$, $\displaystyle \frac{p^2}{\m^2 c^2}$ & $\displaystyle \frac{G \m}{r}$ \\ [3ex] 
\begin{tabular}{@{}c@{}} characteristic\\ dynamics \end{tabular} & bound motion & scattering
\end{tabular}
\caption{Classification of perturbative GR. The speed of light in vacuum $c$ is explicitly shown in the PN column.
} \label{table:PNPM}
\end{table}

The first is the \emph{post-Newtonian} (PN) expansion. In the PN expansion, the ``unperturbed'' dynamics is given as Newtonian description of gravity. Corrections from special and general relativity enter as perturbative corrections to Newtonian gravity, hence the name post-Newtonian. The expression ``relativistic corrections'' usually refers to this expansion, which has been studied as early as from 1938~\cite{Einstein:1938yz}. Formally the expansion can be considered as an expansion in $1/c^2$, where $c$ is the speed of light in vacuum. The dimensionless numbers\footnote{Including spin effects introduces $\frac{S}{r m c}$ as another dimensionless number in the expansion. Since this number scales as $c^{-1}$, one power of spin is formally counted as 0.5 PN.} that characterise this expansion are $\frac{G \mu }{ r c^2}$ and $\frac{p^2 }{ \m^2 c^2}$, where $\m$ is the mass that characterises the system which can be understood as reduced mass $\m = \frac{m_1 m_2}{m_1 + m_2}$ in most contexts.

The second is the \emph{post-Minkowskian} (PM) expansion. In the PM expansion, the ``unperturbed'' dynamics is given as special relativity of free particles. Corrections from GR enter as perturbative corrections to special relativity, hence the name post-Minkowskian. Formally the expansion can be considered as an expansion in the gravitational constant $G$. Since $G$ is the coupling constant of the theory, this expansion is naturally linked to scattering amplitudes which are given as series expansions in coupling constants. A small conceptual caveat is that corrections being added to special relativity do \emph{not} respect the symmetries of the unperturbed theory, and the corrections are written as instantaneous long-distance interactions.

\subsection{Mapping amplitudes to effective Hamiltonians} \label{sec:Amp2Ham}
Consider a non-relativistic spinless two particle system interacting through the potential $V = V (\vec{x}_b - \vec{x}_a)$, which will be later identified as the PM potential of effective two-body Hamiltonian\footnote{Although the system is non-relativistic, relativistic dispersion relation $E^2 = p^2 + m^2$ is used in concordance with the PM expansion.}. Promoting the particles to fields by second quantisation, the interaction is now described by the Hamiltonian
\bl
H_{in} &= \int d\vec{x}_a d\vec{x}_b ~ V (\vec{x}_b - \vec{x}_a) :\phi_a^\dagger (\vec{x}_a) \phi_a(\vec{x}_a) \Phi_b^\dagger (\vec{x}_b) \Phi_b(\vec{x}_b): \,,
\el
where $\phi_a$($\Phi_b$) denotes the positive frequency components of particle species $a$($b$), $\phi_a^\dagger$($\Phi_b^\dagger$) denotes the negative frequency components, and $:
:$ denotes normal ordering. Moving to momentum space by performing a Fourier transform, the Hamiltonian is now expressed as\footnote{The immaterial overall volume factor has been dropped.}
\bl
\bld
H_{in} &= 
\sum_{\vec{k}_a , \vec{k}_b, \vec{q}} V (\vec{q}) ~ a^\dagger_{(\vec{k}_a - \vec{q})} \, a_{\vec{k}_a} ~ b^\dagger_{(\vec{k}_b + \vec{q})} \, b_{\vec{k}_b}
\\ V(\vec{q}) &= \int d\vec{r} ~ V(\vec{r}) ~ e^{- i \vec{q} \cdot \vec{r}} \,,
\eld \label{eq:NREFTintHam}
\el
where $a$ and $a^\dagger$($b$ and $b^\dagger$) are mode operators of particle species $a$($b$)
. The transfer momentum $\vec{q}$ and the displacement vector $\vec{x}_b - \vec{x}_a$ are Fourier duals of each other, which implies that the behaviour of $V(\vec{r})$ at long distances is determined by the behaviour of $V(\vec{q})$ at small transfer momentum.

The interaction Hamiltonian \eqc{eq:NREFTintHam} can be used to construct the S-matrix, e.g. from the Dyson series \eqc{eq:SDyson}. For example, the $2 \to 2$ scattering matrix element can be computed using Born series\footnote{Comparing the leading terms shows that an overall sign difference exists between amplitudes obtained from the series \eqc{eq:SDyson} and the one from the series \eqc{eq:BornAmp}. This sign difference will be neglected and fixed by hand when needed.};
\bl
M_{\vec{k}_a + \vec{k}_b \to \vec{k}_a' + \vec{k}_b'} &= \bra{\vec{k}_a',\vec{k}_b'} H_{in} \sum_{n=0}^\infty ( G_0 (E) H_{in} )^n \ket{\vec{k}_a,\vec{k}_b} \,, \label{eq:BornAmp}
\\ G_0 (E) &= \frac{1}{E - H_0 + i0^+} \,,
\el
where $E$ is the energy of incoming ($\ket{\vec{k}_a , \vec{k}_b}$) and outgoing ($\ket{\vec{k}_a' , \vec{k}_b'}$) states, and $H_0$ is the free Hamiltonian.

The S-matrix constructed from the Hamiltonian \eqc{eq:NREFTintHam} and the S-matrix of the full relativistic theory cannot be the same, but it is possible to find the potential $V$ that gives the ``best fit''. This matching procedure is typically performed using the $2 \to 2$ scattering matrix element \eqc{eq:BornAmp} in the centre of momentum (COM) frame:
\bg\label{Kin1sec2}
\bgd
p_1 = ( E_a, \vec{p} {+} \vec{q} / 2 )\,, \;\;
p_3 = ( E_b, {-} \vec{p}{-} \vec{q} / 2 )\,, \;\;  
\\ p_2 = ( E_a, \vec{p} {-} \vec{q} / 2 )\,, \;\;
p_4 = ( E_b, {-} \vec{p} {+} \vec{q} / 2 ) \,.
\egd
\eg
The approach has been adopted in numerous works to construct the effective potential~\cite{Iwasaki:1971vb,Holstein:2008sw,Holstein:2008sx,Vaidya:2014kza}.

A systematic method to carry out such a matching procedure has been introduced by Cheung, Rothstein, and Solon~\cite{Cheung:2018wkq} where comparison is made at the integrand level. The procedure had been applied to the ansatz for the (classical) PM potential
\bl
V(p,r) &= \sum_{n=1}^\infty c_n(p) \left( \frac{G}{r} \right)^n \,,
\el
to determine the 3PM dynamics of spinless black hole binaries in refs.\cite{Bern:2019nnu,Bern:2019crd}.

\subsection{The classical limit of finite spin particles} \label{sec:Amp2Ham_sm}
While spinless bodies were considered in the previous section, it is also possible to include spin effects in this framework~\cite{Holstein:2008sw,Holstein:2008sx,Vaidya:2014kza,Bern:2020buy}. Most of the results for the spinless case readily generalises to the spinning case, but there is a subtle issue for the classical limit.

Restoration of reduced Planck's constant $\hbar$ has been well reviewed in refs.\cite{Kosower:2018adc,Maybee:2019jus,Bern:2019crd}; the heuristics for $\hbar$ counting can be summarised as follows~\cite{Chung:2019duq}.
\bn
\item Massive particle's mass $m_i$ and momentum $p_i^\m$ scale as $\propto \hbar^0$.
\item Massless particle's momentum $q^\m$ is converted to its wavenumber $\hbar \bar{q}^\m$.
\item Gravitational constant $G$ carries an inverse power of $\hbar$; $G/\hbar$.
\item Spin $S^\m$ is counted in units of Planck's constant $S^\m / \hbar$.
\en
It follows that the ``dimensionless'' combinations in terms of $\hbar$ are
\bl
G\abs{q} \,,\, \abs{q}\abs{S} \,,
\el
and the variables $q^\m$ and $S^\m$ contribute to classical part of the effective Hamiltonian only through these combinations. The subtlety of the classical limit is related to the heuristic involving $S^\m$; spins are measured in units of $\hbar$.

In quantum mechanical treatment of spin, the spin $\vec{S}$ is quantised in half-integral units of $\hbar$. Therefore without any additional scaling behaviour, the size of the spin $|\vec{S}| = s\hbar$ vanishes in the classical limit $\hbar \to 0$. One definition for the classical limit that avoids vanishing of spin would be to fix $s\hbar$ as finite while sending $\hbar \to 0$, which will be referred to as the \emph{classical-spin limit}; this definition has been (implicitly) adopted in refs.\cite{Guevara:2018wpp,Chung:2018kqs,Maybee:2019jus,Guevara:2019fsj,Arkani-Hamed:2019ymq,Chung:2019yfs,Chung:2020rrz,Bern:2020buy}. Although this definition agrees with scaling behaviours of classical spinning objects, the definition is not helpful for doing computations if the limit $s \to \infty$ cannot be implemented in the calculations. For example, the classical limit of a Dirac fermion can only be a spinless particle when classical-spin limit is adopted as the definition for the classical limit.

\bfig
\centering
\includegraphics[width=\linewidth,trim={0 9.4cm 1.7cm 0},clip]{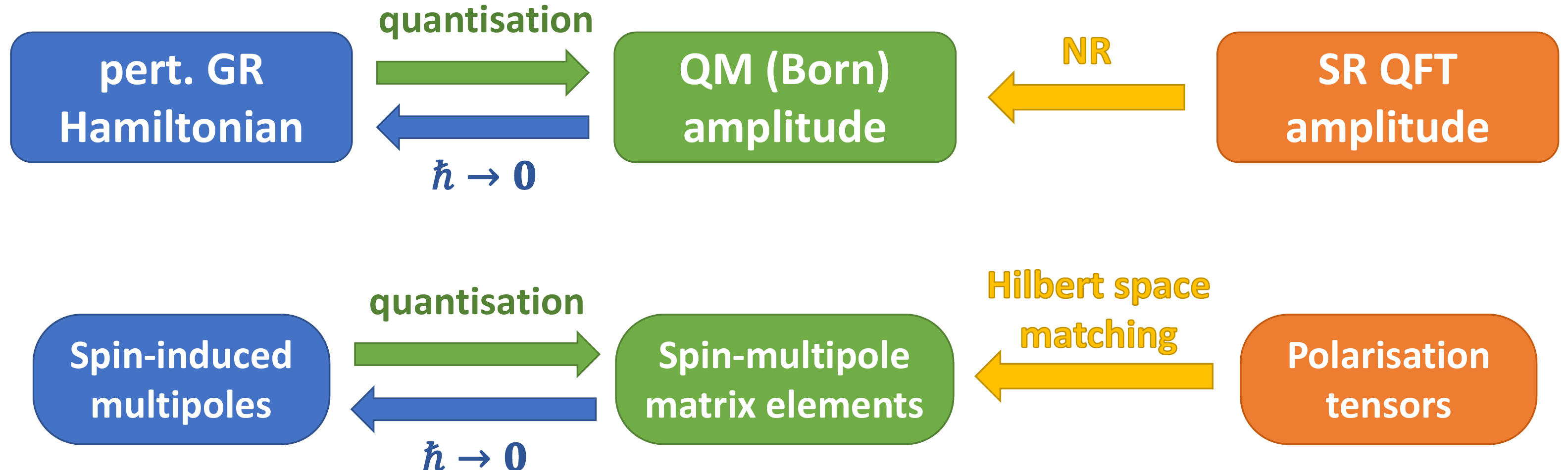}
\caption{Correspondence of spin multipoles moments.} \label{fig:sm}
\efig

An alternative definition for the classical limit is to inspect each spin multipole operator sector individually and then to take the limit $\hbar \to 0$ in each sector. When restricted to the internal space of particle $a$ having spin $s$, the effective Hamiltonian \eqc{eq:NREFTintHam} will become a $(2s+1) \times (2s+1)$ matrix that maps $(2s+1)$-dimensional phase space of the incoming particle to the $(2s+1)$-dimensional phase space of the outgoing particle. This matrix can be decomposed into \emph{spin multipole operators} $S_{(l)}$, defined as the traceless symmetric product of $SU(2) \simeq SO(3)$ spin operators.
\bl
H_{in} &= \sum_{l=0}^{2s} H_{in}^{(l),i_1 \cdots i_l} S_{(l)}^{\, i_1 \cdots i_l} \,, \label{eq:HeffSMexp}
\\ S_{(l)}^{\, i_1 \cdots i_l} &:= \overbrace{S^{(i_1} \cdots S^{\,i_l)}}^{l} - (\text{trace}) \,. \label{eq:SMdefSU2}
\el
The spin $2^l$-pole operator $S_{(l)}$ has $(2l+1)$ degrees of freedom, so the basis \eqc{eq:SMdefSU2} forms a complete basis for the expansion \eqc{eq:HeffSMexp}. The classical limit is defined as taking the limit $\hbar \to 0$ for each $H_{in}^{(l)}$ and identifying $S_{(l)}$ as the classical spin-induced multipole moments. This is the definition adopted in refs.\cite{Donoghue:2001qc,BjerrumBohr:2002ks,Holstein:2006ud,Holstein:2008sw,Holstein:2008sx,Vaidya:2014kza,Chung:2019duq} where spin multipole operators for finite spin (finite $s$) particles were mapped to classical spin-induced multipoles. Also, this definition for the classical limit agrees with the classical-spin limit whenever the limit $s\to\infty$ is available. A subtlety that needs to be addressed when applying this definition to amplitudes of fully relativistic theories is that internal phase space of the incoming and outgoing particles are inequivalent, which is resolved by \emph{Hilbert space matching} introduced in section~\ref{sec:HilbertMatch}. A summary of this definition for the classical limit is given in \fig{fig:sm}.

\chapter{Applications at tree level} \label{chap:tree}

\section{The gravitational three-point amplitude} \label{sec:3pt}
\subsection{Amplitudes from point particle effective action} \label{sec:ppEFT2Amp}
In non-relativistic quantum mechanics, a spin-$s$ particle can have independent spin multipole operators up to $2^{2s}$-pole. Therefore it is necessary to consider massive particles with arbitrarily high spin when exploring the dynamics of spin multipole moments to arbitrary orders. While there were attempts to construct Lagrangians for higher-spin massive particles~\cite{Singh:1974qz,Singh:1974rc}, the resulting Lagrangians were too complicated for practical computations. At first sight, incorporating higher-spin particles into quantum scattering seems to be a formidable task.

On the other hand, it might be possible to write down amplitudes without any Lagrangian description following the old dream of the S-matrix programme. The first stepping stone in this direction would be to write down the building blocks for the S-matrix; the three-point amplitudes. The three-point amplitudes are constrained so heavily by kinematics that it is almost possible to write down the full answer as briefly described in section~\ref{sec:3pt}. When studying gravitational interactions the relevant three-point amplitude is that of graviton coupling, or the spin-$s$\textemdash spin-$s$\textemdash helicity-2 amplitude. The general ansatz is given by \eqc{eq:poshel3pt} and \eqc{eq:neghel3pt}, so the only remaining task is to determine the free parameters $g_i$, $i=1, \cdots, 2s$.

The idea presented in ref.\cite{Chung:2018kqs} was to use the point particle effective action reviewed in section~\ref{sec:ppEFT} as an input. The point particle action \eqc{eq:EFTmin} and \eqc{eq:EFTSI} can be expanded in powers of the background graviton field $h_{\m\n}$ defined by the relation $g_{\m\n} = \eta_{\m\n} + \k h_{\m\n}$. The action \eqc{eq:EFTmin} is then given as
\bl
\bld
S &= \int d\s \left\{ - m\sqrt{u^2} - \half S_{\m\n} \O^{\m\n} + L_{SI} \right\}
\\ &= \int d\s \left\{ \CO^{\text{free}}(\s) - \frac{\k}{2} h_{\m\n} T^{\m\n}(\s) + \CO(h^2) \right\} \,.
\eld
\el

The $\CO(h^0)$ order worldline operator $\CO^{\text{free}}$ corresponds to free Lagrangian in field theory language, which is used to construct asymptotic states for the S-matrix and Feynman rules for the propagators. In the on-shell approach both are given on the get-go, so this term does not carry any new information.

The $\CO(h^1)$ order worldline operator $T^{\m\n}$ carries the information needed to fix the free parameters of three-point amplitude ans\"atze \eqc{eq:poshel3pt} and \eqc{eq:neghel3pt}. The minimal terms contribute the following terms to this operator~\cite{Chung:2018kqs}
\bl
\bld
- m \sqrt{u^2} &\supset - \frac{\k m}{2} h_{\m\n} u^\m u^\n
\\ - \half S_{\m\n} \O^{\m\n} &\supset - \half S_{\m\n} u^\l \G^{\n}_{\l\s} g^{\s\m} \,.
\eld
\el
The contribution from the spin-induced multipole terms $L_{SI}$ can be obtained by linearising the Riemann tensor. The spin tensor $S_{\m\n}$ does not have a well-defined analogue in the on-shell formalism, so it is better to work with spin vector. Imposing the covariant SSC reconstructs the spin tensor from the spin vector by the relation $S^{\m\n} = - \frac{1}{m} \e^{\m\n\l\s} p_\l S_\s$. Combining all the expressions and substituting the graviton field $h_{\m\n}$ by the polarisation tensor $2 \ve_\m \ve_\n$ results in the following expression\footnote{Magnetically coupled terms, i.e. odd $n$ terms, implicitly assumes matching onto spin operators in the spinor basis \eqc{PSDef}~\cite{Chung:2018kqs}.}. 
\bl
h_{\m\n} T^{\m\n} = \sum_{n=0}^{\infty} \frac{m x^{2\eta} }{2} \frac{C_{\text{S}^n}}{n!} \left( - \eta \frac{q \cdot S}{m} \right)^n \,. \label{eq:linhsource}
\el
The $\eta$ denotes the sign of the graviton; $+1$ for $+2$ helicity graviton and $-1$ for negative helicity graviton. The definition for the Wilson coefficients $C_{\text{S}^0} = C_{\text{S}^1} = 1$ has been adopted to simplify the equation. When the graviton is on-shell, $q^2 = 0$, \eqc{eq:linhsource} can be viewed as a multipole expansion in spin-induced multipoles.

One way to map the worldline operator \eqc{eq:linhsource} to the three-point amplitude is to consider it as an operator acting on the in-state, which is described by the polarisation tensor $\ve_{1\{J_s\}}$. The amplitude is then obtained by contracting the resulting polarisation tensor, $ h_{\m\n}T^{\m\n} \ve_{1\{J_s\}}$, with the out-state polarisation tensor $\ve_2^{\ast \{I_s\}}$. This approach leads to the following three-point amplitude~\cite{Chung:2018kqs,Chung:2019duq}, where $x$ denotes the $x$-factor \eqc{eq:x_def}.
\bl
M_s^{2 \eta} =  \ve_2^{\ast \{I_s\}} \left[ \sum_{n=0}^{2s} \frac{\k m x^{2\eta} }{2} \frac{C_{\text{S}^n}}{n!} \left( - \eta \frac{q \cdot S}{m} \right)^n \right] \ve_{1\{J_s\}} \,. \label{eq:1bd3ptAmp1}
\el
This expression can be converted into the spinor basis \eqc{eq:poshel3pt} using the following expression for spin operators in the spinor basis
\bl
\bld
(q\cdot S)_{\a}^{~\b} &= \frac{x}{2} \lambda_{q\a} \lambda_q^\b\equiv \frac{x}{2} |q\rangle\langle q| 
\\ (q\cdot S)^{\dot\a}_{~\dot\b} &=-\frac{\tilde{\lambda}_{q}^{\dot\a} \tilde{\lambda}_{q\dot\b}}{2x}\equiv - \frac{|q][q|}{2x}\,,
\eld \label{PSDef}
\el
and recasting the polarisation tensors as symmetrised $s$-copies of the polarisation vector \eqc{eq:polvecdef}, resulting in the expression
\eqa\label{EFT3pt}
&&M_s^{2
} =\sum_{a+b\leq s}\; \frac{\k m x^{2
} }{2} C_{\text{S}^{a+b}}  n^{s}_{a,b} \la \bold{2} \bold{1} \ra^{s-a} \left( - 
\frac{x \la \bold{2}q \ra \la q\bold{1} \ra}{2m} \right)^a [ \bold{2} \bold{1} ]^{s-b} \left( 
\frac{[ \bold{2}q ] [ q\bold{1} ]}{2mx} \right)^b,\nonumber\\
 &&\quad\quad\quad n^{s}_{a,b}\equiv\frac{1}{m^{2s}} {s \choose a} {s \choose b}
 \,.
\eqae
When matching onto minimal coupling \eqc{eq:mincoup_def}, this amplitude has been shown to reproduce the Wilson coefficients of black holes in the classical-spin limit($s \to \infty$ with $s\hbar$ fixed)~\cite{Chung:2018kqs}, indicating that black holes can be considered minimally coupled to gravitons. Alternative arguments for this statement can be found in refs.~\cite{Guevara:2018wpp,Guevara:2019fsj,Arkani-Hamed:2019ymq,Aoude:2020onz}. The Wilson coefficients corresponding to minimal coupling have been explicitly computed in ref.\cite{Chung:2019duq} as an asymptotic expansion in $\frac1s$.
\bl\label{FiniteSpinWil}
C^{Min,s}_{\text{S}^{n}} &= 1 + \frac{n (n-1)}{4s} + \frac{(n^2-5n+10)n(n-1)}{32s^2} + \CO(s^{-3}) \,.
\el

While intuitive, this approach has the problem that expression \eqc{eq:1bd3ptAmp1} does \emph{not} have the properties of spin operator matrix elements as will be discussed in the next section.

\subsection{Hilbert space matching} \label{sec:HilbertMatch}
The spin vector $S^\m$ is identified with the mass-scaled Pauli-Lubanski pseudo-vector $S^\m = m^{-1} W^\m$. The pseudo-vector is defined as
\bl
W^\m &= - \half \e^{\m\n\l\s} P_\n J_{\l\s} \,,
\el
and commutes with momentum generators.
\bl
[ W^\m , P^\n ] = 0 && \Rightarrow && \bra{p';\{I\}} W^\m \ket{p;\{J\}} = \tilde{\delta}^{(3)} (p'-p) \mathbb{W}^\m_{\{I\},\{J\}} (p) \,. \label{eq:spincon}
\el
In this respect, \eqc{eq:1bd3ptAmp1} cannot be interpreted as a genuine spin operator matrix element\footnote{On first sight there seems to be a tension between operators in \eqc{PSDef} having $SL(2,\IC)$ indices and spin operator in \eqc{eq:spincon} having $SU(2)$ indices. However, both definitions for the action of $(q \cdot S)$ operator on spinors of $p_1$ turns out to be equivalent~\cite{Chung:2019duq}.}; the in-momentum and the out-momentum are in general different, so the operators are sandwiched between states of different momenta. This has been noted in various works~\cite{Chung:2018kqs,Chung:2019duq,Guevara:2019fsj,Arkani-Hamed:2019ymq,Maybee:2019jus,Aoude:2020onz,Bern:2020buy}. The resolution to this problem is to map different momentum states onto a common momentum state, which generates spin effects that cannot be ignored in the classical-spin limit.

Polarization vectors are sufficient for demonstrating how the procedure works. Define the rest momentum $p_0$ as the reference momentum. The corresponding polarization vector then takes the form
\eq
\ve^{\mu}_i (p_0)=\delta^{\mu}_i
\eqe
where the little group index on the polarization vector is aligned with the spatial directions. Polarization vector for a generic momentum $p$ can be obtained by applying the boost that transforms $p_0$ to $p$.
\bl
p^\m = G(p;p_0)^\m_{~\n} p_0^\n \,, \quad \ve_I^\m(p) = G(p;p_0)^\m_{~\n} \ve^\n_I (p_0) \,. \label{eq:BoostDef}
\el
The explicit form of \emph{minimal boost} $G(p;p_0)^{\mu}\,_{\nu}$ is given as:
\bl
\bld
G(p;p_0)^\m_{~\n} &= \delta^\m_{~\n} - \frac{(p +p_0)^\m (p + p_0)_{\n}}{(p \cdot p_0) + m^2} + \frac{2 p^\m p_{0 \n}}{m^2} \,.
\eld \label{eq:BoostExplicit}
\el
By definition, the in- and out-momenta polarisation vectors satisfy the relation
\bl\label{Relate1}
\ve_I (p_{out}) &= G(p_{out};p_0) G(p_{in};p_0)^{-1} \ve_I (p_{in})\,.
\el
The three-point amplitude constructed in the previous section can be put in the abstract form
\eq\label{MAP}
\ve^{\ast \m}_I (p_{out})\, \mathcal{O}^K\,_J \,\ve_{K\m} (p_{in})\,,
\eqe
where the EFT operators $ \mathcal{O}^K\,_J$ act on the little group space. Inserting \eqc{Relate1} into \eqc{MAP} yields
\eqa
\ve^{\ast \m}_I (p_{out})\, \mathcal{O}^K\,_J \,\ve_{K\m} (p_{in})&=&  \e^{\ast \m}_I (p_{in}) \left[ G(p_{in};p_0) G(p_{out};p_0)^{-1} \right] \mathcal{O}^K\,_J \,\ve_{K\m} (p_{in})\nonumber\\
&=&\ve^{\ast \m}_I (p_{in})\, \widetilde{\mathcal{O}}^K\,_J \,\ve_{K\m} (p_{in}) \,,\label{eq:PolContIn}
\eqae
where $\widetilde{\mathcal{O}}^K\,_J$ is the matrix element acting on the little group space of the $in$-state particle. Interpreting the little group operator $\widetilde{\mathcal{O}}^K\,_J$ as spin operators is consistent with the requirement \eqc{eq:spincon}.

The operator inserted in \eqc{eq:PolContIn} can be decomposed into two parts; the Thomas-Wigner rotation factor and the pure boost factor.
\bl
\bld
G(p_{in};p_0) G(p_{out};p_0)^{-1} &= U (p_{in}; p_0, p_{out}) G(p_{in}; p_{out})
\\ U (p_{in}; p_0, p_{out}) &= G(p_{in};p_0) G(p_0;p_{out}) G(p_{out};p_{in}) \,.
\eld \label{eq:HMfactors}
\el
The rotation factor $U$ is generically nontrivial and its effect cannot vanish in the classical-spin limit ($\hbar \to 0$ with fixed $s\hbar$~\cite{Chung:2019duq}); it is a classical effect, and an analogue of this effect also exists in EFT approach~\cite{Levi:2015msa}. This factor is essential for constructing the correct effective Hamiltonian~\cite{Chung:2020rrz}, which will be discussed in detail in section~\ref{sec:rotation}. For discussion of three-point amplitudes, however, the rotation factor is trivial; all reference momentum $p_0$ that can be constructed from momenta of external legs cannot generate nontrivial rotation factor.

The boost term $G(p_{out};p_{in})$ is interesting in that a) its effect is basis-dependent, and b) its effect vanishes for the classical-spin limit when Lorentz tensors are used for polarisations. The basis-dependence can be seen by explicit evaluation of the Lorentz generators;
\begin{gather}
J^i = \left\{ \begin{aligned}
\frac{1}{2} \left( \s^i \right)_{\a}^{~\b} &\quad \text{Chiral}
\\ \frac{1}{2} \left( \s^i \right)^{\dot\a}_{~\dot\b} &\quad \text{Anti-chiral}
\end{aligned} \right. \, , \quad K^i = \left\{ \begin{aligned}
\frac{i}{2} \left( \s^i \right)_{\a}^{~\b} &\quad \text{Chiral}
\\ - \frac{i}{2} \left( \s^i \right)^{\dot\a}_{~\dot\b} &\quad \text{Anti-chiral}
\end{aligned} \right.
\end{gather}
where $\s^i$ are the Pauli matrices, $J^i = \half \e^{ijk} J^{jk}$ are the rotation generators, and $K^i = J^{i0}$ are the boost generators\footnote{Treatment of boost generators as rotation generators in general spacetime dimensions has been given in~\cite{Bautista:2019tdr}.}. The explicit form for the Lorentz group generators in the representation $(\frac{s}{2},\frac{s}{2})$, the \emph{Lorentz tensor representation}, are obtained as a tensor sum of above.
\bl
\bld
J^i &= \half (\s^i \otimes \overbrace{\cdots \otimes \iden}^{2s-1} + \cdots + \overbrace{\iden \otimes \cdots}^{2s-1} \otimes \s^i)
\\ K^i &= \frac{i}{2} (\s^i \otimes \overbrace{\cdots \otimes \iden}^{2s-1} + \cdots + \overbrace{\iden \otimes \cdots}^{s-1} \otimes \s^i \otimes \overbrace{\cdots \otimes \iden}^{s}
\\ &\phantom{=asdfasdfasdfasdf} - \overbrace{\iden \otimes \cdots}^{s} \otimes \s^i \otimes \overbrace{\cdots \otimes \iden}^{s-1} - \cdots - \overbrace{\iden \otimes \cdots}^{2s-1} \otimes \s^i) \,.
\eld
\el
Since little group indices will always be symmetrised, the expressions for generators and their products can be simplified further. For this purpose, let us first fix the normalisations of the spinors for particles of unit mass at rest, where arrows $\uparrow$ and $\downarrow$ are the little group indices.
\bl
\bld
\ket{\bf{0}^\uparrow}_\a = \sket{\bf{0}^\uparrow}^{\dot\a} = {1 \choose 0} &,\quad \ket{\bf{0}^\downarrow}_\a = \sket{\bf{0}^\downarrow}^{\dot\a} = {0 \choose 1}
\\ \bra{\bf{0}^\uparrow}^\a = - \sbra{\bf{0}^\uparrow}_{\dot\a} = - (0 \quad 1) &,\quad \bra{\bf{0}^\downarrow}^\a = - \sbra{\bf{0}^\downarrow}_{\dot\a} = (1 \quad 0)
\eld
\el
The second line follows from the first line by adopting the definition $\e^{\uparrow \downarrow} = +1$. Adopting this normalisation, the generators and their products are simplified as below where $\stackrel{\cdot}{=}$ denotes numerical equivalence when inserted between bra and ket vectors of spin-$s$ states in the rest frame.
\bl
\bld
J^i &\stackrel{\cdot}{=} 2s \times \half \s^i \otimes \overbrace{\cdots \otimes \iden}^{2s-1}
\\ J^i J^j &\stackrel{\cdot}{=} 2s \times \frac{1}{2^2} \s^i \s^j \otimes \overbrace{\cdots \otimes \iden}^{2s-1} + (2s)(2s-1) \times \frac{1}{2^2} \s^i \otimes \s^j \otimes \overbrace{\cdots \otimes \iden}^{2s-2}
\\ K^i &\stackrel{\cdot}{=} 0
\\ K^i K^j &\stackrel{\cdot}{=} - 2s \times \frac{1}{2^2} \s^i \s^j \otimes \overbrace{\cdots \otimes \iden}^{2s-1} - \left[ 2 s (s-1) - 2 s^2 \right] \times \frac{1}{2^2} \s^i \otimes \s^j \otimes \overbrace{\cdots \otimes \iden}^{2s-2}
\eld
\el
The symmetrisation argument can be used to show that $(K)^{2n+1} (J)^m \stackrel{\cdot}{=} 0$, so only even powers of $\vec{K}$ need to be worked out. The contribution with largest $s$ dependence will be the contribution where all Pauli matrices are allotted to different spinor indices, given that $s>n$. The coefficient for such a contribution can be worked out from simple combinatorics.
\bl
\overbrace{J \cdots J}^{2n} &\stackrel{\cdot}{\simeq} \frac{(2s)!}{(2s-2n)!} \frac{1}{2^{2n}} \overbrace{\s \otimes \cdots}^{2n} \otimes \overbrace{\iden \otimes \cdots \otimes \iden}^{2s-2n} + \cdots
\\ \overbrace{K \cdots K}^{2n} &\stackrel{\cdot}{\simeq} 
(2n)! {s \choose n} \frac{1}{2^{2n}} \overbrace{\s \otimes \cdots}^{2n} \otimes \overbrace{\iden \otimes \cdots \otimes \iden}^{2s-2n} + \cdots
\el
Comparing the two yields the relation
\bl
\frac{1}{(2n)!} \left( \vec{\l} \cdot \vec{K} \right)^{2n} &\stackrel{\cdot}{=} \frac{(2s-2n)!}{(2s)!} {s \choose n} \left( \vec{\l} \cdot \vec{J} \right)^{2n} + ( \vec{\l} )^2 F_{2n-2} (\vec{\l} \cdot \vec{J}) \label{eq:Boost2Rot}
\el
where $F_{2m}(x)$ is some even polynomial of degree $2m$. The appearance of the factor $(\vec{\l})^2$ follows from anti-commutator of Pauli matrices; $\s^i \s^j + \s^j \s^i = 2 \delta^{ij}$.

To demonstrate vanishing contributions in the tensor basis, consider the following on-shell three-point kinematics where momentum $\vec{q}$ is complex null; $( \vec{q} )^2 = 0$.
\bg
p_1 = (m, \vec{0}) \,,\quad q = (0, \vec{q}) \,,\quad p_2 = (m, -\vec{q}) \label{eq:3ptKinConfig}
\eg
In the Lorentz tensor basis, the little group matrix element $\ve^\ast_I (\bf{2}) \cdot \ve_J (\bf{1})$ is computed as
\bl
\bld
\ve^\ast_I (\bf{2}) \cdot \ve_J (\bf{1}) &= \ve^\ast_I (\bf{1}) \left[ e^{i \frac{\vec{q}}{m} \cdot \vec{K}} \right] \ve_J (\bf{1}) = \ve^\ast_I (\bf{1}) \left[ \sum_{n=0}^{s} \frac{(-1)^n}{(2n)!} \left( \frac{\vec{q} \cdot \vec{K} }{m} \right)^{2n} \right] \ve_J (\bf{1})
\\ &= \ve^\ast_I (\bf{1}) \left[ \sum_{n=0}^{s} \frac{(-1)^n (2s-2n)!}{(2s)!} {s \choose n} \left( \frac{{q} \cdot {S} }{m} \right)^{2n} \right] \ve_J (\bf{1})
\eld \label{eq:3ptBoost2Rot}
\el
where \eqc{eq:Boost2Rot} has been used to obtain the last line together with the condition $( \vec{q} )^2 = 0$. The coefficient of $(q \cdot S)^{2n}$ scales as $(-4s)^{-n}$ in the limit $s \to \infty$, so they are finite spin effects for $n \neq 0$. Inserting these finite spin pieces into \eqc{eq:1bd3ptAmp1} will give the following result.
\bl
\bld
M_s^{2\eta} &= \frac{\k m x^{2 \eta}}{2} \ve^\ast_I (\bf{1}) \left[ \sum_{i=0}^{s} \frac{(-1)^i (2s-2i)!}{(2s)!} {s \choose i} \left( - \eta \frac{{q} \cdot {S} }{m} \right)^{2i} \right]
\\ &\phantom{=asdfasdfasdfasdfasdfasdf} \times \left[ \sum_{j=0}^{2s} \frac{C_{\text{S}^j}}{j!} \left( - \eta \frac{q \cdot S}{m} \right)^j \right] \ve_J (\bf{1})
\\ &= \frac{\k m x^{2 \eta}}{2} \ve^\ast_I (\bf{1}) \left[ \sum_{n=0}^{2s} \frac{C_{\text{S}^n_{eff}}}{n!} \left( - \eta \frac{q \cdot S}{m} \right)^n \right] \ve_J (\bf{1})\,,
\eld \label{eq:EFT3ptWCeff}
\el
where we've incorporated the Hilbert-space matching terms to define the effective Wilson coefficient $C_{\text{S}^n_{eff}}$. The relation between $C_{\text{S}^n}$ and $C_{\text{S}^n_{eff}}$ is given as:
\bl
\bld
\frac{C_{\text{S}^m_{eff}}}{m!} &= \sum_{i=0}^{\lfloor m/2 \rfloor} \frac{(-1)^i (2s-2i)!}{(2s)!} {s \choose i} \frac{C_{\text{S}^{m-2i}}}{(m-2i)!}
\\ &= \sum_{n=0} \left( \delta_{m,n} - \frac{\delta_{m-n,2}}{4s} + \frac{\delta_{m-n,4} - 4 \delta_{m-n,2}}{32s^2} + \CO(s^{-3}) \right) \frac{C_{\text{S}^n}}{n!}\,. \label{eq:bareWC2eff}
\eld
\el
One can interpret $C_{\text{S}^n_{eff}}$ as the $2^n$-multipole of the particle which would be measured by an observer at infinity.\footnote{Indeed the reference frame $p_0$ chosen here is very similar to the "body-fixed frame" introduced in~\cite{Levi:2015msa}.} Remarkably, substituting the Wilson coefficients for minimal coupling while keeping the finite-spin effects, for example (\ref{FiniteSpinWil}), we find that the effective $C_{\text{S}^n_{eff}}$ turns out to be unity! In other words, \emph{minimal coupling reproduces the Kerr Black hole Wilson coefficients at finite spins once the Hilbert space matching terms are included!}

To prove that minimal coupling reproduces Wilson coefficients of Kerr black holes after Hilbert-space matching, we first note that chiral spinor brackets and anti-chiral spinor brackets can be exchanged when momentum of in-state $p_1$ and out-state $p_1'$ are the same, $p_1 = p_1'$.
\bl
\la \mathbf{1'} \bf{1} \ra = - [\mathbf{1'1}]
\el
Following the kinematical set-up of \eqc{eq:3ptKinConfig}, we go to the frame where $p_1 = p_1'$ is at rest. The Hilbert space matching of minimal coupling becomes
\bl
\bld
\la \bf{21} \ra^{2s} &= \bra{\mathbf{1'}}^{2s} e^{i \frac{q \cdot K}{m}} \ket{\bf{1}}^{2s} = \bra{\mathbf{1'}}^{2s} e^{- \frac{q \cdot S}{m}} \ket{\bf{1}}^{2s} \simeq \ve^\ast ({\mathbf{1'}}) e^{- \frac{q \cdot S}{m}} \ve({\bf{1}})
\\ [ \bf{21} ]^{2s} &= \sbra{\mathbf{1'}}^{2s} e^{i \frac{q \cdot K}{m}} \sket{\bf{1}}^{2s} = \sbra{\mathbf{1'}}^{2s} e^{\frac{q \cdot S}{m}} \sket{\bf{1}}^{2s} \simeq \ve^\ast ({\mathbf{1'}}) e^{ \frac{q \cdot S}{m}} \ve({\bf{1}})
\eld \label{eq:MinKerr}
\el
which is the chiral(anti-chiral) basis version of \eqc{eq:3ptBoost2Rot}. The relations $K = i J = i S$ for chiral spinors and $K = - i J = - i S$ for anti-chiral spinors has been used, and $\simeq$ in the above expression denotes equivalence up to normalisation. In other words, \emph{minimal coupling corresponds to unity Wilson coefficients} $C_{\text{S}^n_{eff}} = 1$. The same conclusion has been reached from heavy particle effective theory (HPET) point of view in~\cite{Aoude:2020onz}.

The map can be generalised to arbitrary $C_{\text{S}_{eff}^n}$. Using the expression for $\frac{q \cdot S}{m}$ in the chiral basis~\cite{Chung:2018kqs},
\bl
\bld
g_i \la \bf{21} \ra^{2s-i} \left( \frac{x \la \bf{2}q \ra \la q \bf{1} \ra}{m} \right)^i &= \frac{2^i (2s-i)! g_i}{(2s)!} \bra{\bf{2}}^{2s} \left( \frac{q \cdot S}{m} \right)^i \ket{\bf{1}}^{2s}
\\ &= \frac{2^i \hat{g}_i}{i!} \bra{\bf{2}}^{2s} \left( \frac{q \cdot S}{m} \right)^i \ket{\bf{1}}^{2s} \,,
\eld
\el
where we have introduced the notation $g_i = {2s \choose i} \hat{g}_i$. The above expression can be matched to \eqc{eq:EFT3ptWCeff}, which expresses the amplitude through $C_{\text{S}_{eff}^n}$.
\bl
\bld
\frac{2^i \hat{g}_i}{i!} \bra{\bf{2}}^{2s} \left( \frac{q \cdot S}{m} \right)^i \ket{\bf{1}}^{2s} &= \frac{2^i \hat{g}_i}{i!} \bra{\mathbf{1'}}^{2s} e^{- \frac{q \cdot S}{m}} \left( \frac{q \cdot S}{m} \right)^i \ket{\bf{1}}^{2s}
\\ &= \bra{\mathbf{1'}}^{2s} \sum_n \frac{C_{\text{S}_{eff}^n}}{n!} \left( - \frac{q \cdot S}{m} \right)^n \ket{\bf{1}}^{2s} \,,
\eld
\el
yielding the relation
\bl
g_i &= {2s \choose i} \hat{g}_i = \frac{1}{2^i} {2s \choose i} \sum_{n} (-1)^n {i \choose n} C_{\text{S}_{eff}^n} \,. \label{eq:WCeff2gi}
\el

\subsection{The residue integral representation}\label{sec:resrep}
We may ask if there is an expression for the three-point amplitude that directly expresses the ampitude in terms of $C_{\text{S}_{eff}^n}$. Such an alternative expression can be found by adopting following definitions.
\bl
\bld
\bar{u}(p_2) u(p_1) &= \frac{ [ \bf{21} ] - \la \bf{21} \ra }{2m}
\\ S_{1/2}^\m &= \half \bar{u}(p_2) \g^\m \g_5 u(p_1) = - \frac{1}{4m} \left( \sbra{\bf{2}} \bar{\s}^\m \ket{\bf{1}} + \bra{\bf{2}} \s^\m \sket{\bf{1}} \right)
\eld
\el
The definition for spin vector $S^\m_{1/2}$, which can be considered as the scaled spin vector $\frac{S_s^\m}{2s}$ of the full spin vector of a spin-$s$ particle $S_s^\m$, has been adopted from Holstein and Ross~\cite{Holstein:2008sx} with a sign choice that matches to our conventions. An extra factor of $\frac{1}{2m}$ has been inserted as a normalisation condition $\bar{u}_I (p) u^J (p) = \delta_I^J$. Setting the momentum conservation condition as $p_2 = p_1 + q$, we propose the following \emph{residue integral representation} of three-point amplitude which expresses spin-$s$ amplitude as $2s$ power of spin-$\half$ amplitude.
\bl
\bld
M_s^{2 \eta} &= \frac{\k m x^{2\eta}}{2} \oint \frac{dz}{2\pi i z} \left( \sum_{n=0}^\infty C_{\text{S}_r^n} z^n \right) \left( \bar{u}(p_2) u(p_1) - \eta \frac{q \cdot S_{1/2}}{m z} \right)^{2s}
\\ &= \frac{\k m x^{2\eta}}{2} \oint \frac{dz}{2\pi i z} \left( \sum_{n=0}^\infty C_{\text{S}_r^n} z^n \right) \left( \frac{ [ \bf{21} ] - \la \bf{21} \ra }{2m} - \frac{\eta}{z} \frac{ [ \bf{21} ] + \la \bf{21} \ra }{2m} \right)^{2s}\,.
\eld \label{eq:EFT3ptNew}
\el
Here the contour encircles the origin, and the contour integral merely serves the auxiliary function of extracting the right combinatoric factors. For positive helicity $\eta = +1$, this expression becomes
\bl
M_s^{+2} &= (-1)^{2s} \frac{\k m x^2}{2} \oint \frac{dz}{2\pi i z} \left( \sum_{n=0}^\infty C_{\text{S}_r^n} z^n \right) \left( \frac{ \la \bf{21} \ra }{m} + \frac{z-1}{z} \frac{x \la \bf{2}3 \ra \la 3 \bf{1} \ra}{2m^2} \right)^{2s} \,.
\el
Using the binomial expansion leaves the following residue integral to be worked out.
\bl
\oint \frac{dz}{2\pi i z} z^n \left( \frac{z-1}{z} \right)^i = (-1)^i \sum_{j=0}^{i} \oint \frac{dz}{2\pi i z} { i \choose j } (-z)^j z^{n-i} = (-1)^n { i \choose n } \,.
\el
One then finds:
\bl
\bld
M_s^{+2} &= (-1)^{2s} \frac{\k m x^2}{2} \sum_{i=0}^{2s} { 2s \choose i} \left[ \frac{1}{2^i} \sum_{n=0}^{\infty} (-1)^n { i \choose n} C_{\text{S}_r^n} \right] \left( \frac{ \la \bf{21} \ra }{m} \right)^{2s-i}\left( \frac{x \la \bf{2}3 \ra \la 3 \bf{1} \ra}{m^2} \right)^{i}
\\ &\equiv (-1)^{2s} \frac{\k m x^2}{2 m^{2s}} \left[ g_0^r \la \bold{2} \bold{1} \ra^{2s} + g_1^r \la \bold{2} \bold{1} \ra^{2s-1} \frac{x \la \bold{2} q \ra \la q \bold{1} \ra}{m} + \cdots + g_{2s}^r \frac{(x \la \bold{2} q \ra \la q \bold{1} \ra)^{2s}}{m^{2s}} \right]
\eld
\el
Comparing the coefficients $g_i$ of \eqc{eq:WCeff2gi} with the above formula, we can conclude that the Wilson coefficients $C_{\text{S}_r^n}$ used in \eqc{eq:EFT3ptNew} is equivalent to $C_{\text{S}_{eff}^n}$; $C_{\text{S}_r^n} = C_{\text{S}_{eff}^n}$.

The representation \eqc{eq:EFT3ptNew} describes a spin-$s$ particle as symmetrised product of $2s$ copies of a spin-$1/2$ particle. One advantage of this representation is that evaluation of cuts become simple; the sum over $2s$ intermediate states can be substituted by a sum over 2 intermediate states powered to order $2s$. This property follows from the following identity.
\bl
\bld
\sum_{\bf{P}^{2s}} \CF_1(\bf{P})^{2s} \CF_2(\bf{P})^{2s} = \left[ \sum_{\bf{P}} \CF_1(\bf{P}) \CF_2(\bf{P}) \right]^{2s}
\eld
\el
In the above identity, $\CF_1$ and $\CF_2$ are arbitrary expressions linear in the massive spinor-helicity variable schematically written as $\bf{P}$. This identity can be proved by writing the sum as the sum over overcomplete basis of spin coherent states.

Another advantage of the expression \eqc{eq:EFT3ptNew} is that it allows us to straightforwardly take the infinite spin-limit and connect to one-particle EFT three-point amplitude. Formally writing $\bar{u}u = 1$ and suppressing the subscript $s$ of $S^\m_s$,
\bl
\bld
\lim_{s \to \infty }M_s^{2 \eta} &= \lim_{s \to \infty} \frac{\k m x^{2\eta}}{2} \oint \frac{dz}{2\pi i z} \left( \sum_{n=0}^\infty C_{\text{S}_r^n} z^n \right) \left( 1 - \frac{1}{2s} \eta \frac{q \cdot S}{m z} \right)^{2s}
\\ &= \frac{\k m x^{2\eta}}{2} \sum_{n=0}^\infty \frac{C_{\text{S}_r^n} }{n!} \left( - \eta \frac{q \cdot S}{m} \right)^n \,,
\eld
\el
which is reminiscent the one-particle EFT amplitude \eqc{eq:1bd3ptAmp1}.

\section{Constructing the 1PM Hamiltonian from amplitudes} \label{sec:1PM_Ham_full}
\subsection{Kinematics for 1PM computation} \label{sec:amp}

\begin{figure}
\centering
\includegraphics[scale=0.5]{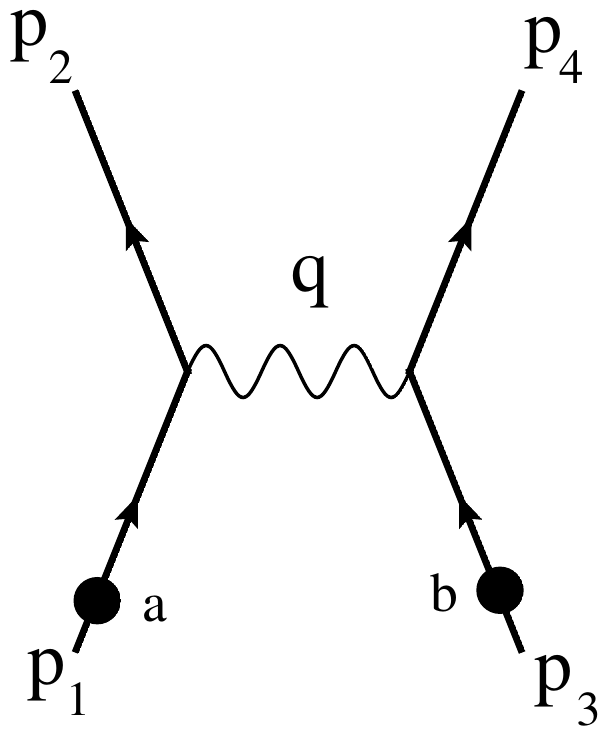}
\caption{The graviton exchange diagram between source $a$ and $b$ that yields the leading $1/q^2$ singularity, which is responsible for the classical potential.}
\label{GraviExchange}
\end{figure}

The 1PM classical potential can be extracted from the singular limit of a single graviton exchange between two compact spinning objects, i.e. the $2\rightarrow 2$ elastic scattering amplitude shown in Fig.~\ref{GraviExchange}. As mentioned in section~\ref{sec:Amp2Ham}, comparison is made in the COM frame \eqc{Kin1sec2} which is reproduced below.
\bg\label{Kin1}
\bgd
p_1 = ( E_a, \vec{p} {+} \vec{q} / 2 ), \;\;
p_3 = ( E_b, {-} \vec{p}{-} \vec{q} / 2 ), \;\;  
\\ p_2 = ( E_a, \vec{p} {-} \vec{q} / 2 ) , \;\;
p_4 = ( E_b, {-} \vec{p} {+} \vec{q} / 2 ) \,,
\egd
\eg
The exchanged momentum $q^\mu=(p_{1}-p_2)^\mu=(0,\vec{q})$ is space-like. In terms of four-vectors, we have
\begin{align}
p_1 = p_a + q/2 \,, \quad p_2 = p_a - q/2 \,, \quad 
p_3 = p_b -q/2 \,, \quad p_4 = p_b + q/2 \,.
\label{kin-scattering}
\end{align}
The momenta $p_a$ and $p_b$ denote the average momentum of each particle, which are not necessarily on-shell.

In the asymptotic region the spinning particles are free and characterized by their momenta and (Pauli-Lubanski) spin vectors. Rescaling them by the masses give the proper velocity and ``spin-length" vectors:
\begin{align}
u_\mu = \frac{1}{m} p_\mu \,, 
\quad 
a_\mu = \frac{1}{m} s_\mu \,.
\end{align}
We denote the Lorentz invariant amplitude by $\mathcal{M}$ and the non-relativistic one by $M$. The two are related by 
\begin{align}
M = \frac{1}{4E_a E_b} \mathcal{M} \,.
\end{align}
We adopt (with slight modification) the kinematic variables of by Bern et al.~\cite{Bern:2019nnu}:
\begin{align}\label{eq:Def}
\begin{split}
&m = m_a + m_b \,, \quad
\nu = m_a m_b /m^2 \,,  
\\
E_{a,b} = \sqrt{\vec{p}^2 + m_{a,b}^2} \,, \quad 
&E = E_a + E_b \,, \quad 
\xi = E_a E_b / E^2 \,, \quad 
\gamma = E/m \,,
\\
&\sigma = \frac{p_a \cdot p_b}{m_a m_b} = u_a \cdot u_b \equiv \cosh\theta \,.
\end{split}
\end{align}
The first and the third lines are Lorentz invariant, whereas the second line is specific to the COM frame. The non-relativistic (NR) limit is characterised by the limits $\sigma\rightarrow 1$ and $\theta\rightarrow 0$.

We will be only interested in the \emph{classical-spin limit} of long-distance physics; we take the limits $\hbar \to 0$ with $s\hbar$ fixed, as explained in section~\ref{sec:Amp2Ham_sm}. The physics is dominated by small $|q|$ effects so we expand in small $|q|$, and since in Lorentzian signature this translate to the zero momentum limit, we will analytically continue to complex (or split signature) momenta. In this case we can have $|q|\rightarrow0$ correspond to null momenta, $q^2=0$. The advantage of such analytic continuation is that with $q^2=0$, the amplitude factorizes into the product of two three-point amplitudes. This approach was introduced by Guevara~\cite{Guevara:2017csg} and named the \emph{holomorphic classical limit} (HCL). The leading order potential is extracted as 
\eqa
V(p,q)\sim\left.\frac{\CM_4(s,q^2)}{4 E_aE_b}\right|_{q^2\rightarrow0}\,.
\eqae 
We write $\sim$ here because for spinning objects, the particles are irreps in distinct little group space as we have seen in section~\ref{sec:HilbertMatch}, and its proper treatment will introduce additional factors which we address in section~\ref{sec:rotation}. The analytically continued HCL kinematics is characterized in a Lorentz invariant way as 
\begin{align}
q^2 = p_a \cdot q = p_b \cdot q = 0\,. 
\end{align}
This implies, for example, 
\begin{align}
\pm i \ve_{\m\n\r\s} p_a^\m p_b^\n q^\r a^\s = m_a m_b (\sinh\theta) (q \cdot a) \,.
\label{pole-strange}
\end{align}
which can be derived by squaring both sides and identifying the determinant of the Gram matrix for the LHS. The sign ambiguity in the above can  also be seen from the definition of $\theta$ in (\ref{eq:Def}), where it is invariant under $\theta\leftrightarrow -\theta$. As we will see later on, our potential will be an even function of $\theta$, and thus the ambiguity is irrelevant.\footnote{The difference for the two choices will be purely imaginary, and is relevant when considering electromagnetic interactions associated with dyons~\cite{Caron-Huot:2018ape} and gravitational dynamics in Taub-NUT space-time~\cite{Huang:2019cja}.} 
\subsection{1PM amplitude} 

The tree-level graviton exchange between two massive scalars is given as 
\begin{align}
\mathcal{M} &= - (16\pi G) \frac{m_a^2 m_b^2}{q^2} (2\sigma^2 - 1) 
= - (16\pi G) \frac{m_a^2 m_b^2}{q^2} \cosh(2\theta) \,.
\end{align}
The spinning analogue was computed by on-shell methods for Kerr black holes in \cite{Guevara:2018wpp,Chung:2018kqs,Guevara:2019fsj} and 
then generalized to general compact spinning bodies in \cite{Chung:2019duq}.
In our conventions, the result of \cite{Chung:2019duq} can be written as 
\begin{align}
\mathcal{M}_{\rm bare} =  - (16\pi G) \left(\frac{m_a^2 m_b^2}{q^2}  \right)
\left[ 
\frac{1}{2} \sum_{s=\pm 1} 
e^{2s\theta} W_a(s\tau_a) W_b(s\tau_b) \right] \,.
\label{spin-amp-bare}
\end{align}
We call it the bare amplitude to emphasize that it is missing the rotation factor alluded to in section~\ref{sec:HilbertMatch}, which we elaborate in section~\ref{sec:rotation}. The factor is missing in \eqc{spin-amp-bare} because the polarisation tensors for asymptotic states bave been systematically stripped off in this expression, as explained in \cite{Guevara:2018wpp,Chung:2018kqs,Guevara:2019fsj,Maybee:2019jus}.

We should also stress that \eqref{spin-amp-bare} is only the leading piece of the amplitude in the $q^2\rightarrow 0$ limit, i.e. the ``leading singularity" of the exchange diagram, which will be sufficient to determine the 1PM potential. The variables $\tau_{a,b}$ in \eqref{spin-amp-bare} are ``off-HCL'' continuation of $q \cdot a_{a,b}$ defined by 
\begin{align}
\tau_{a,b} = i \frac{\ve(q,u_a,u_b,a_{a,b})}{\sinh\theta} \stackrel{\text{HCL}}{\simeq} q \cdot a_{a,b} \,,
\quad 
\ve(a,b,c,d) = \ve_{\m\n\r\s} a^\m b^\n c^\r d^\s \,,
\label{def:tau}
\end{align}
where the spin-length vectors $a_{a,b}^\m \equiv S_{a,b}^\m/m_{a,b}$ 
are regarded as classical variables, even though they originate from quantum operators during the computation of the amplitude. When actually computing the potential we adopt the off-HCL continuation \eqc{def:tau} only when odd powers of $\t$ appear, and for even powers of $\t$ the original expression $q \cdot a$ will be taken. With (\ref{def:tau}), we see that (\ref{spin-amp-bare}) is an even function of $\theta$ as advertised.

Despite its appearance, $\tau_{a,b}$ is not singular in the NR limit $\theta \rightarrow 0$, since the invariant ``area" spanned by the two velocity vectors, $u_a$ and $u_b$, is precisely $\sinh \theta$. In other words, the anti-symmetric tensor,  
\begin{align}
\w_{\m\n} \equiv \frac{1}{\sinh\theta} \ve_{\m\n\r\s} u_a^\r u_b^\s \,,
\end{align}
reflects only the orientation of the 2-plane spanned by the two velocities. 

The functions $W_{a,b}$ encode the gravitational couplings of the action \eqc{eq:EFTSI}, being defined as the generating function: 
\begin{align}\label{GenDef1}
W(\tau) = \sum_{n=0}^\infty \frac{C_{\text{S}^n}}{n!} \tau^n \,.
\end{align}
For a Kerr black hole, $C_{\text{S}^n} = 1$ for all $n$ so $W(\tau) = e^\tau$. The notation $C_n = C_{\text{S}^n}$ will also be used to simplify the equations.

It is sometimes useful to separate the even and odd parts of the generating functions, 
$W_{\pm} = \frac{1}{2}\left[W(\tau) \pm W(-\tau)\right]$,
so that we can write
\begin{align}
\begin{split}
\mathcal{M}_{\rm bare} &= {-} (16\pi G) \left(\frac{m_a^2 m_b^2}{q^2}  \right)\frac{1}{2} \sum_{s=\pm 1} 
e^{2s\theta} W_a(s\tau_a) W_b(s\tau_b) 
\\
&= {-}16\pi G\frac{m_a^2 m_b^2}{q^2}\left[ \cosh(2\theta) (W_{a+}W_{b+} {+} W_{a-}W_{b-}) \right.
\\ &\phantom{=-16\pi G m_a^2 m_b^2 asdf} \left. {+} \sinh(2\theta) (W_{a+}W_{b-} {+} W_{a-}W_{b+})  \right]\,.
\end{split}
\label{spin-odd-even-x}
\end{align}
As noticed by \cite{Guevara:2018wpp,Chung:2018kqs,Guevara:2019fsj}, when both spinning bodies are Kerr black holes, the amplitude 
takes a particularly simple form: 
\begin{align}
\mathcal{M}_{\rm bare}^{\rm (BH)} =  - (16\pi G) \frac{m_a^2 m_b^2}{q^2}  
\cosh\left( 2\theta  + i \frac{\ve(q,u_a,u_b,a_0)}{\sinh\theta} \right) \,, 
\label{spin-amp-BH}
\end{align}
where $a_0^\m = a_a^\m + a_b^\m$ is the total spin-length vector. To extract the classical potential, the above result needs to be dressed by additional factors coming from definition of polarization tensors, 
whose ``appetiser'' has been discussed in section~\ref{sec:HilbertMatch}.

\subsection{Thomas-Wigner rotation} \label{sec:rotation}
The amplitude by definition is a matrix element between distinct (little group) Hilbert spaces, one for each asymptotic state. As discussed in section~\ref{sec:HilbertMatch}, the amplitude generically contains non-trivial rotation factors induced from mapping relations between the distinct Hilbert spaces.  Unlike the three-point kinematics discussed in section~\ref{sec:HilbertMatch}, in the $2\to2$ scattering kinematics the reference momenta $p_0$ are identified as the center of mass momenta, i.e. 
\bl
p_{0,a/b} &= \frac{m_{a/b}}{\sqrt{(p_1 {+} p_3)^2}} (p_1 {+} p_3) \,,
\label{eq:RefMomDef}
\el
where $p_{0,a/b}$ are appropriately normalized for particle $a,b$ respectively. This reference momentum is unique in that only this choice allows analytic continuation of scattering dynamics to bound motion dynamics. Recall that the rotation factor \eqc{eq:HMfactors} is given as
\eqa
U = G(p_{\rm in};p_{0})G(p_0;p_{\rm out}) G(p_{\rm out};p_{\rm in})\,.
\eqae
We now derive its rotation angle. 
Let $u$, $v$, $w$ be 4-velocity vectors; each one is time-like, unit-normalized and future-pointing. 
The inversion of minimal boost $G$, \eqc{eq:BoostExplicit}, exchanges the roles of $u$ and $v$:
\begin{align}
G(u,v)^{-1} = G(v,u) \,.
\end{align}
Now consider a closed loop of three minimal boosts, $G(u,v)G(v,w)G(w,u)$. Since it takes $u$ back to itself, the result should be a rotation on the 3-plane orthogonal to $u$. In a suitably chosen basis, the rotation would be represented by
\begin{align}
[G(u,v)G(v,w)G(w,u)]^\m{}_\n = 
\begin{pmatrix}
1 & 0 & 0 & 0 \\
0 & \cos\alpha & -\sin\alpha & 0 \\
0 & \sin\alpha & \cos\alpha & 0 \\
0 & 0 & 0 & 1 
\end{pmatrix} \,.
\end{align}
A manifestly Lorentz-invariant way to characterize the angle $\alpha$ is 
\begin{align}
\tr [G(u,v)G(v,w)G(w,u)] = 2 + 2 \cos\alpha\,.
\label{tr-cos-a}
\end{align}
Taking the trace explicitly using \eqref{eq:BoostExplicit}, we reproduce a well-known formula for the angle:
\begin{align}
2 + 2 \cos\alpha = 2 \frac{ (1+ u\cdot v+v\cdot w + w\cdot u)^2}{(1+u\cdot v)(1+v\cdot w)(1+w\cdot u)}\,.
\label{tr-cos-b}
\end{align}
We find it useful to rewrite \eqref{tr-cos-b} as 
\begin{align}
1 - \cos\alpha = \frac{- (\ve_{\m\n\r\s} u^\n v^\r \w^\s)^2}{(1+u\cdot v)(1+v\cdot w)(1+w\cdot u)} \,.
\label{alpha-great}
\end{align}
The $(-)$ sign on the RHS reflects the fact that the vector $\ve_{\m}(u,v,w) \equiv \ve_{\m\n\r\s} u^\n v^\r \w^\s$ is space-like when $u$, $v$, $w$ are time-like. The relation \eqref{alpha-great} clearly shows that the angle $\alpha$ vanishes when $u$, $v$, $w$ are linearly dependent. 

\paragraph{Scattering kinematics in the COM frame} 

Let us now specialize to the kinematics of the two body scattering \eqref{kin-scattering}. To compute the rotation angle $\alpha$ for particle $a$, we identify the velocity vectors to be
\begin{align}
u =\frac{p_{\rm in}}{m_a} =\frac{p_1}{m_a} \,,
\quad 
v = \frac{p_a + p_b}{E_a+E_b} \,,
\quad 
w = \frac{p_{\rm out}}{m_a}=\frac{p_2}{m_a} \,.
\label{velo-scattering}
\end{align}

We may insert \eqref{kin-scattering} and \eqref{velo-scattering} into \eqref{alpha-great}. 
For the denominator, we have 
\begin{align}
1 + u\cdot w = 2 \,,
\quad 
1 + u \cdot v = 1 + \frac{E_a}{m_a} =  1 + v \cdot w  \,.
\end{align}
For the numerator, we note that 
\begin{align}
\ve_\m(p_1,p_2,p_a + p_b) =  \ve_\m(p_a + q/2 ,p_a - q/2 ,p_a + p_b) = 
\ve_\m(p_a ,p_b , q) \,.
\end{align}
Combining all the ingredients, we obtain 
\begin{align}
2(1-\cos\alpha) = 4 \sin(\alpha/2)^2 = \frac{- [\ve_\m(p_a ,p_b, q)]^2}{m_a^2 E^2(m_a+E_a)^2} \,. \label{eq:TWfact_angle}
\end{align}
Let $f(x)$ be the inverse function of $2\sin(x/2)$. Clearly, $\tilde{f}(x^2) \equiv (f(x)-x)/x$ is an analytic function of $x^2$ with $\tilde{f}(0)=0$. 
Under the presumption of the HCL kinematics, since 
\begin{align}
(\ve_\m(p_a ,p_b , q))^2 \propto \left[p_a^2 p_b^2 - (p_a\cdot p_b)^2\right] q^2 \approx 0\,,
\end{align}
we may set $\tilde{f}(x^2) \approx 0$ and hence $f(x) \approx x$ in what follows. 

So far, we have worked out the magnitude of the angle $\alpha$ only. We should also find the orientation of the rotation plane. To put the incoming and out-going states on a nearly equal footing, we work in the COM frame. Then, the three 4-vectors $u_a$, $u_b$, $q$ together determine the rotation axis through the $\varepsilon$-tensor. For a spinor in 3d, the rotation is represented by
\begin{align}
U(\pm \hat{m},\alpha) = e^{\pm \frac{i}{2}\alpha (\hat{m}\cdot \vec{\sigma})} = e^{ \pm i \alpha (\hat{m}\cdot \vec{s})} \,. \label{eq:TWfact_param}
\end{align}
We conclude that the rotation factor is
\begin{align}
U_{\rm rotation}^{(a)} = \exp\left[ - i \left(\frac{m_b}{r_a E}\right) \ve(q,u_a,u_b,a_a) \right] \,,
\quad 
r_a \equiv 1 + \frac{E_a }{m_a}  \,.
\label{phase-final}
\end{align}
We have fixed the sign in the exponent of \eqref{phase-final} by matching against our earlier 
work on the leading PN, all orders in spin, computation \cite{Chung:2018kqs,Chung:2019duq}.

\subsection{Complete 1PM potential}
Equipped with the rotation factors $U_{\rm rotation}^{(a)} $ and $U_{\rm rotation}^{(b)}$, we simply dress the bare amplitude in (\ref{spin-amp-BH}) for black holes as 
\begin{align}
M_{\rm dressed}^{\rm (BH)} = -\frac{4\pi G}{q^2} \frac{m_a^2 m_b^2}{E_aE_b} 
\cosh\left( 2\theta  +  i \frac{\ve(q,u_a,u_b,a_0)}{\sinh\theta} \right) U^{(a)} U^{(b)} \,,
\label{BH-dressed}
\end{align}
where we suppress the subscript on $U$. The expression combines the amplitude \eqc{spin-amp-BH} with additional rotation factors $U^{(a)} U^{(b)}$ originating from how polarization tensors are defined.
Setting $q= (0, \vec{q})$ and taking the Fourier transform with $e^{i\vec{q}\cdot \vec{r}}$, we obtain the potential. Since an exponentiated gradient generates a finite translation, 
we can explicitly write the potential as 
\begin{align}
V_{\rm 1PM}^{\rm (BH)} = - \frac{G m_a^2m_b^2}{2E_a E_b} 
\sum_{s=\pm 1} 
e^{2s\theta}
\left| \vec{r} + s\frac{E (\vec{p}\times \vec{a}_0)}{m_a m_b \sinh\theta} - \frac{\vec{p}\times \vec{a}_a}{m_a r_a} - \frac{\vec{p}\times \vec{a}_b}{m_b r_b} 
\right|^{-1} \,. 
\label{V-BH-final}
\end{align}
For general compact spinning bodies with non-minimal Wilson coefficients, 
we dress the general form of the amplitude \eqref{spin-amp-bare} 
with the rotation factors  
to reach the master formula:
\begin{align}
\begin{split}
V_{\rm 1PM}^{\rm (general)} &= - \frac{G m_a^2 m_b^2}{E_a E_b} \int \frac{d^3 \vec{q}}{(2\pi)^3} \frac{4 \pi e^{i\vec{q}\cdot \vec{r}}}{\vec{q}^2}
\left[\half \sum_{s=\pm 1} e^{2s\theta} W_a(s\tau_a) W_b(s\tau_b)\right] U^{(a)} U^{(b)} \,.
\label{V-1PM-master}
\end{split}
\end{align}
We can still perform the Fourier transform, but the result is not as simple as \eqref{V-BH-final}.


The master formula \eqc{V-1PM-master} has been checked to match available LO, NLO, and NNLO in $\vec{p}$ results in the literature~\cite{Levi:2014gsa,Damour:2007nc,Steinhoff:2008zr,Levi:2014sba,Vaidya:2014kza,Levi:2015msa,Hartung:2011te,Hartung:2013dza,Levi:2015uxa,Steinhoff:2007mb,Hartung:2011ea,Levi:2011eq,Porto:2008jj,Steinhoff:2008ji,Hergt:2008jn,Hergt:2010pa,Levi:2015ixa,Levi:2016ofk,Levi:2019kgk}. The details of the matching will be the subject of following sections; the 1PM order potential will be computed explicitly for each spin order in section~\ref{sec:spin-order}, and comparsion with known results will be performed in section~\ref{sec:checks}.

\section{1PM potential at each spin order} \label{sec:spin-order}
In this section, we present an explicit form of the 1PM for potential at each fixed order of spin. The exact result and the LO term will be presented here, while matching at NLO and beyond will be the focus of the next section. To demonstrate the almost complete factorization of the spin-dependence and the momentum dependence, we organize the results using the following notations. 
In writing down the spin($a$)$^m$-spin($b$)$^n$ term $V_{S_a^m S_b^n}$ of the potential \eqc{V-1PM-master}, which describes the interaction between spin-induced $2^m$-pole of particle $a$ and $2^n$-pole of particle $b$,\footnote{The trace part vanishes due to the relation $\nabla^2 r^{-1} = 4 \pi \delta^3$, therefore the symmetric product of spins can be identified as the spin-induced multipole.} we write
\begin{align}
\begin{split}
V_{S_a^m S_b^n} &= \left(\frac{G m_a m_b }{r^{m+n+1}} \right) \left[ F_{(m,n)}(\vec{a}_a,\vec{a}_b,\hat{n}) \right] X_{(m,n)}(\vec{p}^2) 
\qquad (m+n\;\; \mbox{even}) \,,
\\
V_{S_a^m S_b^n} &= \left(\frac{G m_a m_b}{r^{m+n+1}} \right) \left[ \vec{p}\cdot \vec{F}_{(m,n)}(\vec{a}_a,\vec{a}_b,\hat{n}) \right]  X_{(m,n)}(\vec{p}^2) 
\quad (m+n\;\; \mbox{odd}) \,.
\end{split}
\label{def:FvF}
\end{align} 
Explicitly, the spin-dependent factors, $F_{(m,n)}$ and $\vec{F}_{(m,n)}$, are defined by
\begin{align}
\begin{split}
F_{(m,n)} &= r^{m+n} (\vec{a}_a \cdot \nabla)^m (\vec{a}_b \cdot \nabla)^n \left(\frac{1}{r} \right) \,,
\\
\vec{F}_{(m,n)} &= \left\{
\begin{aligned}
\frac{r^{m+n}}{m_a} (\vec{a}_a \times \nabla)  (\vec{a}_a \cdot \nabla)^{m-1} (\vec{a}_b \cdot \nabla)^n \left(\frac{1}{r} \right) && \qquad (m\;\; \mbox{odd}) \,,
\\
\frac{r^{m+n}}{m_b} (\vec{a}_b \times \nabla)  (\vec{a}_a \cdot \nabla)^{m} (\vec{a}_b \cdot \nabla)^{n-1} \left(\frac{1}{r} \right) && \qquad (n\;\; \mbox{odd}) \,.
\end{aligned} \right.
\end{split}
\end{align}
By construction, $F_{(m,n)}$ and $\vec{F}_{(m,n)}$ are homogeneous polynomials of $\vec{a}_a$ and $\vec{a}_b$ of degree $m$ and $n$, respectively. We pulled out an overall factor of masses so as to make $X_{(m,n)}$ dimensionless. 
When we expand the potential in $\vec{p}$, we will use the notation 
\begin{align}
X_{(m,n)} = X_{(m,n)}^{\rm LO} + X_{(m,n)}^{\rm NLO} + \cdots + X_{(m,n)}^{{\rm N}^k{\rm LO} } + \cdots \,,
\end{align}
where $X_{(m,n)}^{{\rm N}^k{\rm LO}}$ is proportional to $(\vec{p}^{2})^k$.

Regardless of the order of expansion in spin or momentum, 
there are two notable differences between our result and those in the literature. 
First, ours results doesn't carry any $(\hat{n}\cdot \vec{p})$ term. 
(Here $\hat{n} = \vec{r}/{r}$ is the unit directional vector between the two bodies.)
In other words, the so-called ``isotropic gauge" is forced upon us 
by the amplitude approach; see \cite{Bern:2019crd} for a related comment.
Second, ours results doesn't carry any $(\vec{a} \cdot \vec{p})$ term either, except 
through a very specific $(\vec{p}\cdot \vec{F})$ structure in \eqref{def:FvF}. 
This is to be contrasted with a typical PN computation which often produces 
a linear combination, 
\begin{align}
\vec{a}_a^2 f_1 (\vec{p}^2) + (\vec{a}_a \cdot \vec{p})^2 f_2(\vec{p}^2)  \,,
\end{align}
with no obvious correlation between the two functions $f_1$ and $f_2$. 

A minor technical remark. To reduce clutter in equations, we introduce a few more short-hand notations such as
$s_\theta = \sinh\theta$, $c_\theta = \cosh\theta$, $c_{2\theta} = \cosh(2\theta)$. 

\subsection{Linear in spin}
Since the Wilson coefficients $C_0=C_1=1$ are universal, at linear order in spin the potential is universal and we may simply work with the black holes. First, from $\mathcal{M}_{\rm bare, BH}$ in (\ref{spin-amp-BH}) the spin-linear term is
\begin{align}
\begin{split}
\cosh\left( 2\theta  +  i \frac{\ve(q,u_a,u_b,a_0)}{\sinh\theta} \right) 
&= \cosh(2\theta) + i  \frac{\ve(q,u_a,u_b,a_0)}{\sinh\theta} \sinh2\theta + \mathcal{O}(a_0^2)
\\
&\approx  \cosh(2\theta) + 2 i (\cosh\theta) \ve(q,u_a,u_b,a_0) \,.
\end{split}
\end{align}
Using the identity,  
\begin{align}
\ve(q,u_a,u_b,a) = \left( \frac{E}{m_a m_b} \right) \vec{p} \cdot (\vec{a} \times \vec{q}) \,,
\end{align}
we find the contribution from $\mathcal{M}_{\rm bare}$ to the spin-linear potential can be written as: 
\begin{align}
V_{\rm bare} = - \frac{2m_a m_b E }{E_a E_b} (\cosh\theta) 
\left[ \vec{p} \cdot (\vec{a}_0 \times \nabla) \right] \left( \frac{G}{r} \right) \,.  
\end{align}
Now from the rotation factor $U^{(a)}$, we find 
\begin{align}
\begin{split}
V_{\rm rotation}
&= \frac{m_a^2m_b^2}{E_a E_b} \frac{1}{m_ar_a } \cosh(2\theta) 
\left[ \vec{p} \cdot (\vec{a}_a \times \nabla) \right] \left( \frac{G}{r} \right) 
\\
&= \frac{m_a m_b^2 }{E_a E_b r_a } \cosh(2\theta) 
\left[ \vec{p} \cdot (\vec{a}_a \times \nabla) \right] \left( \frac{G}{r} \right) 
\,.
\end{split}
\end{align}
Collecting all the terms we obtain 
\begin{align}
V_{S_a^1 S_b^0} = - \frac{m_a m_b E}{E_a E_b} \left(2 c_\theta  - \frac{m_bc_{2\theta}}{Er_a } \right) \left[ \vec{p} \cdot (\vec{a}_a \times \nabla) \right] \left( \frac{G}{r} \right) \,,
\end{align}
or, equivalently, 
\begin{align}
V_{S_a^1 S_b^0} = \left( \frac{G}{r^2} \right)  \frac{m_a m_b E}{E_a E_b} \left(2 c_\theta - \frac{m_bc_{2\theta}}{Er_a } \right) 
[ \vec{p}  \cdot ( \vec{a}_a \times \hat{n} ) ]
\,.
\label{SO-exact}
\end{align}
In the notation of \eqref{def:FvF}, we have
\begin{align}
\vec{F}_{(1,0)} =  \vec{a}_a \times \hat{n} \,,
\quad 
X_{(1,0)} = \frac{m_a E}{E_a E_b} \left(2 c_\theta  - \frac{m_bc_{2\theta}}{Er_a } \right)  \,.
\end{align}
This is the exact linear in spin potential at 1PM. 

\paragraph{LO}
The leading order term in $\vec{p}^2$ can be extracted and given by:
\begin{align}
V_{S_a^1 S_b^0}^{\rm LO} = 
\left( \frac{G}{r^2} \right) [ \vec{p}  \cdot ( \vec{a}_a \times \hat{n} ) ] \left( \frac{4 m_a + 3 m_b}{2}\right)
\,.
\end{align}

\subsection{Quadratic in spin}
\paragraph{Spin-spin couplings}

This term also only utilizes $C_0=C_1=1$ only and thus are universal as well. From the 
bare amplitude $\mathcal{M}_{\rm bare}$
\begin{align}
(V_{\rm bare})_{S_a^1S_b^1} = - c_{2\theta}  \left(\frac{G}{r^3}\right) \frac{m_a^2m_b^2}{E_a E_b} 
 \left[\vec{a}_a\cdot\vec{a}_b  - 3 (\vec{a}_a\cdot \hat{n}) (\vec{a}_b \cdot \hat{n})\right] \,.
\end{align}
Adding up the other two contributions, we obtain 
\begin{align}
\begin{split}
V_{S_a^1 S_b^1} &=  \left(\frac{Gm_a m_b }{r^3}\right) F_{(1,1)} X_{(1,1)}  \,, 
 \quad
 F_{(1,1)} =  - \left[\vec{a}_a\cdot\vec{a}_b  - 3 (\vec{a}_a\cdot \hat{n}) (\vec{a}_b \cdot \hat{n})\right] \,,
 \\
 X_{(1,1)} &= \frac{m_a m_b }{E_a E_b} \left[ c_{2\theta} -\frac{2s_\theta^2 c_{\theta}}{E}
 \left(\frac{m_b}{r_a} +\frac{m_a}{r_b}\right) 
 + \frac{m_a m_b s_\theta^2 c_{2\theta}}{E^2r_a r_b} \right]
\,.
\end{split}
\label{SS-ab}
\end{align}

\paragraph{Spin-squared} 

For the spin-squared piece, one has $C_2$ contribution from $\mathcal{M}_{\rm bare}$, as well as $C_1$ from $\mathcal{M}_{\rm bare}$ times linear expansion of $U$ and the quadratic in spin expansion of $U$. The latter two are once again universal. Adding up all three contributions, we obtain
\begin{align}
\begin{split}
V_{S_a^2 S_b^0} &=  \left(\frac{Gm_a m_b }{r^3}\right) F_{(2,0)} X_{(2,0)}  \,, 
\quad
F_{(2,0)} = - \left[\vec{a}_a^2  - 3 (\vec{a}_a\cdot \hat{n})^2  \right]  \,,
\\
X_{(2,0)} &=  \frac{m_a^2m_b^2}{2 E_a E_b} \left[ C_2^{(a)} c_{2\theta} 
- \frac{4m_b s_\theta^2  c_{\theta}}{Er_a } 
+ \frac{m_b^2 s_\theta^2 c_{2\theta}}{E^2r_a ^2}  \right]
\,.
\end{split}
\label{SS-aa}
\end{align}

\paragraph{LO} To the leading order we have:
\begin{align}
\begin{split}
V^{\rm LO}_{S_a^1 S_b^1} &= - \left(\frac{G m_a m_b}{r^3}\right)  \left[\vec{a}_a\cdot\vec{a}_b  - 3 (\vec{a}_a\cdot \hat{n}) (\vec{a}_b \cdot \hat{n})\right] \,,
\\
V^{\rm LO}_{S_a^2 S_b^0} &= -  \frac{1}{2} C_2^{(a)} \left(\frac{G m_a m_b}{r^3}\right)  \left[\vec{a}_a^2  - 3 (\vec{a}_a\cdot \hat{n})^2  \right] \,.
\end{split}
\end{align}

\subsection{Cubic in spin}
Continuing with the same method, we obtain the formulae for the cubic-in-spin terms. 
For the spin($a$)$^3$ term, we have 
\begin{align}
\begin{split}
V_{S_a^3 S_b^0} &= \left(\frac{Gm_b}{r^4}\right) [\vec{p}\cdot \vec{F}_{(3,0)}] X_{(3,0)}  \,,
\\
\vec{F}_{(3,0)} &= 3 (\vec{a}_a \times \hat{n}) \left[\vec{a}_a^2  - 5 (\vec{a}_a\cdot \hat{n})^2  \right] \,,
\\
X_{(3,0)}&= \frac{m_a E}{E_a E_b} \left[
\frac{1}{3} C_3^{(a)}  c_{\theta} - C_2^{(a)} \frac{m_b c_{2\theta}}{2Er_a } 
+  \frac{m_b^2 \sinh ^2\theta  \cosh \theta }{E^2r_a ^2 }
-\frac{m_b^3 \sinh ^2\theta  \cosh (2\theta) }{6E^3r_a^3 } \right]  \,.
\end{split}
\label{SSS-aaa}
\end{align}
For the mixed spin($a$)$^2$-spin(b)$^1$ term, we have
\begin{align}
\begin{split}
V_{S_a^2 S_b^1} &=  \left(\frac{Gm_b}{r^4}\right) 
[\vec{p}\cdot \vec{F}_{(2,1)} ] X_{(2,1)} \,,
\\
\vec{F}_{(2,1)} &= 3\left\{ (\vec{a}_b \times \hat{n} ) \left[\vec{a}_a^2  - 5 (\vec{a}_a\cdot \hat{n})^2 \right]  -2 (\vec{a}_a\cdot \hat{n}) (\vec{a}_a\times \vec{a}_b)\right\} \,,
\\
X_{(2,1)}&= \frac{m_a  E}{E_a E_b} \left[ C_2^{(a)} \cosh \theta - \frac{c_{2\theta}}{2E}\left(\frac{2m_b}{r_a }+ C_2^{(a)} \frac{m_a}{r_b} \right) 
\right.
\\
&\qquad \qquad\qquad
\left.
+ \frac{m_b s^2_\theta   c_{\theta} }{E^2r_a } 
\left( \frac{m_b  }{r_a } +\frac{2 m_a }{r_b}\right)
-\frac{m_a m_b^2 s^2_\theta c_{2\theta} }{2 E^3r_a ^2 r_b} \right]\,.
\end{split}
\label{SSS-aab}
\end{align}

\paragraph{LO} To the leading order, we find 
\begin{align}
\begin{split}
V_{S_a^3 S_b^0}^{\rm LO} &= 
\left(\frac{G}{r^4}\right) 
\vec{p}  \cdot ( \vec{a}_a \times \hat{n} )   \left[\vec{a}_a^2  - 5 (\vec{a}_a\cdot \hat{n})^2  \right] 
\left( C_3^{(a)} (m_a+m_b) - \frac{3}{4} C_2^{(a)} m_b \right) \,,
\\
V_{S_a^2 S_b^1}^{\rm LO} &=\left(\frac{G}{r^4}\right) \vec{p}\cdot \left\{ ( \vec{a}_b \times \hat{n} ) \left[\vec{a}_a^2  - 5 (\vec{a}_a\cdot \hat{n})^2 \right]  -2 (\vec{a}_a\cdot \hat{n}) (\vec{a}_a\times \vec{a}_b)
\right\}
\\
&\qquad \times 
 \left(  \frac{3}{4} C_2^{(a)}(3m_a + 4m_b)- \frac{3}{2} m_b\right)  \,,
\end{split}
\end{align}
in perfect agreement with the corresponding terms in (3.10) of \cite{Levi:2014gsa}.

\subsection{Quartic in spin} 
We continue to quartic in spin. This is an interesting threshold for black holes from an on-shell perspective, since fundamental particles are only known up to spin-2. It can be shown that beyond spin-2, isolated spinning particle no-longer exists and must either be a bound state or part of an infinite tower of massive states~\cite{Arkani-Hamed:2017jhn, Afkhami-Jeddi:2018apj}.  As a consequence, the gravitational Compton amplitude is no longer unique beyond spin-2, which we review in section~\ref{sec:GravComp}. This ambiguity will be pertinent for 2PM computations.

The spin($a$)-quartic term is 
\begin{align}
\begin{split}
V_{S_a^4 S_b^0} 
&= \left(\frac{Gm_a m_b}{r^5}\right) F_{(4,0)} X_{(4,0)} \,,
\\
F_{(4,0)} &= 3 \left\{ 3\vec{a}_a^4 -30 \vec{a}_a^2 (\vec{a}_a\cdot \hat{n})^2 + 35(\vec{a}_a\cdot \hat{n})^4 \right\}  \,,
\\
X_{(4,0)}&= -\frac{m_a m_b}{24 E_a E_b} \left[ c_{2\theta}
\left(
C_4^{(a)} + 6 C_2^{(a)} \frac{ m_b^2 s^2_\theta  }{r_a ^2 E^2} + \frac{m_b^4 s^4_\theta }{r_a ^4 E^4} \right) \right.
\\ &\phantom{=asdfasdfasdfasdfasdf} \left.
-\frac{8 m_b s^2_\theta c_\theta}{r_a  E}
\left(
C_3^{(a)} + \frac{m_b^2 s^2_\theta}{r_a ^2 E^2} 
\right)
\right]
\,.
\end{split}
\label{S4-aaaa}
\end{align}
The cubic-linear term is
\begin{align}
\begin{split}
V_{S_a^3 S_b^1} 
&= \left(\frac{Gm_a m_b}{r^5}\right)  F_{(3,1)} X_{(3,1)}  \,,
\\
F_{(3,1)} &= 3\left\{ 3\vec{a}_a^2(\vec{a}_a\cdot \vec{a}_b) -15 (\vec{a}_a\cdot\vec{a}_b) (\vec{a}_a\cdot \hat{n})^2 \right.
\\ &\phantom{=asdfasdf} \left. -15 \vec{a}_a^2(\vec{a}_a\cdot \hat{n})(\vec{a}_b\cdot \hat{n}) + 35(\vec{a}_a\cdot \hat{n})^3(\vec{a}_b\cdot \hat{n}) \right\}\,,
\\
X_{(3,1)}&= -\frac{m_a m_b}{6E_aE_b} \left[ c_{2 \theta} \left\{ C_3^{(a)} 
+\frac{3 m_b s^2_\theta }{r_a  E^2} \left(C_2^{(a)} \frac{m_a}{r_b}+\frac{m_b}{r_a}\right)
+\frac{m_a m_b^3 s^4_\theta}{r_a ^3 r_b E^4}
\right\} \right.
\\
&\qquad \qquad \qquad 
\left. -
\frac{2s^2_\theta c_\theta}{E}
\left\{ C_3^{(a)} \frac{ m_a}{r_b} + 3 C_2^{(a)} \frac{m_b}{r_a}+\frac{m_b^2 s^2_\theta }{r_a ^2 E^2} \left(\frac{3 m_a}{r_b}+\frac{m_b}{r_a}\right)\right\} \right]
 \,.
\end{split}
\label{S4-aaab}
\end{align}
The quadratic-quadratic term is the first place where non-trivial Wilson coefficients 
from both spinning bodies contribute together. 
\begin{align}
\begin{split}
V_{S_a^2 S_b^2} 
&=  \left(\frac{Gm_am_b}{r^5}\right) F_{(2,2)} X_{(2,2)}
 \,,
\\
F_{(2,2)} &= 3 \left\{ \vec{a}_a^2 \vec{a}_b^2 + 2(\vec{a}_a\cdot \vec{a}_b)^2 
-5 \vec{a}_a^2  (\vec{a}_b\cdot \hat{n})^2 -5 \vec{a}_b^2  (\vec{a}_a\cdot \hat{n})^2 \right.
\\
&\qquad\qquad\qquad\qquad 
\left. -20 (\vec{a}_a\cdot \vec{a}_b) (\vec{a}_a\cdot \hat{n})(\vec{a}_b\cdot \hat{n}) 
+ 35(\vec{a}_a\cdot \hat{n})^2(\vec{a}_b\cdot \hat{n})^2 \right\} \,,
\\
X_{(2,2)}&= -\frac{m_am_b}{4E_aE_b} \left[ c_{2\theta}
\left\{
\left(C_2^{(a)}+\frac{m_b^2 s^2_\theta}{r_a ^2 E^2}\right) 
\left(C_2^{(b)}+\frac{m_a^2 s^2_\theta}{r_b^2 E^2}\right) 
+ \frac{4 m_a m_b s^2_\theta }{r_a  r_b E^2}
\right\} \right.
\\
&\qquad \qquad \qquad
\left.
- \frac{4s^2_\theta c_\theta}{E} 
\left\{ C_2^{(a)} \frac{m_a}{r_b}+C_2^{(b)} \frac{m_b}{r_a}
+ \frac{m_a m_b s^2_\theta}{r_a  r_b E^2} \left( \frac{m_a}{r_b} + \frac{m_b}{r_a}\right) \right\} \right]
 \,.
\end{split}
\label{S4-aabb}
\end{align}

\paragraph{LO} 
To leading order, we find
\begin{equation}
\begin{split}
V_{S_a^4,S_b^0} &= -\frac{G m_a m_b}{8r^5} C_4^{(a)}\left\{ 3\vec{a}_a^4 -30 \vec{a}_a^2 (\vec{a}_a\cdot \hat{n})^2 + 35(\vec{a}_a\cdot \hat{n})^4 \right\}\\
V_{S_a^3,S_b^1} &= -\frac{G m_a m_b}{2r^5} C_3^{(a)}\left\{ 3\vec{a}_a^2(\vec{a}_a\cdot \vec{a}_b) -15 (\vec{a}_a\cdot\vec{a}_b) (\vec{a}_a\cdot \hat{n})^2 \right.\\
&\qquad\qquad\qquad\qquad\qquad\qquad\quad
\left. -15 \vec{a}_a^2(\vec{a}_a\cdot \hat{n})(\vec{a}_b\cdot \hat{n}) + 35(\vec{a}_a\cdot \hat{n})^3(\vec{a}_b\cdot \hat{n}) \right\}\\
V_{S_a^2,S_b^2} &= -\frac{3G m_a m_b}{4r^5} C_2^{(a)}\left\{ \vec{a}_a^2 \vec{a}_b^2 + 2(\vec{a}_a\cdot \vec{a}_b)^2 
-5 \vec{a}_a^2  (\vec{a}_b\cdot \hat{n})^2 -5 \vec{a}_b^2  (\vec{a}_a\cdot \hat{n})^2 \right.
\\
&\qquad\qquad\qquad\qquad \qquad
\left. -20 (\vec{a}_a\cdot \vec{a}_b) (\vec{a}_a\cdot \hat{n})(\vec{a}_b\cdot \hat{n}) 
+ 35(\vec{a}_a\cdot \hat{n})^2(\vec{a}_b\cdot \hat{n})^2 \right\}
\end{split}
\end{equation}
which is in perfect agreement to (4.4) of \cite{Levi:2014gsa}.

\section{Reproducing 1PM part of PN expansion} \label{sec:checks}

In the previous section, we derived the potential at each spin order that is exact in $\vec{p}$. It is almost trivial to expand the expressions in powers of $\vec{p}^2$. 
Each term in the $\vec{p}^2$ expansion can be compared with the 1PM part of the 
PN computation available in the literature. In this section, we make the comparison explicitly for all spin and momentum orders where the data are available. 

The precise form of the subleading terms in $\vec{p}^2$ depend on the choice of the phase space coordinates $(\vec{r},\vec{p})$. This ``coordinate gauge" ambiguity originates from the general covariance of general relativity. Any two different gauge choices are related to each other by a canonical transformation. We denote by $g$ the generator 
of the canonical transformation, 
\begin{align}
\Delta_\epsilon H = \epsilon \{ H, g \} \,,
\label{cano-generator-epsilon}
\end{align}
where $\epsilon$ is an infinitesimal parameter. In the PN expansion, both $G$ and $1/m$ can be treated as if they were infinitesimal\footnote{While $c^{-2}$ is the proper expansion parameter, this factor only arises through dimensionless combinations $G m / rc^2$ and $p^2/m^2 c^2$. Therefore $G$ and $1/m$ effectively serve as the expansion parameters in the PN expansion.}, so we will use a variant of \eqref{cano-generator-epsilon} without explicitly mentioning the infinitesimal parameter $\epsilon$. 

At NLO in the PN expansion, the only relevant term in $H$ on the right-hand side of \eqref{cano-generator-epsilon} is the Newtonian term $H_{\rm N}$. Since we are 
comparing terms at 1PM only, only the kinetic term of $H_{\rm N}$ contribute. 
\begin{align}
(\Delta H^{\rm NLO})_{\rm 1PM} = \{ H_{\rm N} , g^{\rm NLO} \} = \left\{ \left(\frac{1}{2m_a} + \frac{1}{2m_b} \right) \vec{p}^2 , g^{\rm NLO} \right\} 
+ \mathcal{O}(G^2) \,. 
\label{cano-NLO}
\end{align}
The transformation receives two contributions at NNLO. 
\begin{align}
\begin{split}
(\Delta H^{\rm NNLO})_{\rm 1PM} &= \{ H_{\rm N} , g^{\rm NNLO} \} + \{ H_{\rm 1PN} , g^{\rm NLO} \} + \mathcal{O}(G^2)\,,
\\
H_{\rm 1PN} &= - \left( \frac{1}{8m_a^3} +\frac{1}{8m_b^3} \right)\vec{p}^4 + \mathcal{O}(G) \,.
\end{split}
\label{cano-NNLO}
\end{align}
All canonical transformations to be performed below are based on the elementary Poisson 
algebra: 
$\{ x^i , p_j \} = \delta^i{}_j$. 
The following formula will be used multiple times:
\begin{align}
\begin{split}
&\left\{ \frac{\vec{p}^2}{2}  , \left(\frac{1}{r}\right)^k (\hat{n}\cdot \vec{p})^\ell [\hat{n}\cdot (\vec{p}\times \vec{a})]^m \right\} 
\\
&= 
\left(\frac{1}{r}\right)^{k+1} (\hat{n}\cdot \vec{p})^{\ell -1} [\hat{n}\cdot (\vec{p}\times \vec{a})]^m \left( (k+\ell+m)(\hat{n} \cdot \vec{p})^2 - \ell  \vec{p}^2 \right) 
\,.
\end{split}
\label{cano-master}
\end{align}

\subsection{Linear in spin (up to NNNLO)} 

As explained earlier, our notation for the 1PM and arbitrary PN expansion is 
\begin{align}
V_{S_a^1 S_b^0} = 
\left( \frac{G m_b}{r^2} \right) [ \vec{p}  \cdot ( \vec{a}_a \times \hat{n} ) ] 
\left( X_{(1,0)}^{\rm LO} + X_{(1,0)}^{\rm NLO} + X_{(1,0)}^{\rm NNLO} + X_{(1,0)}^{\rm NNNLO} + \cdots \right)
\,.
\nonumber
\end{align}

\paragraph{NLO and its canonical transformations}

Expanding our formula \eqref{SO-exact}, we find 
\begin{align}
X_{(1,0)}^{\rm NLO} =  
\left( \frac{ 18 m_a^2  + 8 m_a m_b -5 m_b^2 } {8m_a^2 m_b^2} \right)\vec{p}^2 \,.
\label{NLO-SO-iso}
\end{align}
This NLO spin-orbit coupling was computed in the ADM framework in \cite{Damour:2007nc,Steinhoff:2008zr}, 
in the EFT framework in \cite{Levi:2014sba}, and in an amplitude-based approach in \cite{Vaidya:2014kza}. 
The last reference employs the isotropic gauge and the result looks identical to ours. 
It also explains how to use a canonical transformation to check agreement with \cite{Damour:2007nc,Steinhoff:2008zr}. 

Consider a family of Hamiltonians: 
\begin{align}
\begin{split}
H_{\rm SO}^{\rm NLO} &= \left( \frac{G}{r^2} \right) \frac{\hat{n}\cdot(\vec{p}\times \vec{a}_a)}{ 8m_a}
\left[  h_1 \vec{p}^2  + h_2 (\hat{n}\cdot\vec{p})^2  
\right]
\,,
\\
h_k &= h_{k,+} \zeta + h_{k,0} + h_{k,-} \zeta^{-1} \,, \quad (\zeta \equiv m_b/m_a)
\,.
\end{split}
\label{NLO-SO-family}
\end{align}
In this notation, our result \eqref{NLO-SO-iso} amounts to 
\begin{align}
h_1 =  -5 \zeta + 8 + 18 \zeta^{-1} \,,
\quad 
h_2 = 0 \,.
\label{h-so-ours}
\end{align}
Not all parameters are physically meaningful, because some combinations can be altered by 
canonical transformations of the type shown in \eqref{cano-NLO}.
Taking hints from \cite{Vaidya:2014kza}, we take the following ansatz for the generator of the transformation: 
\begin{align}
g_{\rm SO}^{\rm NLO} &= \frac{g_1}{8(1+\zeta^{-1})} \left( \frac{G}{r}\right) [\hat{n}\cdot (\vec{p}\times \vec{a}_a  )]  
(\hat{n}\cdot \vec{p}) \,,
\quad 
g_1 = g_{1,+} \zeta + g_{1,0} + g_{1,-} \zeta^{-1}\,.
\label{gen-SO-NLO}
\end{align}
The factor $1/(1+\zeta^{-1})$ in the generator is inserted to cancel the similar factor in \eqref{cano-NLO}.

Recalling the formula \eqref{cano-master} and setting $k=\ell=m=1$, we can express the changes $\Delta h_k$ in terms of $g_k$:
\begin{align}
\begin{split}
\Delta h_1 = - g_1 \,,
\quad 
\Delta h_2 = 3 g_1 \,.
\end{split}
\label{dh-vs-g}
\end{align} 

Several papers report the NLO spin-orbit potential. For example, (6.22) of \cite{Levi:2015msa}, after being simplified in the COM frame, gives
\begin{align}
h_1 =  -5 \zeta + 8 \zeta^{-1} \,,
\quad 
h_2 =  24 + 30 \zeta^{-1}  \,.
\label{h-so-LS}
\end{align}
Taking the difference, $\Delta h_k = h_k^{\rm old} - h_k^{\rm new}$,  between \eqref{h-so-ours} and \eqref{h-so-LS}, 
we find
\begin{align}
\Delta h_1 = -8 - 10 \zeta^{-1}\,,
\quad 
\Delta h_2 = 24 + 30 \zeta^{-1} \,.
\end{align}
This is compatible with \eqref{dh-vs-g} if we set $g_1 = 8 + 10 \zeta^{-1}$. Thus we have shown that \eqref{NLO-SO-iso} is equivalent to the corresponding term in \cite{Levi:2015msa}.

\paragraph{NNLO its canonical transformations}

\begin{align}
X_{(1,0)}^{\rm NNLO} = 
\left( \frac{- 15 m_a^4 - 15m_a^2 m_b^2  -12 m_a m_b^3 + 7m_b^4}{16m_a^4 m_b^4} \right) \vec{p}^4 \,.
\label{NNLO-SO}
\end{align}
The same term in the Hamiltonian formulation was computed in the ADM framework in \cite{Hartung:2011te,Hartung:2013dza} and in the EFT framework in \cite{Levi:2015uxa}.

Once again, consider the following ansatz for the Hamiltonian: 
\begin{align}
H_{\rm SO}^{\rm NNLO} &= \left( \frac{G}{r^2} \right) \frac{\hat{n}\cdot(\vec{p}\times \vec{a}_a)}{ 16 m_a^2 m_b}
\left[  h_3 \vec{p}^4  + h_4 \vec{p}^2 (\hat{n}\cdot\vec{p})^2 + h_5 (\hat{n}\cdot\vec{p})^4
\right]
\,.
\label{NLO-SO-family2}
\end{align}
In this notation, our result \eqref{NNLO-SO} correspond to 
\begin{align}
h_3 =  7 \zeta^2 - 12 \zeta -15 -15 \zeta^{-2} \,,
\quad 
h_4 = h_5 = 0 \,.
\label{h-so-ours2}
\end{align}
(4.11) of \cite{Levi:2015uxa}, sharing the same convention as \cite{Levi:2015msa}, is translated to our notation as 
\bg
\bgd
h_3 = 7\zeta^2 - 4 \zeta -24 -20 \zeta -12\zeta^{-2} \,,
\quad 
h_4 = -8\zeta -3 + 8\zeta^{-1} \,,
\\ 
h_5 = 60 + 60\zeta^{-1} -15\zeta^{-2}\,.
\egd
\eg
The difference between the two results is then 
\bg
\bgd
\Delta h_3 = 8 \zeta - 9 -20 \zeta + 3 \zeta^{-2} \,,
\quad 
\Delta h_4 = - 3(8\zeta +3 - 8\zeta^{-1}) \,,
\\
\Delta h_5 = 15( 4  + 4\zeta^{-1} -\zeta^{-2})\,.
\label{NNLO-data}
\egd
\eg

Our ansatz for the NNLO generating function is
\begin{align}
g_{\rm SO}^{\rm NNLO} &= \frac{1}{16(1+\zeta^{-1})} \left( \frac{G}{r}\right) 
\frac{[\hat{n}\cdot (\vec{p}\times \vec{a}_a  )]}{m_a m_b} 
\left[ g_2 \vec{p}^2 (\hat{n}\cdot \vec{p}) + g_3 (\hat{n}\cdot \vec{p})^3 \right] \,.
\label{gen-SO-NNLO}
\end{align}
Using \eqref{cano-NNLO} and \eqref{cano-master}, we can easily relate the 
coefficients, 
\bg
\bgd
\Delta h_3 =  (\zeta-1+\zeta^{-1}) g_1 - g_2\,,
\quad 
\Delta h_4 =  3 [ -(\zeta-1+\zeta^{-1}) g_1 +  g_2 - g_3 ] \,,
\\
\Delta h_5 =  5g_3 \,.
\label{NNLO-relation}
\egd
\eg
The value of $g_1$ was already fixed at the NLO order. The difference \eqref{NNLO-data} 
matches the relation \eqref{NNLO-relation} if we set
\begin{align}
g_2 = (1+\zeta^{-1})(11+7\zeta^{-1}) \,,
\quad 
g_3 = 3(4+4\zeta^{-1} - \zeta^{-2}) \,.
\end{align}

\paragraph{NNNLO} 

To the best of our knowledge, the NNNLO spin-orbit coupling has not been computed yet. We simply present the result.
\begin{align}
X_{(1,0)}^{{\rm N}^3{\rm LO}} = 
\left( \frac{84 m_a^6+50 m_a^4 m_b^2+84 m_a^2 m_b^4+80 m_a m_b^5-45 m_b^6}{128 m_a^6 m_b^6} \right) \vec{p}^6 \,.
\label{N3LO-SO}
\end{align}
%

\subsection{Quadratic in spin (up to NNLO)}

Expanding the exact results \eqref{SS-ab} and \eqref{SS-aa} in $\vec{p}^2$, 
we obtain sub-leading corrections. We write down our results explicitly 
up to NNLO and compare them with previous PN computations. 

The NLO spin-spin Hamiltonian was computed in the ADM framework in \cite{Steinhoff:2007mb,Hartung:2011ea} 
and in the EFT framework in \cite{Levi:2011eq,Levi:2014sba}. The equivalence 
between the two approaches was established in \cite{Levi:2014sba}.
The NLO spin-squared coupling was computed in \cite{Porto:2008jj,Steinhoff:2008ji,Hergt:2008jn,Hergt:2010pa,Levi:2015msa}. 
The NNLO spin-squred couplings were computed in
\cite{Levi:2015ixa, Levi:2016ofk}. The equivalence among different approaches 
were established in later references. 

\paragraph{NLO} 
The NLO spin-spin term in our framework is
\begin{align}
X^{\rm NLO}_{(1,1)} = 
\left( \frac{ 2 m_a^2  + 9 m_a m_b + 2 m_b^2  }{4m_a^2 m_b^2} \right) \vec{p}^2 \,.
\label{NLO-SSb-iso}
\end{align}
It can be compared with (6.32) of \cite{Levi:2015msa}. 
Even after reducing to the COM frame, the result of \cite{Levi:2015msa} appears 
to carry many non-vanishing coefficients. 
It is not clear how many of them are gauge invariant. 
According to our result, only two of them are invariant once we take into account 
the exchange symmetry, $m_a \leftrightarrow m_b$. 

The NLO spin-squared term in our framework is
\begin{align}
X_{(2,0)}^{\rm NLO} &= 
\left( \frac{C_2^{(a)} (6m_a^2 +16 m_a m_b + 6 m_b^2) - (8m_a + 7 m_b) m_b }{8m_a^2 m_b^2}
\right) \vec{p}^2 
\,.
\label{NLO-SSa-iso}
\end{align}
It can be compared with (6.45) of \cite{Levi:2015msa}.

\paragraph{Canonical transformation for NLO spin($a$)-spin(b)}

For the spin($a$)-spin(b) interaction term, we parametrize the Hamiltonian by 
\begin{align}
\begin{split}
H_{S_a S_b}^{\text{NLO}}  
= - \frac{1}{4} \left(\frac{G}{r^3}\right) 
& \left[h_1 p^2 (\vec{a}_b\cdot\vec{a}_b) + h_2 p^2 (\vec{a}_a\cdot\hat{n})  (\vec{a}_b\cdot\hat{n}) 
+ h_3 (\vec{p} \cdot \hat{n})^2 (\vec{a}_b\cdot\vec{a}_b)
\right.
\\
&\quad 
+ h_4 (\vec{p} \cdot \hat{n})^2 (\vec{a}_a\cdot\hat{n})  (\vec{a}_b\cdot\hat{n}) 
+ h_5 (\vec{p} \cdot \vec{a}_a)(\vec{p} \cdot \vec{a}_b) 
\\
&\quad\quad \left.
+ \frac{1}{2} (\vec{p} \cdot \hat{n}) \{ h_6 (\vec{p} \cdot \vec{a}_a) (\vec{a}_b \cdot \hat{n})  + \bar{h}_6 (\vec{p} \cdot \vec{a}_b) (\vec{a}_a\cdot\hat{n}) \}\right] \,,
\end{split}
\label{NLO-SSab-family}
\end{align}
and generators of transformation, 
\begin{align}
\begin{split}
g_{S_a S_b}^{\text{NLO}} = - \frac{m_a}{4(1+\zeta^{-1})} \left( \frac{G}{r^2}\right) 
&\left[ g_1 (\vec{p} \cdot \hat{n}) (\vec{a}_b\cdot\vec{a}_b) 
+ g_2  (\vec{p} \cdot \hat{n} ) (\vec{a}_a\cdot\hat{n})  (\vec{a}_b\cdot\hat{n}) 
 \right.
\\
& \quad \qquad 
\left.  
 + \frac{1}{2} \{ g_3 (\vec{p} \cdot \vec{a}_a) (\vec{a}_b \cdot \hat{n}) + \bar{g}_3 (\vec{p} \cdot \vec{a}_b) (\vec{a}_a\cdot\hat{n}) \}
 \right] \,.
\end{split}
\label{gen-SSab}
\end{align}

Our result \eqref{NLO-SSb-iso} amounts to 
\begin{align}
h_1 = 2 \zeta + 9 + 2 \zeta^{-1} \,,
\quad 
h_2 = -3 h_1 \,,
\quad 
h_3 = h_4 = h_5 = h_6 = 0 \,.
\end{align}
In (6.10) of \cite{Levi:2014sba}, the $h$ parameters are
\begin{equation}
\begin{array}{lll}
h_1 = 6 \zeta + 16 +  6 \zeta^{-1} \,,
&
h_2 = -6 \zeta -21  -6 \zeta^{-1} \,,
\\
h_3 = -21 \zeta -12  -12  \zeta^{-1} \,,
&
h_4 = -30 \,, 
\\
h_5 = -6 \zeta -14 -6  \zeta^{-1} \,,
&
h_6 = 12 \zeta + 54 +  24 \zeta^{-1} \,,
\end{array}
\end{equation}
and $\bar{h}_6 = h_6|_{\zeta \rightarrow 1/\zeta}$. 
Taking the difference $\Delta h_k = h_k^{\rm EFT} - h_k^{\rm amp}$, we find
\begin{equation}
\begin{array}{lll}
\Delta h_1 = 4 \zeta + 7 +  4 \zeta^{-1} \,,
&
\;\; \Delta h_2 = 6 \,,
\\
\Delta h_3 = -21 \zeta -12  -12  \zeta^{-1} \,,
&
\;\; \Delta h_4 = -30 \,, 
\\
\Delta h_5 = -6 \zeta -14 -6  \zeta^{-1} \,,
&
\;\; \Delta h_6 = 12 \zeta + 54 +  24 \zeta^{-1} \,,
\end{array}
\label{SSz}
\end{equation}
The canonical transformation at NLO relates $\Delta h_k$ to $g_k$ as
\begin{equation}
\begin{array}{lll}
\Delta h_1 = -g_1 \,,
&
\quad \Delta h_2 = -g_2 \,,
&
\quad \Delta h_3 = 3 g_1 \,,
\\
\Delta h_4 = 5 g_2 \,, 
&
\quad \Delta h_5 =- (g_3 + \bar{g}_3)/2 \,,
&
\quad \Delta h_6 = 3g_3 - 2g_2 \,.
\end{array}
\label{SSw}
\end{equation}
The differences \eqref{SSz} match the relations \eqref{SSw} precisely if we set 
\begin{equation}\label{eq:1408NLO matching}
\begin{array}{lll}
g_1 = -(4 \zeta + 7 +  4 \zeta^{-1}) \,,
&
\quad g_2 = -6 \,,
\\
g_3 = 4 \zeta + 14  + 8  \zeta^{-1} \,,
&
\quad \bar{g}_3 = g_3|_{\zeta \rightarrow 1/\zeta}\,.
\end{array}
\end{equation}

\paragraph{Canonical transformation for NLO spin($a$)-squared} 

For the spin($a$)-squared term, we consider a family of Hamiltonians: 
\begin{align}
\begin{split}
H_{S_a^2}^{\rm NLO} = - \frac{1}{8} \left(\frac{G}{r^3}\right) 
& \left[h_1 \vec{p}^2 \vec{a}_a^2  + h_2 p^2 (\vec{a}_a \cdot \hat{n})^2 
+ h_3 (\vec{p} \cdot \hat{n})^2 \vec{a}_a^2 + h_4 (\vec{p} \cdot \hat{n})^2 (\vec{a}_a \cdot \hat{n})^2\right.
\\
&\quad \left.
+ h_5 (\vec{p} \cdot \vec{a}_a)^2 + h_6  (\vec{p} \cdot \hat{n}) (\vec{p} \cdot \vec{a}_a) (\vec{a}_a \cdot \hat{n})  \right] \,,
\end{split}
\label{NLO-SS-family}
\end{align}
and generators of transformation, 
\begin{align}
g_{S_a^2}^{\rm NLO} &= - \frac{m_a}{8(1+\zeta^{-1})} \left( \frac{G}{r^2}\right) 
\left[ g_1 (\vec{p} \cdot \hat{n}) \vec{a}_a^2 + g_2 (\vec{p} \cdot \hat{n} ) (\vec{a}_a \cdot \hat{n})^2  + g_3 (\vec{p} \cdot \vec{a}_a) (\vec{a}_a \cdot \hat{n})  \right] \,.
\label{gen-SS}
\end{align}
 
Our result \eqref{NLO-SSa-iso} amounts to $(C_{\rm here} = C_2^{(a)}$)
\begin{align}
h_1 = (6 \zeta + 16 + 6 \zeta^{-1})C - (7\zeta +8) \,,
\quad 
h_2 = -3 h_1 \,,
\quad 
h_3 = h_4 = h_5 = h_6 = 0 \,.
\end{align}
This is to be compared with (6.45) of \cite{Levi:2015msa}. 
Reducing it to the COM frame, we obtain a somewhat simplified formula in our notation,
\begin{equation}
\begin{array}{ll}
h_1 = (10\zeta + 18 + 6 \zeta^{-1} )C - (10\zeta + 12)  \,, 
\\
h_2 = - (18\zeta + 42 + 18\zeta^{-1} ) C + 21 \zeta + 24  \,,
\\
h_3 =  -(12 \zeta +6) C + 9 \zeta + 12 \,, 
&
h_4 = -30C \,, 
\\
h_5 = -(4\zeta + 4) C + 10\zeta + 12 \,, 
&
h_6 = (12 \zeta + 24) C - (30\zeta + 36) \,. 
\end{array}
\end{equation}
Taking the difference, $\Delta h_k = h_k^{\rm EFT} - h_k^{\rm amp}$, we find
\begin{equation}
\begin{array}{ll}
\Delta h_1 = (4\zeta +2) C - (3 \zeta + 4) \,, 
&
\quad \Delta h_2 = 6 C \,,
\\
\Delta h_3 = -(12 \zeta +6) C + 9 \zeta + 12 \,,
&
\quad \Delta h_4 = -30C \,, 
\\
\Delta h_5 = -(4\zeta + 4) C + 10\zeta + 12 \,,
&
\quad \Delta h_6 = (12 \zeta + 24) C - (30\zeta + 36) \,.
\end{array}
\label{SSx}
\end{equation}
Performing the canonical transformation, we relate $\Delta h_k$ to $g_k$:
\begin{equation}
\begin{array}{ll}
\Delta h_1 = -g_1 \,,
&
\quad \Delta h_2 = -g_2 \,,
\\
\Delta h_3 = 3 g_1 \,,
&
\quad \Delta h_4 = 5 g_2 \,, 
\\
\Delta h_5 = - g_3 \,,
&
\quad \Delta h_6 = 3g_3 - 2g_2 \,.
\end{array}
\label{SSy}
\end{equation}
The differences \eqref{SSx} match the relations \eqref{SSy} precisely, if we set 
\begin{align}
g_1 = -(4\zeta +2) C + (3 \zeta + 4) \,,
\quad
g_2 = - 6C \,,
\quad 
g_3 = (4\zeta + 4) C - (10\zeta + 12) \,.
\end{align}

\paragraph{NNLO}

The NNLO spin-spin term is
\begin{align}
\begin{split}
X^{\rm NNLO}_{(1,1)} &= 
\left( 
\frac{ - 6 m_a^4 -15 m_a^3 m_b^2 + 4 m_a^2 m_b^2  -15 m_a m_b^3 -6 m_b^4 }{16m_a^4 m_b^4}  \right) \vec{p}^4 \,.
\end{split}
\label{NNLO-SSab}
\end{align}
The NNLO spin-squared term is
\begin{align}
\begin{split}
X^{\rm NNLO}_{(2,0)} &= \left( \frac{ C_2^{(a)} (-5m_a^4 + 18m_a^2m_b^2 -5m_b^4) - 3(7m_a^2 +4 m_a m_b -2 m_b^2) m_b^2 }{16m_a^4m_b^4} \right) \vec{p}^4\,.
\end{split}
\label{NNLO-SSa-iso}
\end{align}
These are to be compared with eqs.(3.3)-(3.4)  of \cite{Levi:2016ofk}.

\paragraph{Canonical transformation for NNLO  spin($a$)-spin(b)}

We parametrize the spin($a$)-spin(b) term of the Hamiltonian at the NNLO order as
\begin{equation}
\begin{split}
H_{S_a S_b}^{\text{NNLO}} 
=&
\frac{1}{16m_a m_b}
\left( \frac{G}{r^3} \right)
\Big[
(\hat{n}\cdot\vec{p})^4 
\left[ h_7 (\vec{a}_a\cdot\vec{a}_b) + h_8 (\hat{n}\cdot \vec{a}_a)(\hat{n}\cdot \vec{a}_b) \right]\\
&+
\frac{1}{2}(\hat{n}\cdot\vec{p})^3 
\left[ h_9 (\hat{n}\cdot\vec{a}_a)(\vec{p}\cdot\vec{a}_b) + \bar{h}_9 (\hat{n}\cdot\vec{a}_b)(\vec{p}\cdot\vec{a}_a) \right]\\
&+
(\hat{n}\cdot\vec{p})^2
\left[h_{10} (\vec{p}\cdot\vec{a}_a)(\vec{p}\cdot\vec{a}_b) + h_{11} \vec{p}^2\vec{a}_a\cdot\vec{a}_b +  h_{12}\vec{p}^2(\hat{n}\cdot \vec{a}_a)(\hat{n}\cdot \vec{a}_b)\right]\\
&+
\frac{1}{2}(\hat{n}\cdot\vec{p})
\left[ h_{13}\vec{p}^2(\hat{n}\cdot\vec{a}_a)(\vec{p}\cdot\vec{a}_b) + \bar{h}_{13}\vec{p}^2(\hat{n}\cdot\vec{a}_b)(\vec{p}\cdot\vec{a}_a) \right]\\
&+
h_{14}\vec{p}^2 (\vec{p}\cdot\vec{a}_a)(\vec{p}\cdot\vec{a}_b) 
+
\vec{p}^4 \left[ h_{15} (\vec{a}_a\cdot\vec{a}_b) + h_{16} (\hat{n}\cdot \vec{a}_a)(\hat{n}\cdot \vec{a}_b) \right]
\Big] \,.
\end{split}
\end{equation}
The NNLO generator is parametrized as
\begin{equation}
\begin{split}
g_{S_a S_b}^{\text{NNLO}} = -& \frac{1}{16m_b(1+\zeta^{-1})}  \left( \frac{G}{r^2}\right) \Big[
(\hat{n}\cdot\vec{p})^3 \left[g_4 (\vec{a}_a\cdot\vec{a}_b)+g_5 (\hat{n}\cdot \vec{a}_a)(\hat{n}\cdot \vec{a}_b)\right] \\
+&
(\hat{n}\cdot\vec{p})^2 \left[ g_6 (\hat{n}\cdot \vec{a}_a) (\vec{p}\cdot \vec{a}_b) + \bar{g}_6 (\hat{n}\cdot \vec{a}_b) (\vec{p}\cdot \vec{a}_a) \right]\\
+&
(\hat{n}\cdot\vec{p}) \left[g_7 (\vec{p}\cdot\vec{a}_a)(\vec{p}\cdot\vec{a}_b) + g_8\vec{p}^2 (\hat{n}\cdot \vec{a}_a)(\hat{n}\cdot \vec{b}_a) + g_9\vec{p}^2 \vec{a}_a \cdot \vec{a}_b\right] \\
+& \vec{p}^2\left[ g_{10} (\hat{n}\cdot \vec{a}_a) (\vec{p}\cdot \vec{a}_b) + \bar{g}_{10} (\hat{n}\cdot \vec{a}_b) (\vec{p}\cdot \vec{a}_a) \right]
\Big] \,.
\end{split}
\end{equation}

Our result \eqref{NNLO-SSab} amounts to 
\begin{equation}
\begin{split}
&
h_7 = h_8 = h_9 = \bar{h}_9 = h_{10} = h_{11} = h_{12} = h_{13} = \bar{h}_{13} = h_{14} = 0, 
\\
&
h_{15} = 6 \zeta ^2 +15 \zeta-4 +15 \zeta^{-1} +6\zeta ^{-2},
\quad 
h_{16} = -3 h_{15}.
\end{split}
\label{SS-NNLO-amp}
\end{equation}
(6.12) of \cite{Levi:2014sba}, translated to our notation, yields
\begin{equation}
\begin{split}
&
h_7 =  0, \quad
h_8 = 210, \quad
h_9 =  -60 ( 5 + 2\zeta^{-1}), \quad 
\bar{h}_9 = h_9|_{\zeta \rightarrow \zeta^{-1}} \,,
\\
&
h_{10} = 12 \left(\zeta + 5 + \zeta^{-1} \right), \quad
h_{11} =- 3 \left(16 \zeta ^2 + 19 \zeta + 14 +19 \zeta^{-1}  +16 \zeta^{-2} \right), \quad
\\
&
h_{12} = 90, \quad
h_{13} = 6 \left(16 \zeta ^2 +19 \zeta +6 +19 \zeta^{-1}  +6 \zeta^{-2}\right), 
\quad
\bar{h}_{13} = h_{13}|_{\zeta \rightarrow \zeta^{-1}} \,,
\\
&
h_{14} = -2(11 \zeta ^2 + 15 \zeta + 4  + 15\zeta^{-1} \zeta + 11\zeta^{-2}), 
\\
&
h_{15} = 22 \zeta ^2 +34 \zeta +10 +34 \zeta^{-1} +22 \zeta^{-2} \,,\\
&
h_{16} = -3\left(6 \zeta ^2 + 15 \zeta + 8 + 15 \zeta^{-1} +6 \zeta^{-2}\right) \,.
\end{split} 
\label{SS-NNLO-EFT}
\end{equation}
The changes in the $h$ parameters are related by \eqref{cano-NNLO} to the $g$ parameters as
\begin{equation}
\begin{split}
&
\Delta h_7 =  - 5g_4, \quad 
\Delta h_8 = - 7g_5, 
\\
&
\Delta h_9 =  2(g_5 - 5g_6), \quad
\Delta \bar{h}_9 =    2(g_5 - 5\bar{g}_6), \quad
\Delta h_{10} = g_6 +\bar{g}_6 - 3g_7, \\
&
\Delta h_{11} =  6\left(\zeta - 1 + \zeta^{-1} \right) g_1 + 3 \left(g_4-g_9\right),
\\
&
\Delta h_{12} =  10 \left(\zeta - 1 + \zeta^{-1} \right) g_2 + (3g_5-5g_8), \\
&
\Delta h_{13} =  2(2 g_6+g_8-3 g_{10}) - 2 \left(\zeta - 1 + \zeta^{-1} \right) \left(2 g_2-3 \bar{g}_3\right),
\\
&
\Delta \bar{h}_{13} = 2(2 \bar{g}_6+g_8-3 \bar{g}_{10} ) - 2 \left(\zeta - 1 + \zeta^{-1} \right) \left(2 g_2-3 g_3\right), \\
&
\Delta h_{14} =   g_7 +g_{10}+ \bar{g}_{10} - \left(\zeta - 1 + \zeta^{-1} \right) \left(g_3+ \bar{g}_{3}\right), 
\\
&
\Delta h_{15} = g_9- 2 \left(\zeta - 1 + \zeta^{-1} \right) g_1,  \quad
\Delta h_{16} =  g_8 - 2\left(\zeta - 1 + \zeta^{-1} \right)  g_2 .
\end{split}
\end{equation}
The difference $\Delta h = h^{\rm EFT} - h^{\rm amp}$ between 
\eqref{SS-NNLO-EFT} and \eqref{SS-NNLO-amp} is accounted for if we choose the $g$ parameters as 
\begin{equation}
\begin{split}
&
g_4 = 0, \quad 
g_5 = -30, \quad 
g_6 = 12 (2 + \zeta^{-1}) , \quad 
\bar{g}_6 = 12 (\zeta +2), \\
& 
g_7 = -4,\quad
g_8 = - 12 \left(\zeta + 2 + \zeta^{-1} \right), \quad 
g_9 = 8 \zeta ^2 + 13 \zeta + 12  + 13 \zeta^{-1} +8\zeta^{-2}, \quad \\
&
g_{10} = -(8 \zeta^2 + 13 \zeta +4 + \zeta^{-1} +2\zeta^{-2}), \quad
\bar{g}_{10} = -(2 \zeta ^2 + \zeta + 4+13 \zeta^{-1} +8\zeta^{-2}).
\end{split}
\end{equation}


\paragraph{Canonical transformation for NNLO spin($a$)-squared} 

We parametrize  the NNLO spin($a$)$^2$ sector Hamiltonian as
\begin{equation}
\begin{split}
 H_{S_a^2}^{\text{NNLO}}
= \frac{1}{16 m_a m_b r^3}  \left( \frac{G}{r^3}\right) & \Big[
(\hat{n}\cdot\vec{p})^4 
\left[ 
h_7\vec{a}_a^2 + 
h_8 (\hat{n}\cdot \vec{a}_a)^2 \right] \\
&+
(\hat{n}\cdot\vec{p})^3 
\left[ h_9(\hat{n}\cdot\vec{a}_a)(\vec{p}\cdot\vec{a}_b)  \right]\\
&+
(\hat{n}\cdot\vec{p})^2
\left[ 
h_{10}(\vec{p}\cdot\vec{a}_a)^2 
+h_{11} \vec{p}^2\vec{a}_a^2
+ h_{12} \vec{p}^2(\hat{n}\cdot \vec{a}_a)^2
\right]\\
&+
(\hat{n}\cdot\vec{p})
\left[ h_{13}\vec{p}^2(\hat{n}\cdot\vec{a}_a)(\vec{p}\cdot\vec{a}_b)  \right]\\
&+
h_{14}\vec{p}^2 (\vec{p}\cdot\vec{a}_a)^2 
+ 
\vec{p}^4 \left[h_{15} \vec{a}_a^2 +  h_{16}(\hat{n}\cdot \vec{a}_a)^2 \right]
\Big] \,.
\end{split}
\end{equation}
The NNLO generator is parametrized as
\begin{equation}
\begin{split}
g_{S_a^2}^{\text{NNLO}} = - \frac{1}{16m_b (1+\zeta^{-1})} \left( \frac{G}{r^2}\right)  \Big[
&(\hat{n}\cdot\vec{p})^3 \left[g_4 \vec{a}_a^2+g_5 (\hat{n}\cdot \vec{a}_a)^2\right] \\
+&
(\hat{n}\cdot\vec{p})^2 \left[ g_6 (\hat{n}\cdot \vec{a}_a) (\vec{p}\cdot \vec{a}_a)  \right]\\
+&
(\hat{n}\cdot\vec{p}) \left[g_7 (\vec{p}\cdot\vec{a}_a)^2 + g_8\vec{p}^2 (\hat{n}\cdot \vec{a}_a)^2+ g_9\vec{p}^2 \vec{a}_a^2 \right] \\
+& \vec{p}^2\left[ g_{10} (\hat{n}\cdot \vec{a}_a)(\vec{p} \cdot \vec{a}_a)   \right]
\Big]
\end{split}
\end{equation}
The changes in the $h$ parameters are related by \eqref{cano-NNLO} to the $g$ parameters as
\begin{equation}
\begin{split}
&
\Delta h_7 =  - 5g_4, \quad 
\Delta h_8 = - 7g_5 , \quad
\Delta h_9 =  2g_5 - 5g_6 , \quad
\Delta h_{10} =   g_6  - 3g_7 , \\
&
\Delta h_{11} =  3\left(\zeta -1 + \zeta^{-1} \right) g_1 + 3 \left(g_4-g_{9}\right),\quad
\Delta h_{12} =  5\left(\zeta - 1 + \zeta^{-1} \right) g_2 + 3g_5-5g_8 , \\
&
\Delta h_{13} =  2 g_6+2g_8-3 g_{10} - \left(\zeta - 1 + \zeta^{-1} \right) \left(2 g_2-3 g_3\right), \\
&
\Delta h_{14} = g_7+g_{10} - \left(\zeta - 1 + \zeta^{-1} \right) g_3, \quad
\Delta h_{15} = g_9 - \left(\zeta - 1 + \zeta^{-1} \right) g_1 , \\
&
\Delta h_{16} =  g_8 - \left(\zeta - 1 + \zeta^{-1} \right) g_2 .
\end{split}
\end{equation}

Our result \eqref{NNLO-SSa-iso} amounts to 
\begin{equation}
\begin{split}
&
h_7 = h_8 = h_9 = h_{10} = h_{11} = h_{12} = h_{13} = h_{14} = 0 ,\\
&
h_{15} = C \left(5 \zeta ^2 -18 + 5  \zeta^{-2} \right)+ (-6 \zeta ^2+12 \zeta +21 ),
\quad 
h_{16} = -3 h_{15} .
\end{split}
\label{SSaa-NNLO-amp}
\end{equation}
This can be compared with (3.4) of \cite{Levi:2016ofk} 
which adopts the same coordinate gauge as \cite{Levi:2015msa}: 
\begin{align}
\begin{split}
&
h_7 = 15\left( \zeta+ 4 + 4 \zeta^{-1}\right) + 15C(3+4\zeta), \quad 
h_8 = 105 C, \\
&
h _9 = - 30 C(4 + 2\zeta) -30(\zeta + 6 + 6\zeta^{-1}) , \\
&
h_{10} =  6(-2\zeta^2-4\zeta+6+10\zeta^{-1}) + 6C(\zeta^2 + 6\zeta + 6) , \\
&
h_{11} = 6C (2\zeta^2 + 6 \zeta + 5 + 2\zeta^{-1}) + 3(5\zeta^2 - 6 \zeta - 33 - 24\zeta^{-1} ) ,\\
&
h_{12} = 30C(2\zeta + 5 + 2\zeta^{-1}) + 15(\zeta + 8 + 8\zeta^{-1}) , \\
&
h_{13} = -6C(3\zeta^2 + 18 \zeta+26+8\zeta^{-1}) - 3(7\zeta^2-27\zeta-50-12\zeta^{-1}) , \\
&
h_{14} = 4C(\zeta^2+ 7 \zeta + 8 + 2\zeta^{-1})+(11\zeta^2-15\zeta-42-12\zeta^{-1}) , \\
&
h_{15} = C(\zeta^2 - 24\zeta-37-4\zeta^{-1}+5\zeta^{-2}) + (-11\zeta^2 + 15 \zeta + 42 + 12\zeta^{-1}) , \\
&
h_{16} = -3C (5\zeta^2 + 4 \zeta -5 + 4\zeta^{-1}+5\zeta^{-2}) - 3(-6\zeta^2 + 13 \zeta + 29 + 8 \zeta^{-1}) .\\
\end{split}
\label{SSaa-NNLO-EFT}
\end{align}
The difference $\Delta h = h^{\rm EFT} - h^{\rm amp}$ between 
\eqref{SSaa-NNLO-EFT} and \eqref{SSaa-NNLO-amp} is accounted for if we choose the $g$ parameters as
\begin{equation}
\begin{split}
&
g_4 = - 3(3+4\zeta)C - 3\left(4 \zeta^{-1} +4 +\zeta \right), \quad 
g_5 = - 15 C, \\
&
g_6 = 6 (2 \zeta +3) C + 2 (3\zeta + 18 + 18 \zeta^{-1}), \quad \\
& 
g_7 = -2C(\zeta^2 + 4\zeta + 3) - 2\left(-2\zeta^2 - 5 \zeta + 4 \zeta^{-1}\right) , \quad\\
&
g_8 = - 3(6\zeta + 11 + 6\zeta^{-1}) C - 3\left(\zeta + 8 + 8 \zeta^{-1}\right) , \quad  \\
&
g_9 = -(8 \zeta^2 + 22\zeta + 21 + 6\zeta^{-1} ) C + 2( -\zeta^2 + 2 \zeta +10  +8 \zeta^{-1} ) , \quad \\
&
g_{10} = 2 (1 + \zeta^{-1}) (\zeta +2) (5 \zeta +3) C + (1 + \zeta^{-1}) (-3 \zeta^2-24 \zeta -16)  .
\end{split}
\end{equation}


%
%

%
%

\chapter{Obstructions and applications at one-loop level} \label{chap:loop}
\section{The gravitational Compton amplitude} \label{sec:GravComp}
The gravitational Compton amplitude, the amplitude for the elastic scattering process of a graviton bouncing off a massive particle, can be used to argue that massive particles with spin $s>2$ cannot be elementary particles. The argument is based on a sick behaviour of the Compton amplitude~\cite{Arkani-Hamed:2017jhn,Chung:2018kqs}, which can be cured at the price of worse high-energy behaviour with some ambiguities~\cite{Chung:2018kqs}. The arguments and the remedy will be reviewed in this section.

\subsection{The na\"ive gravitational Compton amplitude}
The amplitude for $2 \to 2$ elastic scattering of a minimally coupled massive spinning particle and a graviton is given as ($\k = \sqrt{32 \pi G} = 2 / M_{pl}$)
\bl
M_4 (\bf{p}_1^s,k_2^{+2},k_3^{-2},\bf{p}_4^s) &= \frac{(\k/2)^2 \, \sbra{2}p_1\ket{3}^4}{t(s-m^2)(u-m^2)} \left( \frac{[ \mathbf{4} 2 ] \la 3 \mathbf{1} \ra + \la \mathbf{4} 3 \ra [ 2 \mathbf{1} ]}{\sbra{2} p_1 \ket{3}} \right)^{2s} \,, \label{eq:gravComp}
\el
which has been argued based on factorisation properties of scattering amplitudes~\cite{Arkani-Hamed:2017jhn,Chung:2018kqs} and BCFW recursion~\cite{Chung:2019duq,Johansson:2019dnu}. This amplitude does not make sense for $s>2$ because the amplitude develops a \emph{spurious pole} at $\sbra{2} p_1 \ket{3} = 0$; a pole that does not have a corresponding one-particle state. This can be interpreted as a sign that massive particles with spin $s>2$ are necessarily composite particles, rather than elementary~\cite{Arkani-Hamed:2017jhn,Chung:2018kqs}. Although the formula \eqc{eq:gravComp} for the gravitational Compton amplitude nicely extends beyond $s=2$, this amplitude \emph{must} not be trusted for spins $s>2$.

It could be argued that while \eqc{eq:gravComp} develops a spurious pole for $s>2$ and hence incorrect, there is no problem for using it if the spurious pole cannot be reached in the physical region. Unfortunately, the spurious pole can be reached in the physical region; consider the COM frame with following values for the momenta, where $p > 0$.
\bl
\bld
p_1 &= ( \sqrt{m^2 + p^2}, 0, 0, p )
\\ k_2 &= ( p , 0, 0, -p )
\\ k_3 &= ( - p, p \sin\th, 0, p \cos\th )
\\ p_4 &= ( - \sqrt{m^2 + p^2}, - p \sin\th, 0, - p \cos\th )
\eld
\el
All the momenta were considered as incoming. The 
spinor-helicity variables of massless legs 
are given as
\begin{gather}
\bld
\ket{2}_{\a} &= \sqrt{2p} \left( \begin{array}{c} 1 \\ 0 \end{array} \right)
\\ \sbra{2}_{\dot\a} &= \sqrt{2p} \left( \begin{array}{cc} 1 & 0 \end{array} \right)
\eld
\phantom{asdfasdf}
\begin{aligned}
\ket{3}_{\a} &= \sqrt{2p} \left( \begin{array}{c} \cos(\th/2) \\ \sin(\th/2) \end{array} \right)
\\ \sbra{3}_{\dot\a} &= - \sqrt{2p} \left( \begin{array}{cc} \cos(\th/2) & \sin(\th/2) \end{array} \right)
\end{aligned}
\end{gather}
and the Lorentz invariant responsible for the spurious pole becomes
\bl
\bld
\sbra{2} p_1 \ket{3} &= 2mp \left( \sqrt{ 1+ (p/m)^2} + p/m \right) \cos(\th/2) \,.
\eld
\el
The spurious pole is reached by the backscattering condition $\th = \pi$, which lies inside the physical region. Therefore, the formula \eqc{eq:gravComp} cannot be used when the massive particle has spin $s>2$.

\subsection{Resolving the spurious pole of na\"ive Compton amplitude}
The spurious pole appearing in the formula \eqc{eq:gravComp} can be resolved by substituting the expression that is responsible for the pole to another expression that does not possess the pole, which maintains the factorisation property at the physical poles~\cite{Chung:2018kqs}. This approach is based on the idea that tree amplitudes are rational functions of Lorentz invariants, so the amplitude can be uniquely fixed up to polynomials in Lorentz invariants by matching all residues on the poles. The undetermined \emph{polynomial terms} will be linked to tidal response operators of one-particle effective action, whose coefficients are also known as \emph{tidal Love numbers} and their generalisations. It is known that Love numbers vanish in case of non-spinning BHs of GR~\cite{Fang:2005qq,Damour:2009vw,Binnington:2009bb,Gurlebeck:2015xpa}, and their generalisations to spinning BHs are also known to vanish at lowest orders in spin for spinning BHs~\cite{Poisson:2014gka,Pani:2015hfa,Pani:2015nua}. Unfortunately, it is unclear how the information of Love numbers and their generalisations can be disentangled unambiguously from the full amplitude to remove polynomial ambiguities.

The factor that is responsible for the pole can be recast as follows.
\bl
\bld
\frac{ [ \bold{1} 2 ] \la 3 \bold{4} \ra  + \la \bold{1} 3 \ra [ 2 \bold{4} ] }{ \bra{3} p_1 \sket{2} } &= \frac{\bra{\bold{1}} p_1 + k_2 \sket{\bold{4}}}{m^2} - \frac{(s-m^2) \la \bold{1} 3 \ra [ 2 \bold{4} ]  }{m^2 \bra{3} p_1 \sket{2} }
\eld \label{eq:s-chan_recast}
\el
The latter term on the RHS of the above equation vanishes on the factorisation channel $s \to m^2$. The key idea is to \emph{remove} the latter and keep the former, if the factor $\bra{3} p_1 \sket{2}$ in the denominator cannot be cancelled by the same factor on the numerator. This procedure will maintain the factorisation property on the $s$-channel while eliminating the spurious pole problem. The price paid is the introduction of the factor $m^{-2}$, which is a sign of ill-behaviour in the high-energy limit.

Ultimately, all three channels\textemdash $s$-, $t$-, and $u$-channel\textemdash need to be considered and may require introduction of additional pole terms to ensure factorisation. The relation \eqc{eq:s-chan_recast} is unwieldy for this purpose and a different relation that maintains the spirit of \eqc{eq:s-chan_recast} was used for resolving the spurious pole problem of \eqc{eq:gravComp} in ref.\cite{Chung:2018kqs}. The resulting expression is;
\bl
\bld
M_4(s>2)
&= -\frac{\MixLeft{3}{p_1}{2}^4}{(s-m^2) (u-m^2) t M_{pl}^2 } \mathcal{F}^{2s} \\
&\phantom{=}+ \frac{2s \MixLeft{3}{p_1}{2}^3}{t (s-m^2) }\frac{\AB{3 \boldsymbol{4}} \SB{2 \boldsymbol{1}}}{2m^2 M_{pl}^2} \mathcal{F}^{2s-1} - \frac{2s \MixLeft{3}{p_4}{2}^3}{t (u-m^2)}\frac{\AB{3 \boldsymbol{1}} \SB{2 \boldsymbol{4}}}{2m^2 M_{pl}^2} \mathcal{F}^{2s-1} \\
&\phantom{=} + \Bigg\{\frac{\MixLeft{3}{p_1}{2}^2 \AB{3 \boldsymbol{4}}^2 \SB{2 \boldsymbol{1}}^2 }{4m^4 (s-m^2) M_{pl}^2}  \Bigg[\sum_{r=2}^{2s}\binom{2s}{r} \mathcal{F}^{2s-r} \CA_s^{r-2}\Bigg]  \\
 &\phantom{=\{} + \frac{\MixLeft{3}{p_1}{2}^2 \AB{3 \boldsymbol{1}}^2 \SB{2 \boldsymbol{4}}^2 }{4m^4 (u-m^2) M_{pl}^2}  \Bigg[\sum_{r=2}^{2s} \binom{2s}{r} (-1)^{r} \mathcal{F}^{2s-r} \CA_u^{r-2}\Bigg] \Bigg\} \\
&\phantom{=} - \frac{Poly + Poly_{\SB{23} } + Poly_{\AB{23}} }{t M_{pl}^2} \,,
\eld \label{eq:gravComp_resolv}
\el
where definitions for the variables are;
\bl
\bgd
F_1 = \frac{\SB{ \BS{14} }}{m}, \quad F_2 = \frac{ \AB{\BS{4}2}\SB{2\BS{1}} - \AB{\BS{1}2}\SB{2\BS{4}} }{2m^2} \,,
\\ \tilde{F}_1 = \frac{\AB{ \BS{14} }}{m}, \quad \tilde{F}_2 =  \frac{ \AB{\BS{1}3}\SB{3\BS{4}} - \AB{\BS{4}3}\SB{3\BS{1}} }{2m^2} \,,
\\ \mathcal{F} = \frac{1}{2}(F + \tilde{F}), \quad \mathcal{F}_1 = \frac{1}{2}(F_1 + \tilde{F}_1), \quad \mathcal{F}_2 = \frac{1}{2}(F_2 + \tilde{F}_2) \,,
\\ \CA_s = \frac{ - \AB{\boldsymbol{4}3} \SB{32} \AB{2 \boldsymbol{1}}}{4 m^3}  -  \frac{\SB{\boldsymbol{4}3} \AB{32} \SB{2 \boldsymbol{1}} }{4 m^3} \,,
\\ \CA_u = \frac{ -\AB{\boldsymbol{1}3} \SB{32} \AB{2 \boldsymbol{4}}}{4m^3} - \frac{ \SB{\boldsymbol{1}3} \AB{32} \SB{2 \boldsymbol{4}}}{4m^3} \,,
\egd
\el
and the $t$-channel residue terms are given as;
\bl
\bld
Poly &= - \MixLeft{3}{p_1}{2}^2 K^2  \sum_{r=1}^{2s-1} r \binom{2s}{r+1} \mathcal{F}_1^{2s-r-1} \mathcal{F}_2^{r-1} \,,
\\ Poly_{\AB{23}} &= \sum_{r=0}^{ \lceil s \rceil - 3} h_{A}(4 + 2r) ( F_1 - \tilde{F}_1 )^{2r+1} +  \sum_{r=0}^{ \lceil s \rceil - 2} g_{A}(2 + 2r) ( F_1 - \tilde{F}_1 )^{2r}
\\ &\phantom{=asdf} - \frac{ \MixLeft{3}{p_1}{2} K^2 C_{\AB{23}} }{2} \binom{2s}{3} \mathcal{F}_1^{2s-3} \,,
\\ Poly_{\SB{23}} &= \sum_{r=0}^{ \lceil s \rceil - 3} h_{S}(4 + 2r) (\tilde{F}_1 - F_1)^{2r+1} +  \sum_{r=0}^{ \lceil s \rceil - 2} g_{S}(2 + 2r) (\tilde{F}_1 - F_1)^{2r}
\\ &\phantom{=asdf} - \frac{\MixLeft{3}{p_1}{2} K^2 C_{\SB{23}} }{2} \binom{2s}{3} \mathcal{F}_1^{2s-3} \,.
\eld
\el
The functions $h$ and $g$ appearing in the $t$-channel residue terms are defined as;
\bl
\bld
h(n) &\equiv \frac{K^2 C^2}{2^{n-1}} \binom{2s}{n+1} \mathcal{F}_1^{2s-n-1} \,,
\\ g(n) & \equiv  - \frac{K^2 C^2}{2^n}  \sum_{r=1}^{2s-n-1}(2r+1) \binom{2s}{r+n+1} \mathcal{F}_1^{2s-r-n-1} \mathcal{F}_2^{r-1}\\
& \phantom{\equiv asdf} + \Big( \frac{s-u}{2} \Big) \frac{K^3 C}{2^{n-1}}  \sum_{r=1}^{2s-n-1}(r+1) \binom{2s}{r+n+1} \mathcal{F}_1^{2s-n-1-r} \mathcal{F}_2^{r-1} \,,
\eld
\el
with the understanding that $C = C_{\AB{23}}$ for $g_A(n)$ and $h_A(n)$, $C = C_{\SB{23}}$ for $g_S(n)$ and $h_s(n)$, and $g(n \geq 2s-1) = 0$. The definitions for the variables appearing in the $t$-channel residue terms are;
\bl
\bgd
C_{\SB{23}} = \frac{\SB{23}\AB{\boldsymbol{1}3}\AB{3\boldsymbol{4}}}{m}, \quad C_{\AB{23}} = - \frac{\AB{23}\SB{\boldsymbol{1}2}\SB{2\boldsymbol{4}}}{m} \,,
\\ K \equiv  \frac{\AB{3 \boldsymbol{4}} \SB{2 \boldsymbol{1}}}{2m^2} - \frac{\AB{3 \boldsymbol{1}} \SB{2 \boldsymbol{4}}}{2m^2} \,.
\egd
\el

\section{Matter stress tensor for Kerr-Newman black holes} \label{sec:KNBHMST}
The conclusion reached in section~\ref{sec:GravComp} is that massive higher spin particles having spin $s>2$ \emph{cannot} be considered as elementary particles. This begs the question; is it eligible to run composite particles in loops? While there are a plethora of examples where elementary particles running in loops give the correct answer, examples are limited for composite particles running in loops.

It could be argued that running composite particles in loops is fine if degree of freedom counting is done correctly to bar overcounting of independent degrees of freedom, but there is a qualitative difference for massive higher spin particles; while a fictitious elementary particle with the same quantum numbers can be hypothesised for lower spins, this is not available for higher spins. An explicit working example will boost our confidence of the argument, where running massive higher spin particles in loops gives a correct answer.

Such an example must meet two criteria; (a) the tree amplitudes used as building blocks for constructing the loop integrand must be fully under control, and (b) a corresponding classical computation must be available for comparison. Fortunately, the one-loop stress tensor form factor of QED satisfies both criteria, which has been studied in the works refs.\cite{Donoghue:2001qc,Holstein:2006ud}; only three-point amplitudes\textemdash which are fully understood\textemdash are needed, and a corresponding exact classical solution is known for Kerr-Newman BHs. Beginning with classical field theory computations in section~\ref{sec:class_KS}, it will be shown that corresponding QFT computations indeed match the classical field theory computations in section~\ref{sec:QFT_STFF}.

\subsection{Classical computations in Kerr-Schild coordinates} \label{sec:class_KS}
Kerr-Schild representation is a representation for solutions to Einstein's equations which reduces the non-linearities of GR to linear differential equations.\footnote{For a review, consult e.g.~\cite{Stephani:2003tm}.} It is precisely in this form that the double copy relations\textemdash originally found in amplitude contexts\textemdash between classical solutions for Einstein gravity and linearized Yang-Mills was found~\cite{Monteiro:2014cda}. Based on this example, we expect classical solutions having amplitude counterparts to be best represented in Kerr-Schild coordinates.

We will first review the Kerr-Newman solutions in Kerr-Schild coordinates, and then identify the point particle source that produces the Maxwell fields of the solutions. The expression for the source will in later sections be interpreted as the interaction term of the one-particle effective action, and be used to construct the three-point amplitude. Finally, we compute the stress tensor \eqc{eq:Stress_Tensor_Covariant_Form} that will later be compared with QFT computations.

\subsubsection{Kerr-Newman solution in Kerr-Schild coordinates}
The Kerr-Newman solution in Kerr-Schild form is given by\footnote{Following usual conventions of GR literature, we use Gaussian units in this section.}
\bl
\bld
g_{\m\n} &= \eta_{\m\n} - f k_\m k_\n
\\ f &= \frac{G r^2}{r^4 + a^2 z^2} [2 Mr - Q^2]
\\ k_\m &= \left( 1, \frac{rx + ay}{r^2 + a^2}, \frac{ry - ax}{r^2 + a^2}, \frac{z}{r} \right)
\\ A_\m &= \frac{Q r^3}{r^4 + a^2 z^2} k_\m
\eld \label{eq:KNBHinKSform}
\el
where $k^\m$ is null both in $\eta_{\m\n}$ and $g_{\m\n}$, and $r$ is implicitly defined by the relation
\bl
1 &= \frac{x^2 + y^2}{r^2 + a^2} + \frac{z^2}{r^2} \,.
\el
To simplify the expressions, we adopt the following definitions.
\bl
\bgd
x = R^1 \,,\quad y = R^2 \,,\quad z = R^3 \,,
\\ R = \sqrt{x^2 + y^2 + z^2} \,,
\\ \S = \sqrt{(x^2+y^2+z^2-a^2)^2 + 4 a^2 z^2} = \frac{r^4 + a^2 z^2}{r^2} \,.
\egd
\el
When computing the moments of the source, the following relations given in~\cite{Vines:2017hyw} will prove to be useful.
\bl
\bld
\frac{r}{\S} &= \cos (\vec{a} \cdot \vec{\nabla}) \frac{1}{R}
\\ \frac{r}{\S} \frac{\vec{R} \times \vec{a}}{r^2 + a^2} &= \sinh(\vec{a} \times \vec{\nabla}) \frac{1}{R} \,.
\eld \label{eq:sourcematching}
\el

\subsubsection{Source equation for Maxwell fields}\label{subsec:Photon_Effective_Action}
In \eqc{eq:KNBHinKSform} the vector potential $A_\m$ is given as
\bl
A_\m &= \frac{Q r}{\S} k_\m
\el
which does not satisfy the Lorenz gauge condition $\nabla^\m A_\m = 0$. The gauge connection $A'_\m dx^\m$ that satisfies the Lorenz gauge condition can be obtained from $A_\m dx^\m$ by the following gauge transformation
\bl
\bld
A'_\m dx^\m &= A_\m dx^\m - \frac{Q}{2} d \log (r^2 + a^2) 
\\ &= \frac{Qr}{\S} \left[ dt + \frac{a}{a^2 + r^2} \left( y dx - x dy \right) \right]
\eld
\el
which, in dual representation, is
\bl
A'^\m \p_\m &= \frac{Qr}{\S} \left[ \p_t - \frac{a}{a^2 + r^2} \left( y \p_x - x \p_y \right) \right] = \left( \frac{Qr}{\S}, - \frac{Qr}{\S} \frac{\vec{R} \times \vec{a}}{r^2 + a^2} \right) \,. \label{eq:LinEM}
\el
Using \eqc{eq:sourcematching}, the linearised gauge connection in Lorenz gauge \eqc{eq:LinEM} can be written as
\bl
A'^\m \p_\m &= \left( \cos(\vec{a} \cdot \vec{\nabla}) , - \sinh (\vec{a} \times \vec{\nabla}) \right) \frac{Q}{R} \label{eq:MaxwellSol}
\el
which can be covariantised by $u^\m \p_\m = \p_t$~\cite{Vines:2017hyw}
\bl
A'^\m &= \left( u^\m \cos(a \cdot \p) + \e^{\m\n\a\b} u_\n a_\a \p_\b \frac{\sin(a \cdot \p)}{a \cdot \p} \right) \frac{Q}{R} \,. 
\el

From the above, one can determine the effective current on the world-line $j^\m$ as
\bl
\bld
j^\m &= Q \int ds \left[ u^\m \cos(a \cdot \p) + \e^{\m\n\a\b} u_\n a_\a \p_\b \frac{\sin(a \cdot \p)}{a \cdot \p} \right] \delta^4 \left[ x - x_{\text{wl}}(s) \right] 
\\ &= Q \int ds \sum_{n=0}^{\infty} \left[ u^\m \frac{\left( - (a \cdot \p)^2 \right)^n}{(2n)!} + \e^{\m\n\a\b} u_\n a_\a \p_\b \frac{\left( - (a \cdot \p)^2 \right)^n}{(2n+1)!} \right] \delta^4 \left[ x - x_{\text{wl}}(s) \right]\,. 
\eld \label{eq:KNBH_source}
\el
This implies that the world-line couples through the background gauge field as
\bl
S_{int} &= - 4 \pi \int d^4 x A_\m j^\m \,.
\el
 The derivatives on the source can be replaced by derivatives on the gauge field $A_\m$ through integration by parts.
\bl
\bld
S_{int} &= - \int d^4 x ~4 \pi Q \int ds ~\delta^4 \left[ x - x_{\text{wl}}(s) \right]
\\ &\phantom{=asdfasdfasdf} \times \sum_{n=0}^{\infty} \left[ u^\m \frac{\left( - (a \cdot \p)^2 \right)^n}{(2n)!} - \e^{\m\n\a\b} u_\n a_\a \p_\b \frac{\left( - (a \cdot \p)^2 \right)^n}{(2n+1)!} \right] A_\m (x) 
\eld \label{eq:3ptIntVert}
\el
Indeed taking $a\rightarrow 0$ one recovers the minimal electromagnetic coupling in flat space. This expression will later be promoted to one-particle effective action coupled to the Maxwell field, which will be used to determine the on-shell three-point amplitude as in section~\ref{sec:3pt}.

\subsubsection{Stress tensor of Kerr-Newman solution}
The Kerr-Newman BH sources electromagnetic fields, and generated fields contribute to the stress tensor. This contribution is given by the electromagnetic stress tensor, which is defined as
\bl
T_{\m\n} &= - \frac{1}{4 \pi} F_{\m\l} F_{\n}^{~\l} + \frac{1}{16 \pi} \eta_{\m\n} F_{\a\b} F^{\a\b} \,. \label{eq:EMstress}
\el
To compute this quantity, we need to determine the electromagnetic fields that the BH has generated. The computations are easier with vector calculus for 3d Euclidean space, so we will fix the frame to be the rest frame of the BH where BH is at the origin. From \eqc{eq:MaxwellSol} we may deduce the electric and magnetic fields as follows.
\bl
\bld
\vec{E} &= - \vec\nabla \cos(\vec{a} \cdot \vec{\nabla}) \frac{Q}{R}
\\ \vec{B} &= - \vec\nabla \times \sinh (\vec{a} \times \vec{\nabla}) \frac{Q}{R}
\eld
\el
Since $\frac{1}{R}$ is harmonic, $\vec\nabla \times ( \vec{a} \times \vec{\nabla})$ can be considered as $- \vec\nabla (\vec{a} \cdot \vec{\nabla})$ in the above expression. Therefore an equivalent representation for the magnetic field is
\bl
\vec{B} &= \vec\nabla \sin(\vec{a} \cdot \vec{\nabla}) \frac{Q}{R}
\el
This motivates the following definition of holomorphic(anti-holomorphic) electromagnetic fields $\vec{H}$($\vec{\bar{H}}$);
\bl
\bld
\vec{H} &= \vec{E} - i \vec{B} = - \vec\nabla e^{i \vec{a} \cdot \vec{\nabla}} \frac{Q}{R} = - \vec\nabla \frac{Q}{\sqrt{x^2 + y^2 + (z+ia)^2}} = - \vec\nabla Q f(R)
\\ \vec{\bar{H}} &= \vec{E} + i \vec{B} = - \vec\nabla e^{-i \vec{a} \cdot \vec{\nabla}} \frac{Q}{R} = - \vec\nabla \frac{Q}{\sqrt{x^2 + y^2 + (z-ia)^2}} = - \vec\nabla Q \bar{f}(R)
\eld \label{eq:holoEMF}
\el
The funtions $f(R)$ and $\bar{f}(R)$ are harmonic. The components of the stress tensor \eqc{eq:EMstress} are determined from electromagnetic fields \eqc{eq:holoEMF} as follows.
\bl
\bld
T_{00} &= \frac{E^2 + B^2}{8 \pi} = \frac{H \cdot \bar{H} }{8 \pi} = Q^2 \frac{x^2 + y^2 + z^2 + a^2}{8 \pi \S^3}
\\ T_{0i} &= - \frac{1}{4 \pi} \left( \vec{E} \times \vec{B} \right)^i = - \frac{1}{8 \pi i} \left( \vec{H} \times \vec{\bar{H}} \right)^i = - Q^2 \frac{ \left( \vec{a} \times \vec{R} \right)^i }{4 \pi \S^3}
\\ T_{ij} &= \frac{- E^i E^j - B^i B^j + \delta_{ij} (E^2 + B^2) }{8 \pi} = \frac{- H^i \bar{H}^j - \bar{H}^i H^j + \delta_{ij} H \cdot \bar{H}}{8 \pi}
\\ &= Q^2 \left(\frac{- R^i R^j - a^i a^j}{4 \pi \S^3} + \delta_{ij} \frac{x^2 + y^2 + z^2 + a^2 }{8 \pi \S^3} \right)
\eld \label{eq:KNEMstress}
\el
Conversion to momentum space is defined as Fourier transform in 3-space;
\begin{align*} \tilde{f}(\vec{q}) = \int f(\vec{r}) e^{- i \vec{q} \cdot \vec{r}} d^3r \,. \end{align*} 
In momentum space, each component takes the following form.
\bl
\bld
T_{00} &= - \frac{Q^2 \pi}{8} q J_0 (\vec{a} \times \vec{q})
\\ T_{0i} &= - \frac{i Q^2 \pi}{8} q \left[ J_1 (\vec{a} \times \vec{q}) \right]^i = - \frac{i Q^2 \pi}{8} q (\vec{a} \times \vec{q})^i \left[ \frac{J_1 (\vec{a} \times \vec{q}) }{\vec{a} \times \vec{q}} \right]
\\ T_{ij} &= \frac{Q^2 \pi}{8} q \left[ (\vec{a} \times \vec{q})^i (\vec{a} \times \vec{q})^j \right] \left[ \frac{J_2 (\vec{a} \times \vec{q})}{(\vec{a} \times \vec{q})^2} \right] + \frac{Q^2 \pi}{8} \frac{q^i q^j - q^2 \delta_{ij}}{q} \left[ \frac{J_1 (\vec{a} \times \vec{q}) }{\vec{a} \times \vec{q}} \right]
\eld
\el
The functions $\frac{J_n (\vec{a} \times \vec{q})}{(\vec{a} \times \vec{q})^n}$ are obtained by substituting powers of $x^2$ in $\frac{J_n (x)}{x^n}$ by powers of $(\vec{a} \times \vec{q})^2$, where $J_n(x)$ are Bessel's functions of the first kind. In the above expression $q^2$ should be understood as $(\vec{q})^2$. To covariantise the expression, let us introduce the following definitions.
\bg\label{EDef}
u^\m = \frac{P^\m}{m} = (1, \vec{0}) \,,\quad E^\m = - \frac{1}{m^2} \e^{\m\n\l\s}P_\n S_\l q_\s = (0, \vec{a} \times \vec{q}) \,.
\eg
This sets
\bl
\bld \label{eq:Stress_Tensor_Covariant_Form}
\frac{8 \, T_{\m\n}}{Q^2 \pi \sqrt{-q^2}} &= - u_\m u_\n \left[ J_0 (\vec{a} \times \vec{q}) + \frac{J_1 (\vec{a} \times \vec{q}) }{\vec{a} \times \vec{q}} \right] + E_\m E_\n \left[ \frac{J_2 (\vec{a} \times \vec{q})}{(\vec{a} \times \vec{q})^2} \right]
\\ &\phantom{=asdfasdf} + \left( \frac{q_\m q_\n - q^2 \eta_{\m \n}}{-q^2} + 2 i u_{(\m} E_{\n)} \right) \left[ \frac{J_1 (\vec{a} \times \vec{q}) }{\vec{a} \times \vec{q}} \right]
\eld
\el
where $q^2 = \eta_{\m\n} q^\m q^\n = - ( \vec{q} )^2$. Note that the functions $\frac{J_n (\vec{a} \times \vec{q})}{(\vec{a} \times \vec{q})^n}$ can be expressed covariantly by modified Bessel's functions; $\frac{J_n (\vec{a} \times \vec{q})}{(\vec{a} \times \vec{q})^n} = \frac{I_n (E^\m) }{(E^\m)^n}$.

\subsection{Quantum field theory computations} \label{sec:QFT_STFF}
We now move on to evaluating the corresponding QFT computation of \eqc{eq:Stress_Tensor_Covariant_Form}; the stress tensor form factor \eqc{eq:QFT_ST}. The stress tensor form factor can be computed as the amplitude with \emph{massive} external legs~\cite{Arkani-Hamed:2017jhn}, where the mass for spin-2 particle corresponding to $T_{\m\n}(q)$ is given as $m^2 = q^2$. This ``amplitude'' can be expanded on scalar triangle and bubble integrals, and the classical part that corresponds to \eqc{eq:Stress_Tensor_Covariant_Form} is given as triangle integral contributions. Generalised unitarity reviewed in section~\ref{sec:GenUnit} can be used to compute the coefficient of the triangle integral from the triple-cut \eqc{eq:On-shell_amplitude}.

First we determine the three-point amplitude for Kerr-Newman BHs to couple to photons, by promoting the source equation \eqc{eq:KNBH_source} to interaction terms in the action \eqc{eq:3ptIntVert}. The resulting three-point amplitude turns out to be minimal coupling, which has been argued in ref.\cite{Arkani-Hamed:2019ymq} based on double copy relations; because Kerr BHs couple minimally to gravitons~\cite{Chung:2018kqs,Guevara:2018wpp,Guevara:2019fsj,Arkani-Hamed:2019ymq,Aoude:2020onz}, its ``single copy''\textemdash which also is minimal coupling\textemdash would be the amplitude for Kerr-Newman BHs coupling to photons.

Next, we evaluate the stress tensor form factor corresponding to the classical result \eqc{eq:Stress_Tensor_Covariant_Form} from the following three-particle cut, which is given by the product of two electric minimal coupling and one stress-tensor form factor with photons as external states:
\eq\label{eq:On-shell_amplitude}
\vcenter{\hbox{\includegraphics[scale=0.45]{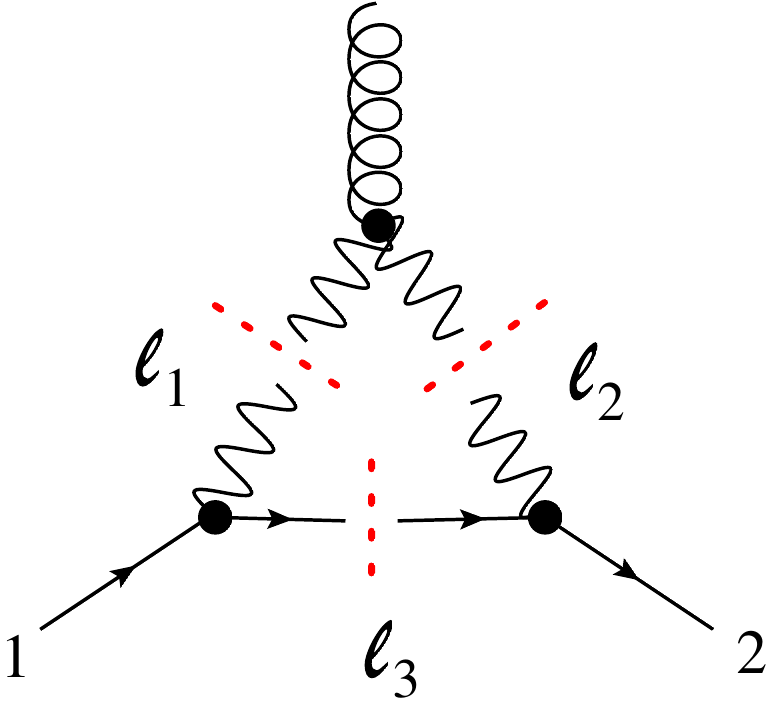}}}\;= \bra{\gamma_{\ell_1}} T_{\m\n} (q) \ket{\gamma_{\ell_2}}\otimes A_3(1^s\ell_1\ell^s_3)\otimes A_3(\ell^s_3\ell_22^s)\,,
\eqe
where each of these three-point coupling are uniquely determined kinematically. To simplify the computations, we take the holomorphic classical limit (HCL)~\cite{Guevara:2017csg} and the classical-spin limit directly on the cut. As a result, the two minimal couplings exponentiate, as discussed in~\cite{Arkani-Hamed:2017jhn}, and the extraction of triangle coefficient is greatly simplified.

The cut will be computed using the following parametrisation for the form factor, which was inspired from the classical result \eqc{eq:Stress_Tensor_Covariant_Form}.
\bl
\bra{p_2} T_{\m\n} (q) \ket{p_1} =\frac{1}{\sqrt{4 E_1 E_2}} \left[  F_1 P_{\mu}P_{\nu} + 2 F_2 P_{(\mu}E_{\nu)} + F_3(q_{\mu}q_{\nu} - \eta_{\mu\nu} q^2) + F_4 E_{\mu}E_{\nu}\right] \,. \label{eq:STFF_basis}
\el
This basis has the advantage that it is applicable to any spins, unlike the original computations for spin-$\frac{1}{2}$ and spin-$1$ particles~\cite{Donoghue:2001qc,Holstein:2006ud} where the bases were constructed on a case-by-case basis.

\subsubsection{The on-shell three-point amplitude}
The interaction term \eqc{eq:3ptIntVert} can be promoted to one-particle effective worldline action by promoting $a^\m$ to mass-scaled spin operators. Following the same steps as in section~\ref{sec:ppEFT2Amp}, we arrive at the epxression below for the three-point amplitude,
\bl
M_s^{\eta} = \e^\ast(\bf{2})^{\{\dot{\beta}_s\} \{\beta_s\}} \left[ \frac{4 \pi Q x^{\eta} }{\sqrt{2}} \sum_{n=0}^{\infty} \frac{1}{n!} \left( - \eta \frac{q \cdot S}{m} \right)^n \right]_{\{\beta_s\} \{\dot{\beta}_s\}}^{\{\dot{\alpha}_s\}\{\alpha_s\}} \e(\bf{1})_{\{\alpha_s\}\{\dot{\alpha}_s\}} \label{eq:EM3ptEFT}
\el
which has a very similar structure to EFT graviton three-point amplitude \eqc{eq:1bd3ptAmp1} with Wilson coefficients $C_{\text{S}^n} = 1$. Taking into account the Hilbert space matching effects considered in section~\ref{sec:HilbertMatch}, it is sensible to understand the above expression as minimal coupling \eqc{eq:mincoup_def}, i.e.,
\bg
M^{+1}_s = \frac{4 \pi Q x}{\sqrt{2}} \frac{\la \bf{21} \ra^{2s}}{m^{2s-1}} \,,\quad M^{-1}_s = \frac{4 \pi Q}{\sqrt{2} x} \frac{[ \bf{21} ]^{2s}}{m^{2s-1}} \,.
\eg
In other words, Kerr-Newman BHs couple minimally to photons.

\subsubsection{Kerr-Newman stress tensor from form factors}
In this section, we begin the evaluation of the classical piece of the stress tensor via \eqref{eq:On-shell_amplitude}. As in \cite{Chung:2018kqs, Chung:2019duq, Guevara:2017csg}, the classical component only requires the calculation of the triangle coefficient, which can be captured by the triangle cut \eqref{eq:On-shell_amplitude}. It should be stressed again that $q^2 \neq 0$ to capture the form factors, so that the stress tensor is coupled to an \emph{on-shell massive spin-2 particle}:
\begin{equation}
\LAB{p_2}T_{\mu\nu}\RAB{p_1} = \frac{\Delta_{\mu\nu}(q, p_1, -p_2)}{2m}
\end{equation}
with 
\begin{equation}
\Delta_{\mu\nu}(q, p_1, -p_2) = J^{-1} \times \text{LS} \times I_{\Delta} \Big|_{q^2 \rightarrow 0}
\end{equation}
where $\text{LS}$ is defined in \eqref{eq:LS}, $I_{\Delta}$ is the scalar triangle coefficient, $J^{-1}$ is the inverse Jacobian factor from solving the delta functions in \eqref{eq:LS} and $q^2 \rightarrow 0$ indicates that we are taking the leading order in $q^2$ expansion. The overall factor of momentum conservation $\delta(p_2-p_1-q)$ has been dropped for simplicity. It is shown in \cite{Chung:2018kqs} that the $J^{-1}$ and $I_{\Delta}$ cancel with each other in leading order of $q^2$ expansion, so that 
\begin{equation}
\Delta_{\mu\nu}(q, p_1, -p_2) = \text{LS} 
\end{equation}
Here, we aim to capture the stress tensor in all orders in spins by evaluating the amplitude of finite spin particles in the classical-spin limit, which will ultimately lead to \eqref{eq:Stress_Tensor_Covariant_Form}.\footnote{Note that in principle one also needs to conduct Hilbert space matching of section~\ref{sec:HilbertMatch}. However, for three-point kinematics in the classical-spin limit its effects are irrelevant. } 

\subsubsection{Kinematical set-up}
In this paper, we follow the parametrization of momenta introduced in \cite{Guevara:2017csg}:
\begin{equation}
\begin{split}
p_1 &= \left(E, -\vec{q}/2 \right) = \RAB{\lambda}\LSB{\eta} + \RAB{\eta}\LSB{\lambda}  \\
p_2 &= \left(E, +\vec{q}/2 \right) =\beta \RAB{\lambda}\LSB{\eta} + \frac{1}{\beta}\RAB{\eta}\LSB{\lambda} + \RAB{\lambda}\SB{\lambda}
\end{split}
\end{equation}
where spinors are normalised as $\la \l \eta \ra = [\l \eta] = m$ and the momentum transfer:
\begin{equation}
q = p_2 - p_1 = \left(0, \vec{q}\right) = \RAB{\lambda}\SB{\lambda} + \CO(\beta-1) = K + \CO(\beta-1)
\end{equation}
in the {holomorphic classical limit} (HCL) introduced in \cite{Guevara:2017csg} where $q^2 = 0$ without $|\vec{q}| \rightarrow 0$. The triangle coefficient of \eqref{eq:On-shell_amplitude} is captured by the following integral:
\begin{equation}\label{eq:LS}
\begin{split}
\text{LS}
&= \frac{1}{4} \sum_{h_1,h_2 = \pm 1}\int_{\Gamma_{LS}} d^4 L \delta(L^2-m^2) \delta(k_3^2) \delta(k_4^2) \\
&\qquad \qquad \qquad \times M_3(q, -k_3^{h_1}, -k_4^{h_2})\times M_3(1^s, k_3^{-h_1}, -L)\times M_3(-2^s, k_4^{-h_2}, L)\\
&=\sum_{h_1, h_2}\frac{\beta}{16(\beta^2-1)m^2} \int_{\Gamma_{LS}} \frac{dy}{y} M_3(q, -k_3^{h_1}, -k_4^{h_2})\times M_3(1^s, k_3^{-h_1}, -L^s)
\\ &\phantom{=asdfasdfasdfasdf} \times M_3(-2^s, k_4^{-h_2}, L^s)
\end{split}
\end{equation}
where the loop momenta $k_3$ and $k_4$ are given as
\begin{equation}
\begin{split}
k_3 &=
\frac{1}{\beta + 1} \left( \RAB{\eta}(\beta^2-1)y - \frac{1}{y} \RAB{\lambda}(1+\beta y) \right)
\\ &\phantom{=asdfasdfasdf} \times
\frac{-1}{\beta + 1} \left( \LSB{\eta}(\beta^2-1)y + \LSB{\lambda}(1+\beta y) \vphantom{\frac{1}{y}}\right) \,, \\
k_4 &=
\frac{1}{\beta + 1} \left( \frac{1}{\beta} \RAB{\eta}(\beta^2-1) + \frac{1}{y} \RAB{\lambda}(1 - y) \right)
\\ &\phantom{=asdfasdfasdf} \times
\frac{-1}{\beta + 1} \left( - \beta \LSB{\eta}(\beta^2-1)y + \LSB{\lambda}(1 - \beta^2 y) \vphantom{\frac{1}{y}}\right) \,.
\end{split}
\end{equation}
It will be convenient to define following vectors as a basis for expanding other vectors and tensors.
\bg
\bgd
\mathcal{U}^{\mu} = \frac{1}{2}\MixLeft{\eta}{\sigma^{\mu}}{\lambda} \,,\, v^{\mu} = \frac{1}{2}\MixLeft{\lambda}{\sigma^{\mu}}{\eta} \,, \\
K^{\mu} = \frac{1}{2}\MixLeft{\lambda}{\sigma^{\mu}}{\lambda} \,,\, R^{\mu} = \frac{1}{2}\MixLeft{\eta}{\sigma^{\mu}}{\eta} \,.
\egd \label{eq:UvKR_basis}
\eg

The two three-point amplitudes corresponding to Kerr-Newman BHs coupling to photons are minimally coupling three-point amplitudes as pointed out in section~\ref{subsec:Photon_Effective_Action}.
\begin{equation}
\begin{split}
M_3(-2^s, k_4^{\eta}, L^s) &=
x_2^{\eta} \oint \frac{dz}{2\pi i z} \left( \sum_{n=0}^\infty  z^n \right) \left( \frac{ [ \bf{2L} ] - \la \bf{2L} \ra }{2m} + \frac{\eta}{z} \frac{ \sbra{\bf{2}} k_4 \ket{\bf{L}} + \bra{\bf{2}} k_4 \sket{\bf{L}} }{4m^2} \right)^{2s} \\
M_3(1^s, k_3^{\eta}, -L^s) &=
x_1^{\eta} \oint \frac{dz}{2\pi i z} \left( \sum_{n=0}^\infty z^n \right) \left( \frac{ [ \bf{L1} ] - \la \bf{L1} \ra }{2m} + \frac{\eta}{z} \frac{ \sbra{\bf{L}} k_3 \ket{\bf{1}} + \bra{\bf{L}} k_3 \sket{\bf{1}} }{4m^2} \right)^{2s}
\end{split}
\end{equation}
where $\eta$ is the helicity of the massless particle and the $x$-factors are given by
\begin{equation}
x_4 = \frac{\sbra{k_4} L \ket{\z_1} }{\la k_4 \z_1 \ra m} \,, \quad 
x_3 = \frac{ \sbra{k_3} p_1 \ket{\z_2} }{\la k_3 \z_2 \ra m } \,.
\end{equation}
The three-point amplitudes will take an exponential form in the classical-spin limit~\cite{Arkani-Hamed:2019ymq,Chung:2019duq}, and this property will also be shown to hold for the triangle cut.

The amplitude $M_3(q, -k_3^{h_1}, -k_4^{h_2})$ is uniquely fixed by three-point kinematics~\cite{Arkani-Hamed:2017jhn}:
\begin{equation}
M_3(q,-k_3^{+1},-k_4^{-1}) = - \frac{\lambda_{k_4}^4\SB{k_3 k_4}^{2}}{M_{pl} M^2} \,, \quad 
M_3(q,-k_3^{-1},-k_4^{+1}) = - \frac{\lambda_{k_3}^4\SB{k_3 k_4}^{2}}{M_{pl} M^2} \,.
\end{equation}
where $M^2=q^2$. For our purposes, we want these amplitudes to be in the symmetric basis:
\begin{equation}
\begin{split}
M_3^{\text{sym}}(q,-k_3^{+1},-k_4^{-1}) 
&= - \frac{\lambda_{k_4, \a}\lambda_{k_4, \b} \lambda_{k_4}^{\g}\lambda_{k_4}^{\delta} \SB{k_3 k_4}^{2}}{M_{pl} M^2}\frac{q_{\gamma\dot{\a}}}{M}\frac{q_{\delta\dot{\b}}}{M}
\\&= - \frac{\lambda_{k_4, \a} \tilde{\lambda}_{k_3,\dot{\a}} \lambda_{k_4, \b} \tilde{\lambda}_{k_3,\dot{\b}} }{M_{pl}}\\
&\rightarrow - \frac{\MixLeft{k_4}{\s_\m}{k_3}\MixLeft{k_4}{\s_\n}{k_3}}{M_{pl}} \,,
\\
M_3^{\text{sym}}(q,-k_3^{-1},-k_4^{+1}) 
&= - \frac{\lambda_{k_3, \a}\lambda_{k_3, \b} \lambda_{k_3}^{\g}\lambda_{k_3}^{\delta}\SB{k_3 k_4}^{2}}{M_{pl} M^2} \frac{q_{\gamma\dot{\a}}}{M}\frac{q_{\delta\dot{\b}}}{M}
\\ &= -\frac{ \lambda_{k_3, \a}\tilde{\lambda}_{k_4,\dot{\a}} \lambda_{k_3, \b} \tilde{\lambda}_{k_4,\dot{\b}} }{M_{pl}}\\
&\rightarrow - \frac{\MixLeft{k_3}{\s_\m}{k_4}\MixLeft{k_3}{\s_\n}{k_4}}{M_{pl}} \,.
\end{split}
\end{equation}

\subsubsection{Evaluating the integrand}
Combining all the intermediate results, the integrand can be separated into a spin-independent part and a spin-dependent part where wanted terms are picked out from the residue integral.
\begin{equation}
(M_3)^3 
= ( M_3^{s=0} )^3 \oint \frac{dz_1}{2 \pi i z_1} \frac{dz_2}{2 \pi i z_2} \left( \sum_{n=0}^\infty  z_1^n \right) \left( \sum_{n=0}^\infty  z_2^n \right) F (- \bf{2}, k_4, k_3, \bf{1})^{2s}\\ 
\end{equation}
The spin-independent part is given as
\begin{equation}
\begin{split}
( M_3^{s=0} )^3 &= m^2 \alpha_q^2 \left( \frac{\sbra{k_4} L \ket{\z_1} }{\la k_4 \z_1 \ra m} \right)^{\eta_1} \left( \frac{ \sbra{k_3} p_1 \ket{\z_2} }{\la k_3 \z_2 \ra m } \right)^{\eta_2} M_3 (- k_4^{- \eta_1}, - k_3^{- \eta_2}, \bf{q})\\
\end{split}
\end{equation}
with 
\begin{equation}
\begin{split}
M_3 (- k_4^{+}, - k_3^{-}, \bf{q}) &= - \frac{\bra{k_3} \s_\m \sket{k_4} \bra{k_3} \s_\n \sket{k_4}}{M_{pl}}\\
M_3 (- k_4^{-}, - k_3^{+}, \bf{q}) &= - \frac{\bra{k_4} \s_\m \sket{k_3} \bra{k_4} \s_\n \sket{k_3}}{M_{pl}}
\end{split}
\end{equation}
The residue integrand for the spin-dependent part is
\begin{equation}
\begin{split}
F (- \bf{2}, k_4, k_3, \bf{1}) &= \left( \frac{ [ \bf{2L} ] - \la \bf{2L} \ra }{2m} + \frac{\eta_1}{z_1} \frac{ \sbra{\bf{2}} k_4 \ket{\bf{L}} + \bra{\bf{2}} k_4 \sket{\bf{L}} }{4m^2} \right)
\\ &\phantom{=asdf} \times \left( \frac{ [ \bf{L1} ] - \la \bf{L1} \ra }{2m} + \frac{\eta_2}{z_2} \frac{ \sbra{\bf{L}} k_3 \ket{\bf{1}} + \bra{\bf{L}} k_3 \sket{\bf{1}} }{4m^2} \right)\\
&= \bar{u}(2) u(1) - \left( \frac{\eta_1}{z_1} k_4^\m + \frac{\eta_2}{z_2} k_3^\m \right) S^\m_{1/2}
\\ &\phantom{=a}
 + \frac{k_3^\m}{2m} \left( \frac{p_1^\m + p_2^\m}{2m} \bar{u}(2) u(1) - \frac{q^\n}{2m} \bar{u}(2) \g_{\m\n} u(1) \right)
\\ &\phantom{=a} - \left( \frac{\eta_1}{z_1} + \frac{\eta_2}{z_2} \right) \left( \frac{q^2}{8m^2} \bar{u}(2) \g^5 u(1) + \frac{k_4^\m k_3^\n}{4m^2} \bar{u}(2) \g_{\m\n} \g^5 u(1) \right)
\\ &\phantom{=a} - \frac{\eta_1 \eta_2}{z_1 z_2} \left( \frac{q^2}{8m^2} \bar{u}(2) u(1) + \frac{k_4^\m k_3^\n}{4m^2} \bar{u}(2) \g_{\m\n} u(1) \right)
\end{split}
\end{equation}
\noindent
where $\g^5 = i \g^0 \g^1 \g^2 \g^3$ and $\g^{\m\n} = \half [\g^\m, \g^\n]$. The definition for spin-$\half$ spin vector $S^\m_{1/2}$ is given as
\begin{equation}\label{eq:Spin vec def}
\begin{split}
\bar{u}(p_2) u(p_1) &= \frac{ [ \bf{21} ] - \la \bf{21} \ra }{2m}\\
S_{1/2}^\m &= \half \bar{u}(p_2) \g^\m \g_5 u(p_1) = - \frac{1}{4m} \left( \sbra{\bf{2}} \bar{\s}^\m \ket{\bf{1}} + \bra{\bf{2}} \s^\m \sket{\bf{1}} \right)
\end{split}
\end{equation}
In the HCL only the first two terms of $F$ survive:
\begin{equation}
\begin{split}
F 
&\stackrel{HCL}{\to} 1 - \left( \frac{\eta_1}{z_1} k_4^\m + \frac{\eta_2}{z_2} k_3^\m \right) \frac{S_{1/2}^\m }{m} \\
&= 1 - \eta \left( \frac{k_4^\m}{z_1}  - \frac{k_3^\m}{z_2}  \right) \frac{S_{1/2}^\m }{m} 
= 1 + \eta \left[ \frac{(y-1)^2}{4 z_1 y} + \frac{(y+1)^2}{4 z_2 y} \right] \frac{K \cdot S_{1/2}}{m} + \CO(\b-1)
\end{split}
\end{equation}
where $\eta = \eta_1$ is defined as the helicity of the $k_4$ photon on the 2 massive amplitude leg. Taking the classical-spin limit $s \to \infty$ and adopting the definition $\frac{1}{2s} S^\m = S^\m_{1/2}$, the $F$ term can be written as
\bl
\lim_{s\rightarrow \infty}F^{2s} = \exp{\left\lbrace \eta \left[ \frac{(y-1)^2}{4 z_1 y} + \frac{(y+1)^2}{4 z_2 y} \right] \frac{K \cdot S}{m} \right\rbrace} \equiv \exp\left[\eta f\times (a\cdot q)\right]
\el
where 
\begin{equation}
f \equiv \frac{(y-1)^2}{4 z_1 y} + \frac{(y+1)^2}{4 z_2 y}
\end{equation}
Summarising, the integrand becomes
\bl\label{eq:Partial Integrand}
(M_3)^3 &\stackrel{HCL}{\to} ( M_3^{s=0} )^3 \oint dZ \exp\left[\eta f\times (a\cdot q)\right]
\el
where we adopted the definition
\bl
\oint dZ \equiv \oint \frac{dz_1}{2 \pi i z_1} \frac{dz_2}{2 \pi i z_2} \left( \sum_{n=0}^\infty  z_1^n \right) \left( \sum_{n=0}^\infty z_2^n \right)
\el
for simplicity.

\subsubsection{Comparison with classical computations}
Based on the basis \eqc{eq:STFF_basis} for the stress tensor form factor, we evaluate the following quantity.
\begin{equation}
\Delta_{\mu\nu} = 2m\LAB{p_2}T_{\mu \nu} \RAB{p_1} = F_1 P_{\mu}P_{\nu} + 2 F_2 P_{(\mu}E_{\nu)} + F_3(q_{\mu}q_{\nu} - \eta_{\mu\nu} q^2) + F_4 E_{\mu}E_{\nu}
\end{equation}
We can extract the $F_1$, $F_2$, $F_3$, $F_4$ coefficients by contracting the integrand with different vectors or tensors to get a set of linear combinations of these four coefficients. By solving the equations, we show that the solution is consistent with the HCL version of \eqref{eq:Stress_Tensor_Covariant_Form} where $(\vec{a} \times \vec{q})^2 = |\vec{a}|^2|\vec{q}|^2 - (\vec{a} \cdot \vec{q})^2 \stackrel{\text{HCL}}{\rightarrow} - (\vec{a} \cdot \vec{q})^2$. In this limit, the Bessel's function expressions of \eqref{eq:Stress_Tensor_Covariant_Form} are converted to modified Bessel's function expressions.
\bl
\frac{J_n (\vec{a} \times \vec{q})}{(\vec{a} \times \vec{q})^n} \stackrel{\text{HCL}}{\to} \frac{I_n (a \cdot q)}{(a \cdot q)^n}
\el

The kinematical set-up we are using implies the conditions
\begin{equation}
P\cdot q = P \cdot E = 0
\end{equation}
so that we will be solving the following system of linear equations:
\begin{equation}\label{eq:Linear equations}
\begin{split}
\Delta_{\mu\nu}P^\m P^\n &= m^4 F_1 - m^2 q^2 F_3 \\
\Delta_{\mu\nu}P^\m E^\n &= m^2 (\vec{a}\cdot \vec{q})^2 F_2 \\
\Delta_{\mu\nu}E^\m E^\n &= (\vec{a} \cdot \vec{q})^2 \left(F_4 (\vec{a}\cdot\vec{q})^2 - q^2 F_3\right) \\
\Delta_{\mu\nu} \eta^{\mu\nu} &= m^2 F_1 - 3q^2 F_3 + (\vec{a} \cdot \vec{q})^2 F_4 = 0
\end{split}
\end{equation}
where the last line is the traceless condition of the stress tensor. In the following sections, we explicitly show the computations of $\Delta_{\mu\nu}P^\m P^\n$, $\Delta_{\mu\nu}E^\m P^\n$ and $\Delta_{\mu\nu}E^\m E^\n$.

\paragraph{Computing $\Delta_{\m\n} P^\m P^\n$}$\phantom{123}$\newline
Contracting the stress tensor with $P_{\mu}P_{\nu}$  can help us read out  out the terms proportional to $P_\m P_\n$ and the $\eta_{\mu\nu}q^2$ part
\begin{equation}\label{eq:AngPSqr}
\begin{split}
\bra{k_3} P \sket{k_4} &= -( \b - 1 ) m^2\\
\bra{k_4} P \sket{k_3} &= -\frac{\b - 1}{\b} m^2
\end{split}
\end{equation}
where $P = (1-\a) P_3 + \a P_4$. Note that the above results are independent of $\a$\footnote{Overall factor of $\frac{\alpha_q^2}{M_{pl}}$ temporarily suppressed}. Therefore in the HCL the integrand will take the form
\begin{equation}
\begin{split}
(M_3)^3 
= 
2 m^6 (\b - 1)^2 \oint dZ \cosh \left[f\times (a\cdot q)\right] + \CO(\beta-1)^3
\end{split}
\end{equation}
To compare with the classical computation \eqref{eq:Stress_Tensor_Covariant_Form}, we need to expand the integrand as a series in $\left(a \cdot q \right)^{2}$:
\begin{equation}
\begin{split}
\Delta_{\mu\nu}P^{\mu}P^{\nu} 
&= \frac{\b}{16m^2 (\beta^2 - 1)} \int\frac{dy}{y} (M_3)^{3} 
\\ &= \frac{(\b-1)  \b m^4}{8 (\b+1)}  \sum_{n=0}^{\infty} \int \frac{dy}{y}  	\oint dZ \frac{f^{2n}}{(2n)!}\left(a \cdot q \right)^{2n}\\
&= - \frac{(\b-1) \b m^4}{8 (\b+1)} \sum_{n=0} \frac{1}{2^{2n}(n!)^2} \left(a \cdot q \right)^{2n}
\\&= - \frac{(\b-1) \b m^4}{8 (\b+1)} I_0 (a \cdot q)
\end{split}
\end{equation}
where $I_0$ is the modified Bessel's functions.

\paragraph{Computing $\Delta_{\m\n} P^\m E^\n$}$\phantom{123}$\newline
The vector $E_{\mu}$ can be spanned using the $\mathcal{U}_{\mu}$, $v_{\mu}$, $K_{\mu}$, $R_{\mu}$ basis \eqc{eq:UvKR_basis}.
\begin{equation}
E^{\mu} = \frac{i}{m^2} \left[\left(\mathcal{U}^{\mu} - v^{\mu} + \frac{\b^2-1}{\b}R^{\mu} \right)(K\cdot S) - K^{\mu} S \cdot \left( u-v + \frac{\b^2-1}{\b}R \right)\right]
\end{equation}
Contracting $E_{\mu}$ with $\MixLeft{k_4}{\sigma^{\mu}}{k_3}$ and $\MixLeft{k_3}{\sigma^{\mu}}{k_4}$ yields
\begin{equation}\label{eq:AngESqr}
\MixLeft{k_4}{E}{k_3} =  \frac{i  \left(y^2+1\right)}{2 y} (q\cdot S) (\beta -1) + \CO(\beta-1)^2 = -\MixLeft{k_3}{E}{k_4} 
\end{equation}
Combining with \eqref{eq:AngPSqr}, the integrand  in the HCL will take the form
\begin{equation}
\begin{split}
(M_3)^3 
&=
-\frac{i  m^5 \left(y^2+1\right)(q\cdot a)}{y}  (\beta -1)^2  \oint dZ\sinh \left[ f \times (q\cdot a) \right]
\end{split}
\end{equation}
Evaluating the $y$, $z_1$ and $z_2$ integrals then:
\begin{equation}
\begin{split}
\Delta_{\mu\nu}P^{\mu}E^{\nu} 
& = \frac{\b}{16m^2(\beta^2-1)}\int \frac{dy}{y}(M_3)^3  \\
&= \frac{-i  m^3 \beta (\beta -1)}{16(\beta + 1)} \int \frac{dy}{y}\frac{ \left(y^2+1\right)}{y}   \sum_{n=0}^{\infty} \oint dZ  \frac{f^{2n+1}}{(2n+1)!}(q\cdot a)^{2n+2}\\
&= \frac{im^3 \b (\b - 1)}{8(\b + 1)} \sum_{n=0}^{\infty} \frac{ (a\cdot q)^{2n+2} }{2^{2n+1}n!(n+1)!}
= \frac{im^3 \b (\b - 1)}{8(\b + 1)} (a\cdot q)I_1 (a \cdot q)
\end{split}
\end{equation}
Solving for $F_2$ yields
\begin{equation}
F_2 
= \frac{i m \b (\b - 1)}{8(\b + 1)} \sum_{n=0}^{\infty} \frac{(a\cdot q)^{2n}}{2^{2n+1}n!(n+1)!}  
= \frac{i m \b (\b - 1)}{8(\b + 1)} \frac{I_1 (a \cdot q)}{a \cdot q}
\end{equation}

\paragraph{Computing $\Delta_{\m\n} E^\m E^\n$}$\phantom{123}$\newline
From \eqref{eq:AngESqr}
\begin{equation}
\MixLeft{k_4}{E}{k_3}^2 = \MixLeft{k_3}{E}{k_4}^2  = -\frac{ \left(y^2+1\right)^2}{4 y^2} (q\cdot S)^2 (\beta -1)^2 + \CO(\beta-1)^3
\end{equation}
The integral becomes
\begin{equation}
\begin{split}
\Delta_{\m\n} E^\m E^\n 
& = \frac{\b}{16m^2(\beta^2-1)}\int \frac{dy}{y}(M_3)^3  \\
&= - \frac{\beta(\beta-1)m^2}{32(\beta+1)}(q \cdot a)^2 \int \frac{dy}{y} \frac{\left(y^2+1\right)^2}{ y^2}  \oint dZ\cosh\left[f\times (a\cdot q)\right]\\
&=  \frac{\beta(\beta-1)m^2}{16(\beta+1)} \sum_{n=0}^{\infty} \frac{2n+1}{2^{2n}n!(n+1)!}(a \cdot q)^{2n+2}\\
&= \frac{\beta(\beta-1)m^2}{8(\beta+1)} \left[ (a\cdot q)^2 I_2(a\cdot q) + (a\cdot q) I_1(a\cdot q)\right]
\end{split}
\end{equation}

\paragraph{Solving for the coefficients $F_1$, $F_2$, $F_3$, $F_4$}$\phantom{123}$\newline
The solution for \eqref{eq:Linear equations} is 
\begin{itemize}
\item {}$F_1$
\begin{equation}
\begin{split}
F_1 &=  \frac{1}{m^2}\frac{\Delta_{\mu\nu}E^{\mu}E^{\nu} }{(a\cdot q)^2}+ \frac{2}{m^4} \Delta_{\mu\nu}P^{\mu}P^{\nu}
\\ &= -\frac{1}{m^2} \frac{(\beta - 1)\beta m^2 }{8(\beta+1)}  \left[ \frac{I_1(a\cdot q)}{(a\cdot q)} + I_0(a\cdot q) \right]
\end{split}
\end{equation}
\item {}$F_2$
\begin{equation}
\begin{split}
F_2 
= \frac{i m^2 \b (\b - 1)}{8(\b + 1)} \frac{1}{m} \frac{I_1 (a \cdot q)}{a \cdot q}
\end{split}
\end{equation}
\item {}$F_3$
\begin{equation}
\begin{split}
F_3 &= \frac{1}{q^2}  \left( \frac{\Delta_{\mu\nu}E^{\mu}E^{\nu}}{ (a\cdot q)^2} + \frac{\Delta_{\mu\nu}P^{\mu}P^{\nu}}{m^2}\right) 
= - \frac{1}{q^2} \frac{(\beta - 1)\beta m^2 }{8(\beta+1)} \frac{I_1(a\cdot q)}{(a\cdot q)}
\end{split}
\end{equation}
\item {}$F_4$
\begin{equation}
\begin{split}
F_4 &=  \frac{2 \Delta_{\mu\nu}E^{\mu}E^{\nu} }{(a\cdot q)^4}+ \frac{1}{m^2} \frac{\Delta_{\mu\nu}P^{\mu}P^{\nu}}{(a\cdot q)^2} = \frac{(\beta - 1)\beta m^2 }{8(\beta+1)} \frac{I_2(a\cdot q)}{(a \cdot q)^2}
\end{split}
\end{equation}

\end{itemize}
In summary, the form factor from calculating $\Delta_{\mu\nu}$ is 
\begin{equation}
\begin{split}
\LAB{p_2}T_{\mu\nu} \RAB{p_1}
=
\frac{|\vec{q}| }{32} \frac{\alpha_q^2}{M_{Pl}}
&\left\lbrace 
- \frac{P_{\mu}P_{\nu}}{m^2}\left[I_0(a \cdot q) + \frac{I_1(a \cdot q)}{(a \cdot q)} \right] + E_{\mu}E_{\nu} \left[ \frac{I_2(a \cdot q)}{(a \cdot q)^2} \right] \right.\\
& \left.
\quad + \left( \frac{q_{\mu}q_{\nu} - \eta_{\mu\nu}q^2}{-q^2} + 2i\frac{P_{(\mu}E_{\nu)}}{m} \right) \left[ \frac{I_1(a \cdot q)}{(a \cdot q)} \right]
\right\rbrace
\end{split} \label{eq:QFT_ST}
\end{equation}
which indeed reproduces the HCL version of \eqref{eq:Stress_Tensor_Covariant_Form}. The difference in overall normalisation is due to inclusion of gravitational coupling $M_{Pl}^{-1}$ and difference in units; $\a_q = \frac{e^2}{4 \pi}$ and $e = 4 \pi Q$.

\section{One-loop amplitude for 2PM spinning BH Hamiltonian at HCL} \label{sec:Amp42PM}
The 2PM effective Hamiltonian by definition scales as $G^2$. Writing the interaction Hamiltonian as $V = V^{(1)} + V^{(2)}$ where $V^{(i)}$ is the $i$-th order PM Hamiltonian, the amplitude up to $G^2$ order obtained from the Born series \eqc{eq:BornAmp} can be schematically represented as follows.
\bl
\bld
M^{(1)}_{i \to f} &= \bra{f} V^{(1)} \ket{i} \,,
\\ M^{(2)}_{i \to f} &= \bra{f} V^{(1)} G_0 V^{(1)} \ket{i} + \bra{f} V^{(2)} \ket{i} \,.
\eld \label{eq:BornAmp2PM}
\el
In the full relativistic theory, the equivalent $G^{2}$ order amplitude is given by one-loop amplitudes. This means the 2PM potential $V^{(2)}$ can be obtained as
\bl
\bra{f} V^{(2)} \ket{i} &= \CM^{\text{1-loop}}_{i \to f} - \bra{f} V^{(1)} G_0 V^{(1)} \ket{i} \,,
\el
where $\CM^{\text{1-loop}}$ is the one-loop amplitude of the full relativistic theory, following the notation of section~\ref{sec:1PM_Ham_full}. However, the full one-loop amplitude is not needed when specialising to \emph{classical} part of the effective Hamiltonian; the cut diagram topologies illustrated in fig.~\ref{fig:2PMcut} and their external leg permutations are enough~\cite{Cheung:2018wkq,Bern:2019crd}.

\bfig
\centering
\includegraphics[width=0.90\linewidth]{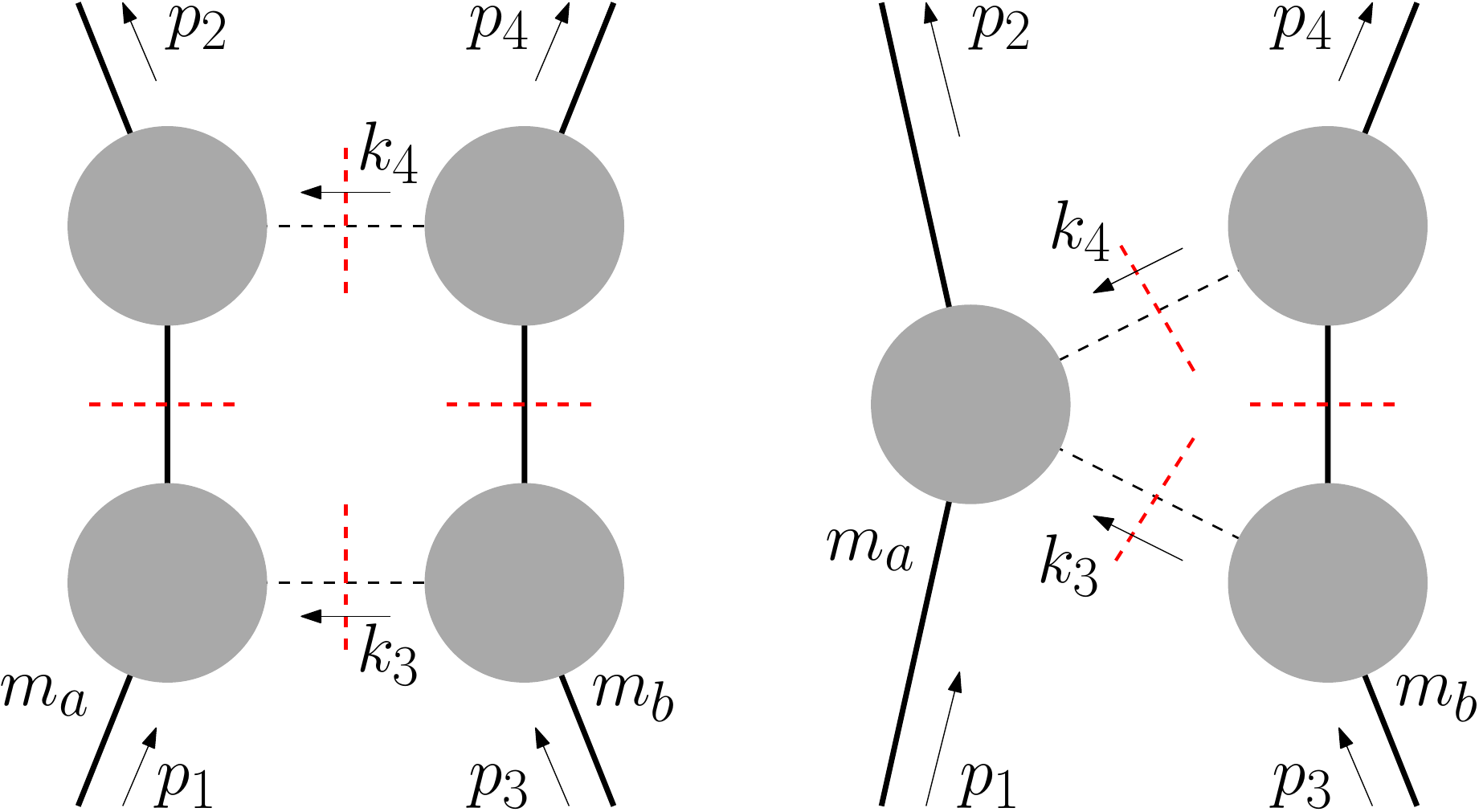}
\caption{Cut diagrams relevant for 2PM effective Hamiltonian. Thick solid lines represent massive particles while thin dashed lines represent massless particles. LHS: The quadrupole cut for the scalar box. 
The crossed box is obtained by exchanging legs $p_1$ and $p_2$. 
RHS: The $b$-topology triangle cut for the scalar triangle. 
The $a$-topology triangle cut is obtained by exchanging the labels $a \leftrightarrow b$.} \label{fig:2PMcut}
\efig

Of the two topologies, classical physics is encoded in the scalar triangle contribution and the scalar box contribution is the so-called \emph{superclassical terms} corresponding to iterated effects of tree-order classical physics~\cite{Holstein:2004dn,Neill:2013wsa,Guevara:2017csg,Bern:2019crd}. Based on these arguments, ref.\cite{Chung:2018kqs} only focused on the contribution from the scalar triangle integral to the 2PM effective Hamiltonian. However, the computations were done in HCL which could not probe certain terms that survive in the classical limit, and the computations were limited to leading PN order.

Improved understanding of the PM expansion accumulated since ref.\cite{Chung:2018kqs} allows extension of HCL computations to full 2PM order. This section will focus on the ``bare'' part of the one-loop amplitude\textemdash the part of the amplitude without the Thomas-Wigner rotation factor $U^{(a)}U^{(b)}$ of section~\ref{sec:rotation}\textemdash contributions in the HCL, which are relevant for obtaining the 2PM effective Hamiltonian. When putting back the rotation factors the factors must be appended to the \emph{left} of the bare amplitude, because operator ordering may matter at 2PM order when full $q^2$ dependence is considered~\cite{Bern:2020buy}.

Specifically, this section will present computations for the cut diagram topologies of fig.~\ref{fig:2PMcut} in terms of spin multipole moments in the HCL. Only the amplitudes will be given for reference, and extension to full $q^2$ dependence and computation of iteration terms will be addressed in a future work presenting the full 2PM spinning Hamiltonian~\cite{toappear}.
 

The definitions for the variables used in this section follow that of section~\ref{sec:1PM_Ham_full}.

\subsection{Kinematical set-up}
The spinor variable parametrisation of external momenta introduced in ref.\cite{Guevara:2017csg} is used\footnote{While it was implicitly assumed that $\b = \b'$ in \cite{Guevara:2017csg}, unless $m_a = m_b$ this does not hold true in general. However, the difference can be ignored at leading order in the HCL.} to compute one-loop contributions relevant to the 2PM effective Hamiltonian in the HCL
.
\bl
\bgd
p_1 
= \sket{\hat\eta} \bra{\hat\l} + \sket{\hat\l}\bra{\hat\eta}
\,,\, p_2 
= \b' \sket{\hat\eta} \bra{\hat\l} + \frac{1}{\b'} \sket{\hat\l}\bra{\hat\eta} + \sket{\hat\l} \bra{\hat\l}
\\ p_3 
= \sket{\eta} \bra{\l} + \sket{\l}\bra{\eta}
\,,\, p_4 
= \b \sket{\eta} \bra{\l} + \frac{1}{\b} \sket{\l}\bra{\eta} + \sket{\l} \bra{\l}
\egd
\el
The HCL is defined as the limit $\b \to 1$. The normalisation for the spinors are $\la \hat\l \hat\eta \ra = [\hat\l \hat\eta] = m_a$ and $\la \l \eta \ra = [\l \eta] = m_b$. The triangle cut conditions for the $b$-topology, given on the RHS of fig.~\ref{fig:2PMcut}, determine the internal momenta $k_3$ and $k_4$ as
\begin{equation}
\begin{split}
k_3 &=
\frac{1}{\beta + 1} \left( \RAB{\eta}(\beta^2-1)y - \frac{1}{y} \RAB{\lambda}(1+\beta y) \right)
\\ &\phantom{=asdfasdfasdf} \times
\frac{1}{\beta + 1} \left( \LSB{\eta}(\beta^2-1)y + \LSB{\lambda}(1+\beta y) \vphantom{\frac{1}{y}}\right) \,, \\
k_4 &=
\frac{1}{\beta + 1} \left( \frac{1}{\beta} \RAB{\eta}(\beta^2-1) + \frac{1}{y} \RAB{\lambda}(1 - y) \right)
\\ &\phantom{=asdfasdfasdf} \times
\frac{1}{\beta + 1} \left( - \beta \LSB{\eta}(\beta^2-1)y + \LSB{\lambda}(1 - \beta^2 y) \vphantom{\frac{1}{y}}\right) \,.
\end{split}
\end{equation}
The remaining loop momentum variable $y$ serves as the dummy variable $y$ of the residue integral \eqc{eq:tricut_coeff}. The mirror version of the computation, the $a$-topology, can be computed by exchanging the labels $a \leftrightarrow b$. Details of the triple-cut computations can be found in refs.\cite{Guevara:2017csg,Chung:2018kqs}. The box cut, given on the LHS of fig.~\ref{fig:2PMcut}, can be computed by solving the condition $p_1 \cdot k_3 = 0$ for $y$. For the crossed box cut, the condition changes to $p_1 \cdot k_4 = 0$ instead.

One simplification of the HCL is that the kinematics of the full amplitude separates into two copies of three-point kinematics. As noted in ref.\cite{Guevara:2017csg}, this allows full parametrisation of the amplitude in terms of spinor projections $\sket{\l} \sbra{\l}$ and $\sket{\hat\l} \sbra{\hat\l}$ when external states are described by anti-chiral spinors;
\bg
\CM_4 \stackrel{\text{HCL}}{\to} A(s) \CM_{3,a} \times \CM_{3,b} \,,
\\ \bld
\CM_{3,a} &= \left( \frac{\sbra{\bf{2}}}{\sqrt{m_a}} \right)^{2s_a} \left[ \iden + \hat{c}_1 \frac{\sket{\hat\l} \sbra{\hat\l}}{m_a} + \cdots \right] \left( \frac{\sket{\bf{1}}}{\sqrt{m_a}} \right)^{2s_a} \,,
\\ \CM_{3,b} &= \left( \frac{\sbra{\bf{4}}}{\sqrt{m_b}} \right)^{2s_b} \left[ \iden + c_1 \frac{\sket{\l} \sbra{\l}}{m_b} + \cdots \right] \left( \frac{\sket{\bf{3}}}{\sqrt{m_b}} \right)^{2s_b} \,.
\eld \label{eq:HCL3pt}
\eg
The above series can be understood as a series expansion in operators $q \cdot a_{a,b}$, although Hilbert space matching effects must be taken into account. The spinor projections appearing in \eqc{eq:HCL3pt} are interpreted as~\cite{Chung:2018kqs};
\bl
\bld
{2s_a \choose n} \left( \frac{\sket{\hat\l} \sbra{\hat\l}}{m_a} \right)^{n} &= \frac{2^n}{n!} \left( \frac{q \cdot S_a}{m_a} \right)^n \,,
\\ {2s_b \choose n} \left( \frac{\sket{\l} \sbra{\l}}{m_b} \right)^{n} &= \frac{(-2)^n}{n!} \left( \frac{q \cdot S_b}{m_b} \right)^n \,.
\eld
\el
After the above identification has been made, the result is augmented by boost factors \eqc{eq:MinKerr} from Hilbert space matching. Inclusion of these factors were argued as effects of rewriting in terms of polarisation tensors in ref.\cite{Chung:2018kqs}, while the same factors were argued by introduction of \emph{generalised expectation value} in refs.\cite{Guevara:2018wpp,Guevara:2019fsj}.  

\subsection{The box integral contribution at HCL}
The coefficients for the scalar box integral can be evaluated using the four-particle cut on the LHS of fig.~\ref{fig:2PMcut}. All on-shell subamplitudes appearing on the LHS of fig.~\ref{fig:2PMcut} are three-point amplitudes of a massive spinning particle coupled to a single graviton, which are fully fixed for any spin, e.g. by \eqc{eq:EFT3ptNew}. Therefore, box integral contributions can be computed to arbitrary spin multipole order. In the HCL, the coefficient for the box integral is
\bl
\bld
i \CM_{\square}^{{\rm bare}} &= 64 \pi^2 G^2 \left[ \cosh \left( 4\theta  + i \frac{\ve(q,u_a,u_b,a_0)}{\sinh\theta} \right) + \cosh \left( q \cdot a_0 \right) \right]
I_\square \,.
\eld
\el
Interestingly, the coefficients of the box integral $I_\square$ and the crossed box integral $I_{\triangleright \hskip -1pt \triangleleft}$ turns out to be the same in this limit. 
\bl
\bld
i \CM_{\triangleright \hskip -1pt \triangleleft}^{{\rm bare}} &= 64 \pi^2 G^2 \left[ \cosh \left( 4\theta  + i \frac{\ve(q,u_a,u_b,a_0)}{\sinh\theta} \right) + \cosh \left( q \cdot a_0 \right) \right]
I_{\triangleright \hskip -1pt \triangleleft} \,.
\eld
\el
As already explained, the expressions do \emph{not} include Thomas-Wigner rotation factor $U^{(a)}U^{(b)}$. The box integral is listed in the appendix of ref.\cite{Holstein:2008sw} as;
\bl
\bld
I_\square &= \frac{- i}{8 \pi^2} \frac{\log (-t)}{t} \frac{1}{\sqrt{\Lambda}} \left[ \log \abs{\frac{\sqrt{\Lambda} - (s - s_0)}{-\sqrt{\Lambda} - (s - s_0)}} + i \pi \th ( s - s_0 ) \right] \,,
\\ \Lambda &:= (s-s_0)(s - s_0 + 4 m_a m_b) \,,
\eld
\el
where the Mandelstam variables are defined as $t = q^2 = - \vec{q}^2$ and $s = (p_1 + p_3)^2$, $s_0$ is the rest energy squared $s_0 = (m_a + m_b)^2$, and $\th(x)$ is the step function. The crossed box integral $I_{\triangleright \hskip -1pt \triangleleft}$ is obtained from the box integral $I_\square$ by changing the Mandelstam variable $s$ to Mandelstam variable $u = 2 (m_a^2 + m_b^2) - s - t$.

\subsection{The triangle integral contribution at HCL}
The coefficient for the triangle integral can be computed using the three-particle cuts; the $b$-topology given on the RHS of fig.~\ref{fig:2PMcut} and the $a$-topology given by exchange of labels $a \leftrightarrow b$. The major difference from the scalar box is that one of the subamplitudes appearing in three-particle cuts is the gravitational Compton amplitude, which is only unambiguously fixed up to massive spin-2 particles. Since well-defined subamplitudes are needed for reliable results, only the results for spin orders accessible by massive spin-2 particles will be presented in this section. The resolved gravitational Compton amplitude for massive higher spins \eqc{eq:gravComp_resolv} yield the same results for the spin orders accessible by massive spin-2 particles, but polynomial term ambiguities render higher spin orders inaccessible.

The bare amplitude contribution from these two topologies in the HCL will be parametrised as
\bl
\bld
i \CM_{\bigtriangleup + \bigtriangledown}^{{\rm bare}} &= \frac{\pi^2 G^2 m_a^2 m_b^2}{\sqrt{-t}} \sum_{i,j=0}^{4} A_{i,j} \left( \frac{q \cdot S_a}{m_a} \right)^i \left( \frac{q \cdot S_b}{m_b} \right)^j \,,
\eld
\el
where Hilbert space matching effects of \eqc{eq:MinKerr} has been included. Just like the box integral result, the Thomas-Wigner rotation factor $U^{(a)}U^{(b)}$ 
is not included in the above expression. Since $A_{j,i}$ can be obtained from $A_{i,j}$ by exchanging labels $a \leftrightarrow b$, only the coefficients $A_{i,j}$ for $i \le j$ will be presented. The spin-independent and spin-linear part are given as;
\bl
A_{0,0} &= 3 (5 \cosh (2 \theta )+3) \left(m_a+m_b\right) \,,
\\ A_{0,1} &= \frac{(3 \cosh (\theta )+5 \cosh (3 \theta )) \left(3 m_a+4 m_b\right)}{2 \sinh (\th)} \,.
\el
A crucial difference between the two expressions is that the factor $\sinh (\th)$ appears in the denominator for $A_{0,1}$. This factor vanishes in the static limit $E_a + E_b \to m_a + m_b$, therefore $A_{0,1}$ diverges in this limit. This implies that the triangle coefficient contains terms that correspond to second order Born series. Ref.\cite{Chung:2018kqs} presented the results by interpreting removal of poles as removal of second order Born series, but the prescription will not be adopted here and the full expression will be presented.

Another interesting observation is that \emph{all} coefficients related to spin dependence, $A_{i,j}$ with $i+j \ge 1$, contains a $\sinh^{2} (\th)$ factor in the denominator. While the divergent factor seems to be $\frac{1}{\sinh (\th)}$ for $i+j$ odd this is an illusion, as the Lorentz-invariant combination of variables $\e(u_a,u_b,q,a_{a,b}) \propto \sinh (\th) ( q \cdot a_{a,b})$ \emph{must} appear at odd powers in spin\footnote{This condition is forced by requiring the effective Hamiltonian to be parity-even.}. Further $A_{i,j}$ coefficients accessible by massive spin-2 particles will be listed below.

The spin-quadratic part are given as;
\bl
A_{0,2} &= \frac{\left(m_a (60 \cosh (4 \theta )+4)+m_b (-28 \cosh (2\theta )+95 \cosh (4 \theta )-3)\right)}{32 \sinh^2 (\th)} \,,
\\ A_{1,1} &= - \frac{(\cosh (2 \theta )-5 \cosh (4 \theta )) \left(m_a+m_b\right)}{\sinh^2 (\th)} \,.
\el
The spin-cubic part are given as;
\bl
A_{0,3} &= \frac{\left(m_a (5 \cosh (2 \theta )+1)+m_b (9 \cosh (2 \theta )-1)\right) \cosh (\th)}{2 \sinh (\th)} \,,
\\ A_{1,2} &= \frac{\left(80 m_a \cosh (2 \theta )+m_b (95 \cosh (2 \theta )-7)\right) \cosh (\th)}{8 \sinh(\th)} \,.
\el
The spin-quartic part are given as;
\bl
A_{0,4} &= \frac{\left(96 m_a (5 \cosh (2 \theta )-3) \cosh ^2(\theta )+m_b (-44 \cosh (2 \theta )+239 \cosh (4 \theta )-3)\right)}{768 \sinh^2 (\th)} \,,
\\ A_{1,3} &= \frac{\left(4 m_a (\cosh (2 \theta )+5 \cosh (4 \theta ))-3m_b (\cosh (2 \theta )-9 \cosh (4 \theta ))\right)}{24 \sinh^2 (\th)} \,,
\\ A_{2,2} &= \frac{ (95 \cosh (4 \theta )+1) \left(m_a+m_b\right)}{64 \sinh^2 (\th)} \,.
\el
Some higher spin order results are also available, which is listed below.
\bl
A_{2,3} &= \frac{\cosh (\theta ) \left(m_a (95 \cosh (2 \theta )+7)+108 m_b \cosh (2 \theta )\right)}{48 \sinh (\th)} \,,
\\ A_{2,4} &= \frac{2 m_a (28 \cosh (2 \theta )+95 \cosh (4 \theta )-3)+m_b (239 \cosh (4 \theta )+1)}{1536 \sinh^2 (\th)} \,,
\\ A_{3,3} &= \frac{(\cosh (2 \theta )+9 \cosh (4 \theta )) \left(m_a+m_b\right)}{48 \sinh^2 (\th)} \,,
\\ A_{3,4} &= \frac{\cosh (\theta ) \left(24 m_a (9 \cosh (2 \theta )+1)+m_b (239 \cosh (2 \theta )+11)\right)}{1152 \sinh (\th)} \,,
\\ A_{4,4} &= \frac{(44 \cosh (2 \theta )+239 \cosh (4 \theta )-3) \left(m_a+m_b\right)}{18432 \sinh^2 (\th)} \,.
\el

\section{Discussions and outlook for one-loop order}
In section~\ref{sec:1PM_Ham_full}, it was shown that scattering amplitudes of massive higher-spin particles at small momentum transfer can be used to construct the 1PM effective Hamiltonian involving all-orders-in-spin effects \eqc{V-1PM-master}. The aim of this chapter was to show conceptual obstructions for applying the same tools to one-loop amplitudes, specialising to classical physics of black holes.

From the gravitational Compton amplitude of minimally coupled massive spinning particles, the main argument of section~\ref{sec:GravComp} was that massive particles with spins $s>2$ cannot be regarded as elementary; the gravitational Compton amplitude cannot be extended beyond $s=2$. The conclusion raises two questions; a) is there an eligible expression for gravitational Compton amplitude of minimally coupled massive higher-spin particles? b) are loop computations valid when non-elementary massive higher-spin particles run inside the loops? The former has been answered in the affirmative, although the resolution comes at a price of worse high-energy behaviour and uncontrolled polynomial terms.

As an attempt to answer the latter, section~\ref{sec:KNBHMST} looked for a all-orders-in-spin computation that can be done both in classical and quantum field theories. The stress tensor of the ambient electromagnetic field generated by a Kerr-Newman black hole turned out to be the wanted example; classical field theory computations are readily available, and quantum field theory computations turned out to be computation of stress tensor form factors which could be computed from available three-point amplitudes of massive higher-spin particles. Although one working example is never enough to constitute a proof, its existence does suggest that this is a valid approach.

Section~\ref{sec:Amp42PM} was given as a reference for future works targeting the 2PM effective Hamiltonian to all-orders-in-spin. The coefficients for the box and triangle integrals of the one-loop amplitude, which are the integrals that contribute to classical physics of black hole scattering, were computed at the HCL to full PM order.



\appendix
\chapter{Conventions}
The conventions used in this dissertation are summarised in this appendix. The conventions for perturbative gravity are;
\bl
\k &= \sqrt{32 \pi G} = 2 / M_{Pl}
\\ g_{\m\n} &= \eta_{\m\n} + \k h_{\m\n}
\el

\section{Invariant tensors}
Conventions for invariant tensors
\bl
\eta_{\m\n} &= \text{diag}(+1, -1, -1, -1)
\\ \e^{0123} &= +1
\\ \e^{ab} &= -\e_{ab} \,, \, \e^{01} = + 1
\\ (\s^\m)_{\a \dot{\b}} &= (\iden, \vec{\s})
\\ (\bar{\s}^\m)^{\dot{\a} {\b}} &= (\iden, -\vec{\s})
\el
The indices $ab$ of $\e^{ab}$ stands for chiral indices $\a$, anti-chiral indices $\dot\a$, and little group indices $I$. Usual definitions for the Pauli matrices $\vec{\s}$ are adopted.
\bl
\vec{\s} = \left( \left( \begin{array}{cc} 0 & 1 \\ 1 & 0 \end{array} \right), \left( \begin{array}{cc} 0 & -i \\ i & 0 \end{array} \right), \left( \begin{array}{cc} 1 & 0 \\ 0 & -1 \end{array} \right) \right)
\el
Algebraic relations involving $\s$ tensors
\bl
\bar{\s}^{\m \dot{\a} \a} &= \e^{\a \b} \e^{\dot{\a} \dot{\b}} \s^\m_{\b \dot{\b}}
\\ \s^\m_{\a \dot{\a}} \s_{\m \b \dot{\b}} &= 2 \e_{\a \b} \e_{\dot{\a} \dot{\b}}
\\ \left( \s^\m \bar{\s}^{\n} + \s^\n \bar{\s}^{\m} \right)_{\a}^{~\b} &= 2 \eta^{\m\n} \delta_{\a}^{~\b}
\el

\section{Analytic continuation and complex conjugation}
Conventions for analytic continuation
\bl
\ket{-p}_{\a} &= \ket{p}_{\a}
\\ \sket{-p}^{\dot{\a}} &= - \sket{p}^{\dot{\a}}
\el
Conventions for Lorentz transformations and Lorentz invariance
\bl
\bld
\L^{\m}_{~\n} &\equiv \left[ e^{- \frac{i}{2} \e_{\m\n} J^{\m\n}} \right]^{\m}_{~\n}
\\ g_{\a}^{~\b} &\equiv \left[ e^{- \frac{i}{2} \e_{\m\n} J^{\m\n}} \right]_{\a}^{~\b}
\\ \s^\m_{\a \dot{\a}} &= \L(g)^\m_{~\n} g_{\a}^{~\b} \s^\n_{\b\dot\b} \left( g^\dagger \right)^{\dot\b}_{~\dot\a}
\eld
\el
Relations among massless spinor-helicity variables ($ \l_{\a} := \ket{p}_{\a} $ and $ \bar{\l}^{\dot{\a}} := \sket{p}^{\dot{\a}}$)
\bl
\l^{\a} &:= \bra{p}^{\a} = \e^{\a \b} \ket{p}_{\b}
\\ \bar{\l}_{\dot{\a}} &:= \sbra{p}_{\dot{\a}} = \e_{\dot{\a} \dot{\b}} \sket{p}^{\dot{\b}}
\\ \la \l \m \ra &:= \l^\a \m_\a = \bra{\l}^\a \ket{\m}_\a
\\ [ \bar{\l} \bar{\m} ] &:= \bar{\l}_{\dot{\a}} \bar{\m}^{\dot{\a}} = \sbra{\bar\l}_{\dot\a} \sket{\bar\m}^{\dot\a}
\\ ( p \cdot \s)_{\a \dot{\b}} &= \ket{p}_{\a} \sbra{p}_{\dot{\b}}
\\ ( p \cdot \bar{\s} )^{\dot{\a} {\b}} &= \sket{p}^{\dot{\a}} \bra{p}^{\b}
\el
Relations among massive spinor-helicity variables
\bl
\l_{\a}^{~I} \bar{\l}_{\dot{\b} I} &= - \l_{\a I} \bar{\l}_{\dot{\b}}^{~I} = ( p \cdot \s)_{\a \dot{\b}}
\\ \bar{\l}^{\dot{\a} I} \l^{\b}_{~I} &= - \bar{\l}^{\dot{\a}}_{~I} \l^{\b I} = - ( p \cdot \bar{\s} )^{\dot{\a} {\b}}
\\ \l_{\a I} &= \e_{IJ} \l_{\a}^{~J}
\\ \bar{\l}^{\dot{\a}}_{~I} &= \e_{IJ} \bar{\l}^{\dot{\a} J}
\\ \left( (p \cdot \bar{\s}) \l^I \right)^{\dot{\a}} &= m \bar{\l}^{\dot{\a} I}
\\ \left( (p \cdot {\s}) \bar{\l}^I \right)_{{\a}} &= m {\l}_{\a}^{~I}
\el
Complex conjugation relations for massless spinors
\bg
(\l_{\a})^\ast = \text{sgn}(p^0) \bar{\l}_{\dot{\a}}
\eg
Complex conjugation relations for massive spinors
\bg
(\l_{\a}^{~I})^\ast = \text{sgn}(p^0) \bar{\l}_{\dot{\a} I}, ~(\l^{\a I})^\ast = \text{sgn}(p^0) \bar{\l}^{\dot{\a}}_{~I}
\\ (\l_{\a I})^\ast = - \text{sgn}(p^0) \bar{\l}_{\dot{\a}}^{~I}, ~ (\l^{\a}_{~I})^\ast = - \text{sgn}(p^0) \bar{\l}^{\dot{\a} I}
\eg

\section{Numerical implementations and the high-energy limit} \label{sec:HE}
The lowercase roman indices $a$, $b$, $c$, etc., and round brackets $\fbra{p}$ and $\fket{p}$ will be reserved for numerical relations. Define following normalised unit vectors as follows.
\bl
\fket{{n} +} &:= n^+_a = \left( \begin{gathered} \cos \frac{\th}{2} \\ e^{i \phi} \sin \frac{\th}{2} \end{gathered} \right)
\\ \fket{{n} -} &:= n^-_a = \left( \begin{gathered} - e^{-i \phi} \sin \frac{\th}{2} \\ \cos \frac{\th}{2} \end{gathered} \right)
\el
$n$ denotes a unit vector in $\IR^3$, with $\th$ and $\phi$ respectively denoting its polar and azimuthal angles in polar coordinates. These vectors are solutions to the following eigenvalue problem.
\bl
\vec{n} \cdot \vec{\s} \, \fket{n\pm} = \pm \fket{n\pm}
\el
The vectors $\fket{n}$ and covectors $\fbra{n}$ are related by Hermitian conjugation. Lowering and raising indices by the $\e$ tensor gives the following relation.
\bl
n^{\pm a} := \e^{ab} n^{\pm}_b = \mp \left( \fbra{n \mp} \right)^a
\el

Setting $n$ to be parallel to the 3-momentum $\vec{p}$, the numerical values of contraction between 4-momentum and $\s$ tensors can be evaluated as follows.
\bl
(p \cdot \s)_{ab} &= E - \vec{p} \cdot \vec{\s} = (E - p) n^+_a n^+_b + (E + p) n^-_a n^-_b
\\ (p \cdot \bar{\s})_{ab} &= E + \vec{p} \cdot \vec{\s} = (E + p) n^+_a n^+_b + (E - p) n^-_a n^-_b
\el
Computation of square roots is straightforward, assuming $E > 0$.
\bl
\sqrt{p \cdot \s} &= \frac{m}{\sqrt{E + p}} \, \fket{n+} \fbra{n+} + \sqrt{E + p} \, \fket{n-} \fbra{n-}
\\ \sqrt{p \cdot \bar{\s}} &= \sqrt{E + p} \, \fket{n+} \fbra{n+} + \frac{m}{\sqrt{E + p}} \, \fket{n-} \fbra{n-}
\el
The relation $ E - p = m^2 / (E + p)$ was used to make the high-energy limit manifest. The helicity basis is defined by the relations\footnote{$0 \leftrightarrow +$ and $1 \leftrightarrow -$ will be used interchangably for $I$ indices in this basis.}
\bl
(\l^\pm)_\a &= \sqrt{E \mp p} \, \fket{n\pm} \label{eq:shvarnumdef1}
\\ (\bar{\l}^\pm)^{\hat{\a}} &= \sqrt{E \pm p} \, \fket{n\pm} \label{eq:shvarnumdef2}
\el
which manifestly shows that spinors $\l^-$ and $\bar{\l}^+$ survive in this limit, while the spinors $\l^+$ and $\bar{\l}^-$ vanish.

\section{Dirac spinors}
Massive spinor-helicity variables are mapped to Dirac spinors by the relations
\bl
u^I (p) := \left( \begin{gathered} \l_{\a}^{~I} \\ \bar{\l}^{\dot{\a} I} \end{gathered} \right), ~ v^I (p) := \left( \begin{gathered} \l_{\a}^{~I} \\ - \bar{\l}^{\dot{\a} I} \end{gathered} \right) \,.
\el
An outgoing anti-particle is associated with the spinor wavefunction $v$, and flipping its momentum will continue it to an incoming particle associated with the spinor wavefunction $u$. The choice for analytic continuation $\bar{\l}(-p)^{\dot\a} = - \bar{\l}(p)^{\dot\a}$ follows from this reasoning\footnote{An alternative sign choice results when basing the reasoning on spinor wavefunctions $\bar{u}$ and $\bar{v}$, however.}. Expressions for conjugate Dirac spinors follow from the definition $\bar{\psi} := \psi^\dagger \g^0$.
\bl
\bar{u}_I (p) := \left(
\begin{aligned} - \l^{\a}_{~I} && \bar{\l}_{\dot{\a} I} \end{aligned}
\right), ~ \bar{v}_I (p) := \left(
\begin{aligned} \l^{\a}_{~I} && \bar{\l}_{\dot{\a} I} \end{aligned}
\right) \,.
\el
The above relations will \emph{define} conjugate Dirac spinors for complex momenta, since spinors and their conjugates are independent variables for complex momenta.

\section{Polarisation vectors}
The following definition for the polarisation vector has been adopted.
\bl 
e^{IJ,\m}(p) &:= \frac{1}{\sqrt{2} m} \bra{p^{\{ I}} \s^\m \sket{p^{J\} }} = \frac{1}{2 \sqrt{2} m} \left( \bra{p^{I}} \s^\m \sket{p^{J}} + \bra{p^{J}} \s^\m \sket{p^{I}} \right) \,. \label{eq:polvecdef}
\el
Due to complex conjugation relations, its complex conjugate is;
\bl 
\left( e^{IJ,\m}(p) \right)^\ast &:= - \frac{1}{\sqrt{2} m} \bra{p_{\{ I}} \s^\m \sket{p_{J\} }} 
\,.
\el
Similar to conjugate Dirac spinors, the above relation will \emph{define} complex-conjugated polarisation vectors for complex values of momenta for the same reasons. They are orthonormal in the sense that  
\bl
\varepsilon^{IJ} \cdot (\varepsilon^{KL})^\ast = - \half(\delta^{I}_{K} \delta^{J}_{L} + \delta^{I}_{L} \delta^{J}_{K}) \,,
\quad
\sum_{I,J} \varepsilon^{IJ}_\m (\varepsilon^{IJ}_\n)^* = - \left( \eta_{\m\n} - \frac{p_\m p_\n}{m^2} \right) \,.
\el
The massless limit is reached by parametrising vanishing spinors $(\l^+)_\a$ and $(\bar{\l}^-)^{\dot{\a}}$ as $m (\z^-)_\a$ and $m (\bar{\z}^+)^{\dot{\a}}$, together with substituting $m$ in the denominator as $m \la \l \z \ra$ or $m [\l \z]$. Ultimately, the definitions take the following form.
\bl
\e^{+}_{\m}(k) &:= \frac{\sbra{k} \bar{\s}_\m \ket{\z} }{\sqrt{2} \la k \z \ra}
\\ \e^{-}_{\m}(k) &:= \frac{\bra{k} {\s}_\m \sket{\z} }{\sqrt{2} [ k \z ]}
\el
Upper index denotes helicity, while $\z$ parametrises the gauge redundancy of the polarisation vector. Alternatively, polarisation vectors can be expressed in the following form.
\bl
\e^{+}(k) &= \sqrt{2} \frac{\sket{k} \bra{\z} }{\la k \z \ra}
\\ \e^{-}(k) &= \sqrt{2} \frac{\ket{k} \sbra{\z} }{[ k \z ]}
\el
The polarisation vectors satisfy the normalisation $\e^{\pm} \cdot (\e^{\pm})^\ast = -1$ and $\e^{\pm} \cdot (\e^{\mp})^\ast = 0$.

\section{BOLD notation}
For a fixed massive particle, the $SU(2)$ Little group is always completely symmetrized. 
Ref.~\cite{Arkani-Hamed:2017jhn} introduced the \bf{BOLD} notation, which suppresses the $SU(2)$ little group indices by means of an auxiliary parameter for each particle. 
For instance, for particle 1, 
\bl
(\l_1)_\a{}^I (t_1)_I  = \ket{1^I} (t_1)_I = \ket{\bf{1}} \,, 
\\
(\tilde{\l}_1)^{\dot{\a}I} (t_1)_I  = \sket{1^I} (t_1)_I = \sket{\bf{1}} \,. 
\el
It is clear how to reinstate the SU(2) index if needed.

\bibliographystyle{JHEP}
\bibliography{mybib}

\keywordalt{산란진폭, 포스트-뉴턴 전개, 포스트-민코프스키 전개}
\begin{abstractalt}
LIGO나 VIRGO와 같이 밀집쌍성계(compact binary coalescence)를 주요 중력파원으로 삼는 중력파 검출기에서는 효과적인 중력파 검출을 위해 다양한 이론적 도구를 필요로 한다. 그러한 이론적 도구 중 하나는 충분한 거리를 두고 공전하고 있는 쌍성계의 동역학에 대한 해석적 묘사로, 포스트-뉴턴 전개(post-Newtonian expansion)란 이름으로 널리 알려져 있다. 이러한 이체 문제의 유효 해밀토니안을 구하는 방법 중 하나는 양자역학적 산란진폭을 이용하는 것이다.

이 학위논문은 이체 문제의 유효 해밀토니안을 구하는 문제에 응용될 수 있도록 양자역학적 산란진폭에 포함된 회전하는 물체의 고전물리 중 스핀에 의해 생성된 고차 다중극(multipole moment) 효과의 이해를 목표로 한다. 이 논문에서는 임의의 회전하는 밀집성에 대해 스핀의 모든 차수 효과를 고려한 일차 포스트-민코프스키 전개 (뉴턴상수 $G$에 대해 선형이며 상대운동량 $p^2$에 대해 모든 차수를 포함) 유효 해밀토니안을 제시한다. 다음으로 이 방법론을 이차 포스트-민코프스키 전개로 확장할 때의 장애와 전망에 대해 동등한 차수의 양자장론 계산을 통해 논한다.

이 학위논문은 \cite{Chung:2018kqs,Chung:2019duq,Chung:2019yfs,Chung:2020rrz}의 논문을 바탕으로 한다.

\end{abstractalt}

\acknowledgement

가장 먼저 학위과정동안 격려와 지원을 아끼지 않아주신 가족에게 감사를 표합니다. 또한 학위과정동안 방황할 때 지도와 조언을 아끼지 않아주신 지도교수님과 김석 교수님께 감사를 표합니다. 여러분께 받은 도움이 없었다면 학위과정을 무사히 마치지 못했을 것입니다. 다음으로 산란진폭으로부터 블랙홀의 고전역학을 구하는 문제를 제안해 주신 National Taiwan University의 Yu-Tin Huang 교수님께 감사를 표합니다. 이 제안이 없었더라면 학위논문 주제를 찾기 위한 방황이 더욱 길어졌을 것입니다. 학위과정동안 생활에 걱정이 없도록 지원을 아끼지 않아주신 관정 이종환 교육재단 관계자 여러분께도 깊은 감사를 표합니다.

논문을 쓰는 동안 여러 기관에 방문할 기회가 있었습니다. National Taiwan University를 방문할 때마다 많은 도움을 받았던 Ming-Zhi Chung에게 감사를 표합니다. 또한 Niels Bohr Institutet에 초청해주신 Mich\`ele Levi에게 감사를 표합니다. 그 곳에서의 논의가 없었더라면 제 포스트-뉴턴 전개에 대한 이해는 보다 피상적인 상태로 남아있었을 것입니다. KIAS에서 제 논문 주제에 대해 알릴 기회를 제공해주신 송재원 교수님과 여러 학회에서 발표할 기회를 제공해주신 김성수 교수님께도 감사를 표합니다. KIAS에 방문할 때마다 조언을 아끼지 않아주신 이기명 교수님과 이필진 교수님, 여러가지 주제에 대해 흥미로운 논의를 나눠주신 윤정기 박사님, 그리고 APCTP에 방문할 때마다 제 연구에 관심을 가져주신 김낙우 교수님과 강동민 박사님께 감사를 표합니다.

이 논문은 Nima Arkani-Hamed, Rafael Porto, Jan Steinhoff, Justin Vines, Kays Haddard, Andrea Cristofoli와의 논의가 없었더라면 많은 부분이 완성되지 못했을 것입니다. 또한 심사위원으로서 졸업논문 주제에 대해 익숙하지 않은 사람들이 가질 수 있는 의문들에 대해 의견을 내어주신 이원종 교수님과 김형도 교수님, 그리고 졸업논문 초고를 감수해 주시고 조언을 아끼지 않아주신 강궁원 교수님께 감사를 표합니다.

마지막으로 학위과정동안 동고동락한 연구실 동료들\textemdash Prarit Agarwal 박사님, 영빈, 동욱, 준, 기홍, 선진, 호진, 승엽, 윤석, 승연\textemdash과 경희대 및 포항공대의 동료들\textemdash 세진, 명보, 예인, 태환, 성준, 형주\textemdash, 술 생각이 날 때 같이 잔을 기울여준 친구들\textemdash 준규, 다운, 진호, 기혁, 봉규, 수빈, 진혁, 홍준, 선재, 황영, 현, 경연, 기영, 희태, 연범, 우현\textemdash, 그리고 같은 취미를 공유하며 웃을 수 있었던 동호회 여러분과 멀리서도 익명으로 격려를 아끼지 않아주신 여러분께 감사를 표합니다.

\end{document}